\documentclass[onecolumn,amsmath,amssymb,nofootinbib,12pt]{article}
\pdfoutput=1 
\usepackage{jheppub} 
\usepackage{amsmath,amssymb,braket}

\DeclareMathOperator{\arcsinh}{arcsinh}

\DeclareMathOperator{\arcsech}{arcsech}

\usepackage{graphicx,hyperref,color}
\usepackage[subrefformat=parens]{subcaption}
\usepackage{here}
\usepackage[hang,small,bf]{}
\hypersetup{
    bookmarks=true,         
    unicode=false,          
    pdftoolbar=true,        
    pdfmenubar=true,        
    pdffitwindow=false,     
    pdfstartview={FitH},    
    pdfnewwindow=true,      
    colorlinks=true,       
    linkcolor=blue,          
    citecolor=blue,        
    filecolor=blue,      
    urlcolor=blue           
}

\usepackage{graphicx,subcaption}
\graphicspath{ {figures/} }

 \usepackage{multirow}

\newcommand{\de}{\partial}
\newcommand{\be}{\begin{equation}}
\newcommand{\ba}{\begin{eqnarray}}
\newcommand{\ea}{\end{eqnarray}}
\newcommand{\ee}{\end{equation}}

\newcommand{\s}{\sqrt}

\newcommand{\ti}{\tilde}

\newcommand{\no}{\nonumber \\}
\newcommand{\la}{\langle}
\newcommand{\lb}{\rangle}
\newcommand{\bea}{\begin{eqnarray}}
\newcommand{\eea}{\end{eqnarray}}
\newcommand{\bes}{\begin{equation*}}
\newcommand{\beas}{\begin{eqnarray*}}
\newcommand{\eeas}{\end{eqnarray*}}
\newcommand{\bas}{\begin{array*}}
\newcommand{\eas}{\end{array*}}
\newcommand{\ees}{\end{equation*}}

\newcommand{\ep}{\epsilon}

\newcommand{\eg}{{\it e.g.,}\ }
\newcommand{\ie}{{\it i.e.,}\ }

\newcommand{\mt}[1]{\textrm{\tiny #1}}
\renewcommand{\(}{\left(}
\renewcommand{\)}{\right)}

\newcommand{\CO}{{\cal O}} 
\newcommand{\uend}{{u_{\mathrm{end}}}} 
\newcommand{\vend}{{v_{\mathrm{end}}}} 

 \def\CM{{\cal M}}

\title{ Zoo of holographic moving mirrors}

\author[a]{Ibrahim Akal,}
\author[a]{Taishi Kawamoto,}
\author[a]{Shan-Ming Ruan,}
\author[a,b,c]{Tadashi Takayanagi}
\author[a]{and Zixia Wei}

\affiliation[a]{Center for Gravitational Physics and Quantum Information,
	Yukawa Institute for Theoretical Physics,
	Kyoto University,\\
	Kitashirakawa Oiwakecho, Sakyo-ku, Kyoto 606-8502, Japan}

\affiliation[b]{Inamori Research Institute for Science,\\
	620 Suiginya-cho, Shimogyo-ku,
	Kyoto 600-8411 Japan}
\affiliation[c]{Kavli Institute for the Physics and Mathematics of the Universe (WPI),
	University of Tokyo,\\
	Kashiwa, Chiba 277-8582, Japan}

\emailAdd{ibrakal@yukawa.kyoto-u.ac.jp}
\emailAdd{taishi.kawamoto@yukawa.kyoto-u.ac.jp}
\emailAdd{ruan.shanming@yukawa.kyoto-u.ac.jp}
\emailAdd{takayana@yukawa.kyoto-u.ac.jp}
\emailAdd{zixia.wei@yukawa.kyoto-u.ac.jp}

\abstract{We systematically study moving mirror models in two-dimensional conformal field theory (CFT). By focusing on their late-time behavior, we separate the mirror profiles into four classes, named type A (timelike) mirrors, type B (escaping) mirrors, type C (chasing) mirrors, and type D (terminated) mirrors. We analytically explore the characteristic features of the energy flux and entanglement entropy for each type and work out their physical interpretation. Moreover, we construct their gravity duals for which end-of-the-world (EOW) branes play a crucial role. Depending on the mirror type, the profiles of the EOW branes show distinct behaviors. In addition, we also provide a criterion that decides whether the replica method in CFTs computes entanglement entropy or pseudo entropy in moving mirror models.}

\begin{document} 
	
	\begin{flushright}
		YITP-22-42
		\\
		IPMU22-0023
		\\
	\end{flushright}

	\maketitle
	\flushbottom

\section{Introduction}

\begin{figure}[t!]
	\centering		
	\includegraphics[width=0.35\textwidth]{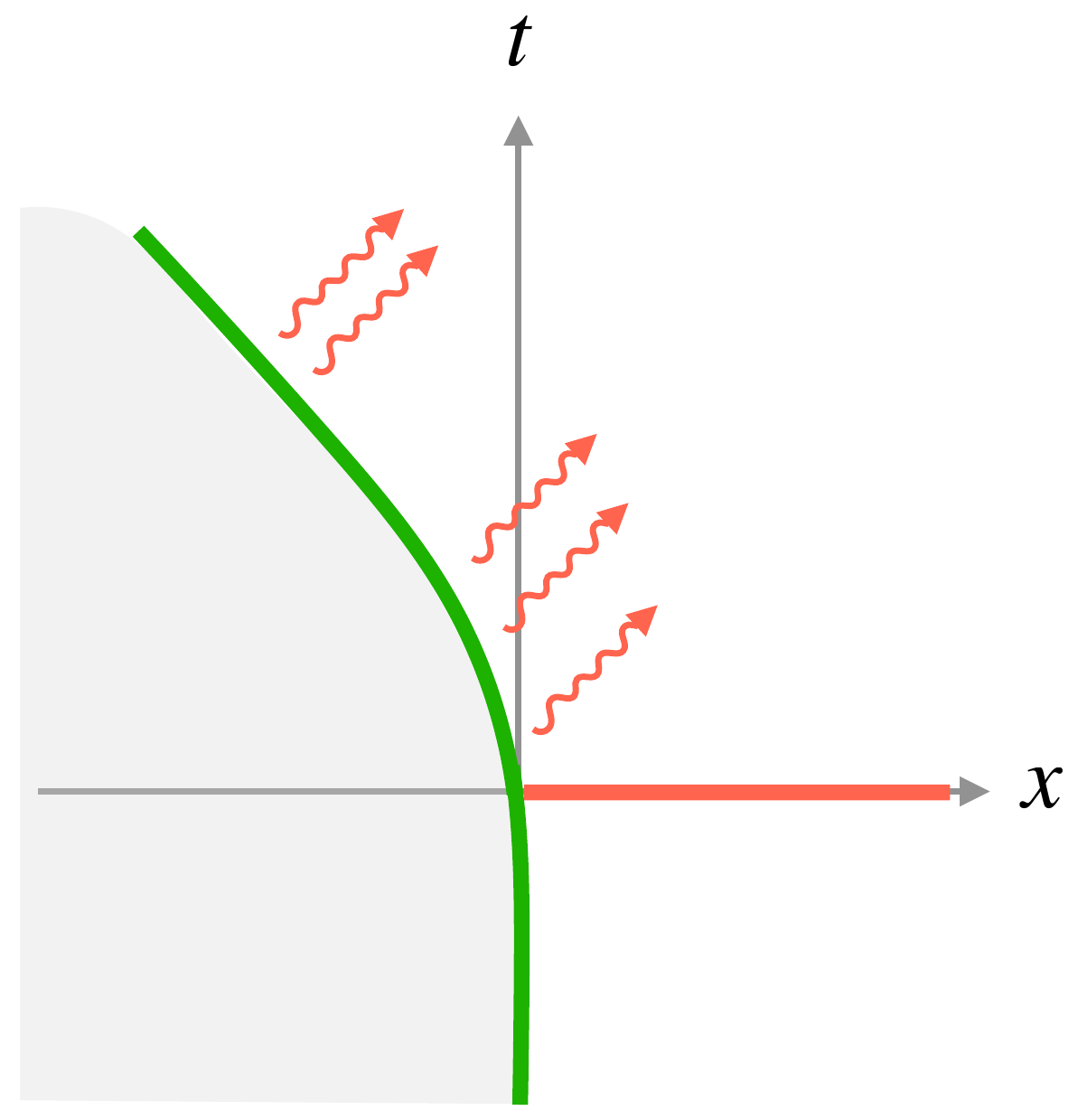}
	\caption{A moving mirror (green curve), being the boundary of the quantum system (red line) under consideration, accelerates to the left while approaching the null line. This generates radiation in form of right moving quanta (red, wiggly arrows).}
	\label{fig:MM}
\end{figure}

Moving mirrors \cite{Davies:1976hi,Birrell:1982ix} have led to interesting and tractable classes of time-dependent backgrounds in quantum field theories (QFTs). A moving mirror model is described by a QFT defined on a spacetime $\Sigma$ with a time-dependent boundary $\de \Sigma$, which is identified with the mirror trajectory as depicted in Fig.~\ref{fig:MM}. When conformal field theories (CFTs) are considered, we can impose boundary conformal invariance on the mirror trajectory, leading to a boundary CFT (BCFT) \cite{Cardy:2004hm}. Until now, lots of efforts have particularly been made for applying moving mirrors to modeling Hawking radiation emanating from black holes \cite{Hawking:1974sw}, see e.g. \cite{Davies:1977yv,Ford:1982ct,Carlitz:1986nh,Wilczek:1993jn,Raval:1996vt,Bianchi:2014qua,Hotta:2015huj,Hotta:2015yla,Chen:2017lum,Good:2019tnf,Akal:2020twv,Akal:2021foz,Reyes:2021npy}. 
However, moving mirrors have also been subject to studies of the (dynamical) Casimir effect \cite{Jaekel:1997hr,Plunien:1999ba,Romualdo:2019eur,Brevik:2000zb}, cosmological expansion \cite{Casadio:2002dj,Good:2020byh,Cotler:2022weg}, entanglement harvesting \cite{Cong:2018vqx,Cong:2020nec}, as well as quantum energy inequalities \cite{Fewster:2004nj}. Experimental setups involving \textit{moving mirrors} have, for instance, been discussed in \cite{Chen:2015bcg}. The mini-cosmology experiment focusing on dynamical chiral edge states in a quantum Hall system \cite{Hotta:2022aiv} is also closely related to moving mirror models.

Recently, gravity duals of moving mirrors in two-dimensional CFTs have been constructed in \cite{Akal:2020twv,Akal:2021foz} by employing the AdS/BCFT correspondence \cite{Takayanagi:2011zk,Fujita:2011fp,Karch:2000gx}, which is an extension of the AdS/CFT duality \cite{Ma} to the case where the CFT is defined on a manifold with a boundary. On the gravity side, the mirror trajectory is dual to a so-called end-of-the-world (EOW) brane that extends into the AdS bulk spacetime. We can calculate entanglement entropy in a geometrical way by using the dual gravity solutions \cite{Ryu:2006bv,Ryu:2006ef,Hubeny:2007xt}. This allows to derive the Page curve \cite{Page:1993df,Page:1993wv} for the entanglement entropy of Hawking radiation, which is a sign of unitary dynamics of black hole evaporation. In this example, the EOW brane takes a characteristic profile which is dual to black hole evaporation in two-dimensional gravity via brane-world holography \cite{Karch:2000ct,Randall:1999ee,Randall:1999vf}
and the island proposal \cite{Penington:2019npb,Almheiri:2019psf,Almheiri:2019hni}. 
Other quantum informational quantities are also analyzed recently in moving mirror setups \cite{Sato:2021ftf, Kawabata:2021hac, BasakKumar:2022stg}. 
Another scenario aiming at restoring unitary dynamics in a global black hole spacetime is given by the black hole final state proposal \cite{Horowitz:2003he}. In \cite{Akal:2021dqt}, it has recently been shown that the evolution of pseudo entropy \cite{Nakata:2021ubr} resembles the Page curve when a partially spacelike mirror profile acts as a final state projection. 

Motivated by these developments, the main goal of the present paper is to extensively study a whole class of moving mirrors in two-dimensional CFTs, namely, going beyond those which have been motivated for modeling Hawking radiation, both from conformal field theoretic and holographic perspectives. Focusing on the causal behavior at late times and assuming that there is only a single mirror, we separate the moving mirror models into four different classes named type A, B, C, and type D. As will be seen, models in each of the classes show characteristic properties for the energy stress tensor and entanglement entropy.
Moreover, we present the corresponding gravity duals based on the AdS/BCFT correspondence. 
Especially, by studying various types of mirror setups, we observe a peculiar behavior of the dual EOW branes when the mirror trajectory approaches a lightlike one. Though, at first sight, the gravity duals look puzzling for certain types, we are able to work out a reasonable interpretation in general.

The remainder of the paper is organized as follows. In section~\ref{sec:2}, we first discuss the field theory description of moving mirrors in two-dimensional CFTs. We then group various mirror profiles into four classes, where each type is further divided into three subclasses. 
In section~\ref{sec:3}, we explain the holographic description of moving mirrors via the AdS/BCFT correspondence. We also present the calculation of entanglement entropy. 
In section~\ref{sec:4}, we study the gravity dual for each mirror type by especially focusing on the profile of the EOW brane. In section~\ref{sec:5}, we summarize the main results of this paper. In appendix \ref{app:A}, we present an analysis of a timelike-spacelike-timelike moving mirror model. In appendix \ref{sec:LIandDM}, having a closer look at the path integral formalism, we give an argument explaining why our conformal map method can successfully reproduce entanglement entropy and why analogous computations give pseudo entropy \cite{Nakata:2021ubr} in general.

\section{Classifying moving mirrors}
\label{sec:2}

In the following, we explore various moving mirrors in two-dimensional spacetime. Particularly, we restrict our attention to moving mirrors that satisfy the following simple criteria. 
\begin{enumerate}
	\item Mirror trajectory starts from the past timelike infinity, $i^-$ (\ie $t=-\infty$). 
	\item Mirror trajectory is continuous (\ie no piecewise trajectories).
	\item Mirror cannot move faster than the speed of light (\ie no spacelike trajectories).
\end{enumerate}
As a result of these assumptions, we can distinguish various moving mirrors by inspecting their endpoints. In this section, it is shown that all moving mirrors under the conditions listed above can be classified into four basic types whose endpoints are given by (A) future timelike infinity $i^+$, (B) right future null infinity $\mathcal{I}^+_{\mt{R}}$, (C) left future null infinity $\mathcal{I}^+_{\mt{L}}$, and (D) a termination point in bulk spacetime. These classes are further subdivided into three subclasses described by adding the subscripts 
$+,0$, and $-$, depending on their late-time behavior.\footnote{In the classification scheme above, we have ignored spacelike mirrors because the trajectory of any physical object cannot move faster than the speed of light. However, it is possible that we interpret the spacelike mirror as a projective measurement as discussed in \cite{Rajabpour:2015uqa,Rajabpour:2015xkj,Numasawa:2016emc,Akal:2021dqt}. Motivated by this possibility, we study in appendix \ref{app:A} an example where the mirror is initially timelike and becomes spacelike afterwards.}

\subsection{Four types of holographic moving mirrors}
\begin{figure}[h!]
	\centering		
	\includegraphics[width=3in]{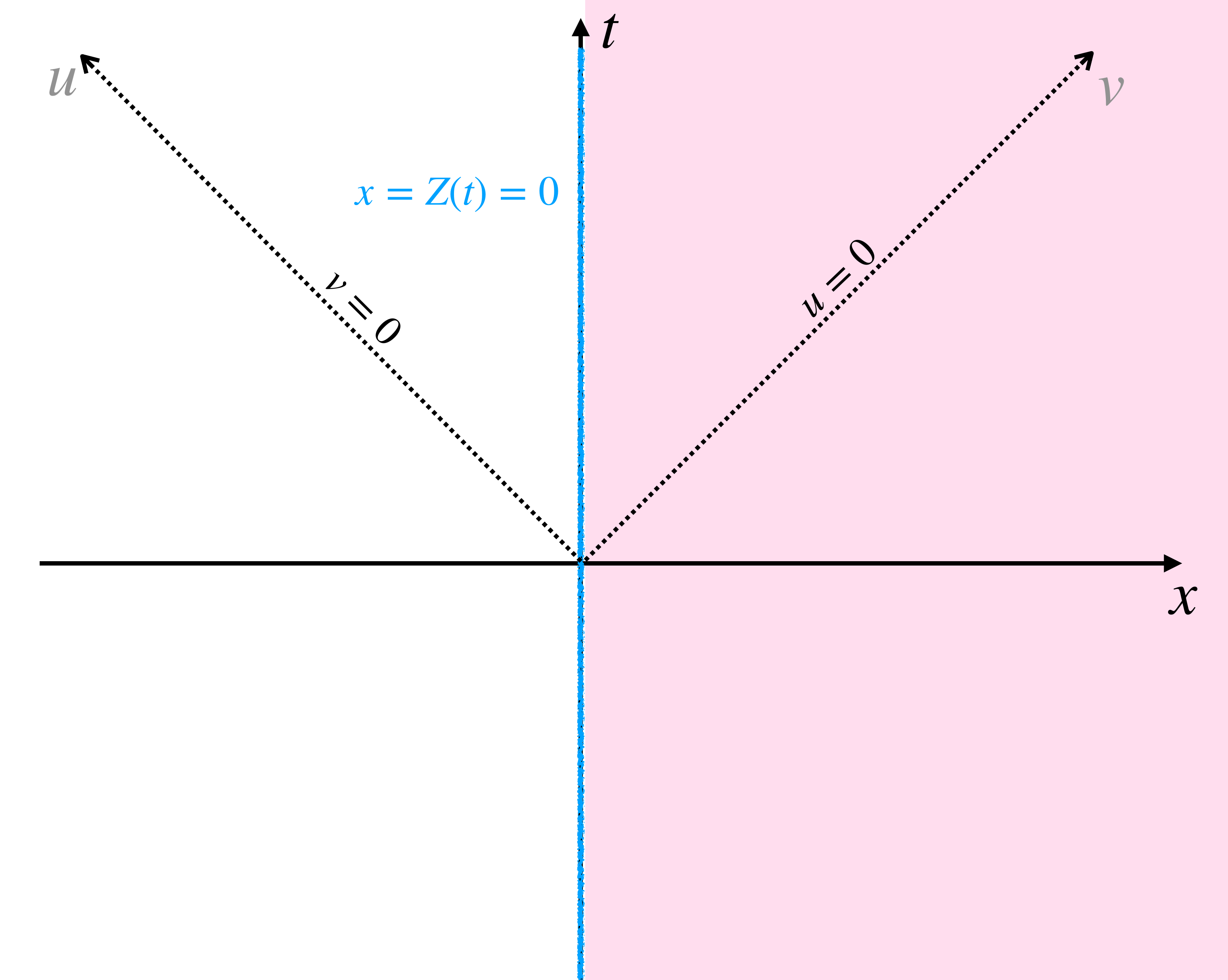}\,
	\includegraphics[width=2.4in]{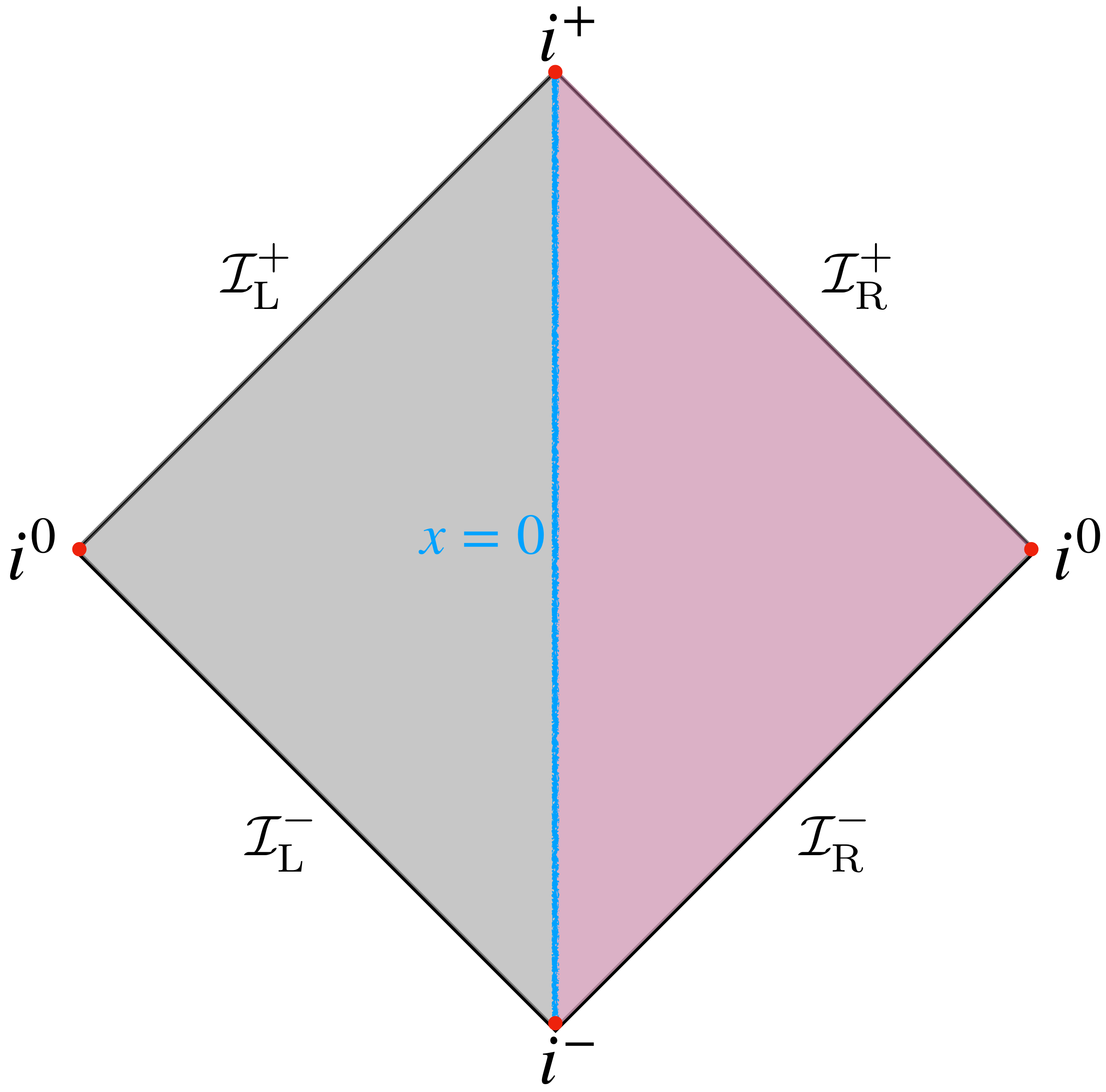}
	\caption{Minkowski spacetime with a static mirror located at $x=0$.}
	\label{fig:Min}
\end{figure}

We are interested in studying a $\text{CFT}_2$ with a moving mirror. Considering a two-dimensional spacetime defined by 
\begin{equation}\label{eq:define2Dmetric}
ds^2= -dt^2 + dx^2= -dudv\,, \qquad \text{with} \qquad  u =t- x, v=t+x\,,
\end{equation}
we assume that the trajectory of the moving mirror is given by $x=Z(t)$. For example, the simplest trajectory is the static mirror case:
\begin{equation}
x=Z(t) = 0  \,. 
\end{equation}
This is nothing but the standard setup of BCFT$_2$ where the boundary is located at $x=0$. Although there are infinitely many trajectories for moving mirrors, we do not need to deal with all of them one by one. Instead, we can map the original configuration to the simple setup of a BCFT$_2$ with a static mirror by employing conformal transformations.

We start from a vacuum state with vanishing energy stress tensor $T_{uu}=T_{vv}=T_{uv}=0$ in a two-dimensional flat spacetime with 
\begin{equation}\label{eq:simple2D}
ds^2 = -d\tilde{t}^2 + d \tilde{x}^2= -d \tilde{v} d\tilde{u} \,, \qquad  \tilde{u}= \tilde{t} - \tilde{x}\,, \tilde{v}= \tilde{t} + \tilde{x}\,,
\end{equation}
where $\tilde{v}, \tilde{u}$ denote advanced and retarded (null) coordinates, respectively.
The static mirror as the boundary is a timelike straight line located at $ \tilde{x}=0$, \ie $\tilde{u}=\tilde{v}$. We are interested in studying various mirrors defined in the physical two-dimensional spacetime described in terms of the coordinates $(t,x)$ given in eq.~\eqref{eq:define2Dmetric}. 
Assuming that the coordinates are related by a chiral conformal transformation as follows 
\begin{equation}\label{eq:chiraltrans}
\tilde{u} = p(u)\,, \qquad \tilde{v}= q(v)=v\,,  
\end{equation}
one can also map the static mirror with $\tilde{u} = \tilde{v}$ to a moving mirror at $x=Z(t)$ by using the function
\begin{equation}\label{eq:Zt}
 v=p(u)\,, \qquad  \text{or} \qquad 
\begin{cases}
t(u)= \frac{p(u)+u}{2},\\
x(u)= Z(u)= \frac{p(u)-u}{2}\,.\\
\end{cases}
\end{equation}
It is straightforward to find that the line element on the moving mirror reduces to 
\begin{equation}\label{eq:dsmirror}
ds^2 \big|_{\rm{Mirror}} =  - (dt(u))^2 + (dx(u))^2 = - p'(u) du^2 =  - p'(u)  \( \frac{2}{p'(u) +1}  \)^2 dt^2 \,,
\end{equation}
which simply implies that the moving mirror is timelike or spacelike when $p'(u)>0$ or $p'(u) <0$, respectively. We remark that the mirror is null when either $p'(u)=0$ or $p'(u) \to \infty$, where the velocity of the mirror is determined by $p'(u)$. In the main body of this paper, the mirror velocity is assumed to be not exceeding the speed of light, \ie timelike or null moving mirrors with $p'(u) \ge 0$.  Due to the importance and specialty of the zero point and divergence of $p'(u)$, we call a trajectory function normal if there is no zero point or divergence. 

\begin{figure}[h!]
	\centering		
	\includegraphics[width=2.7in]{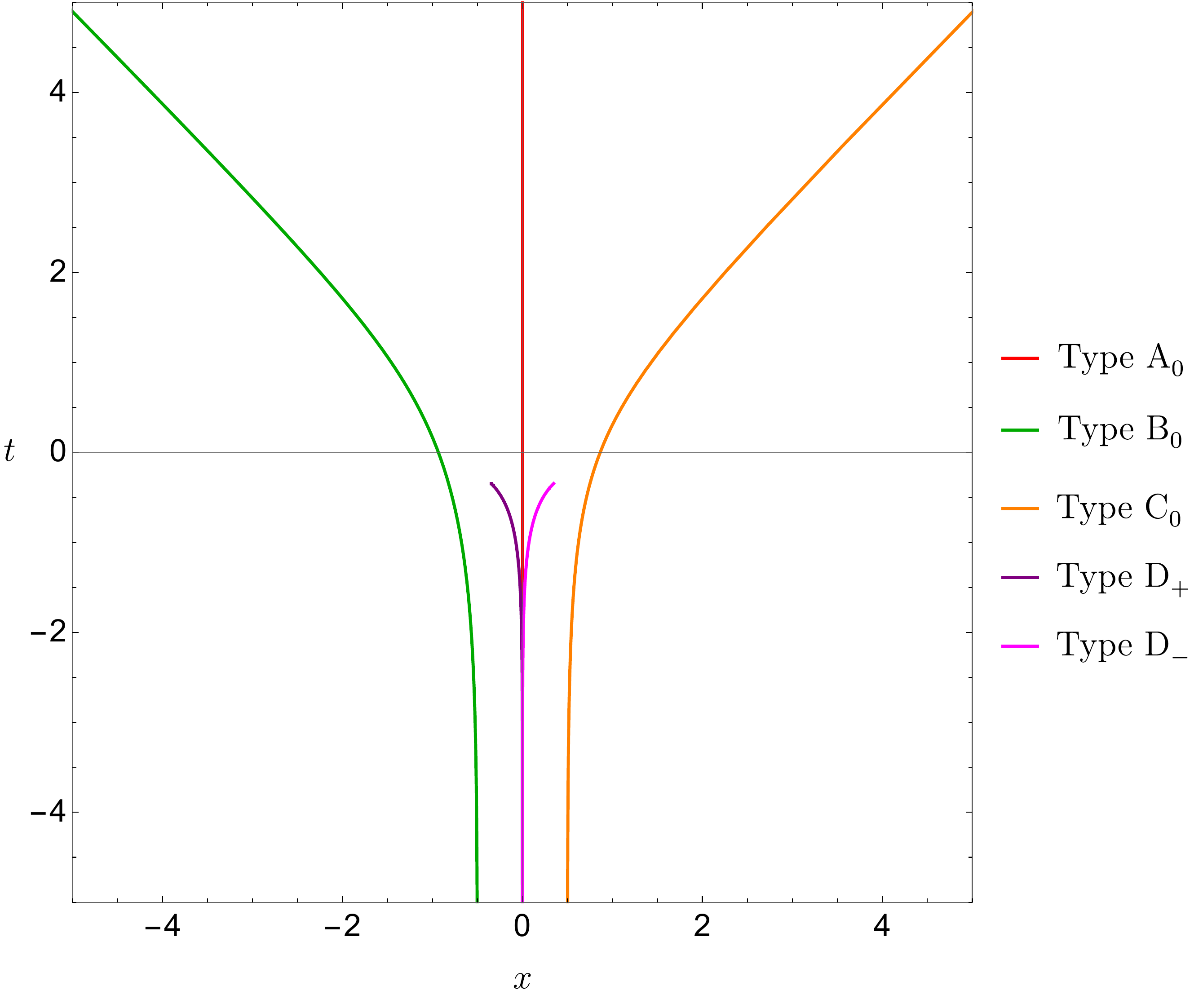}
	\includegraphics[width=3.1in]{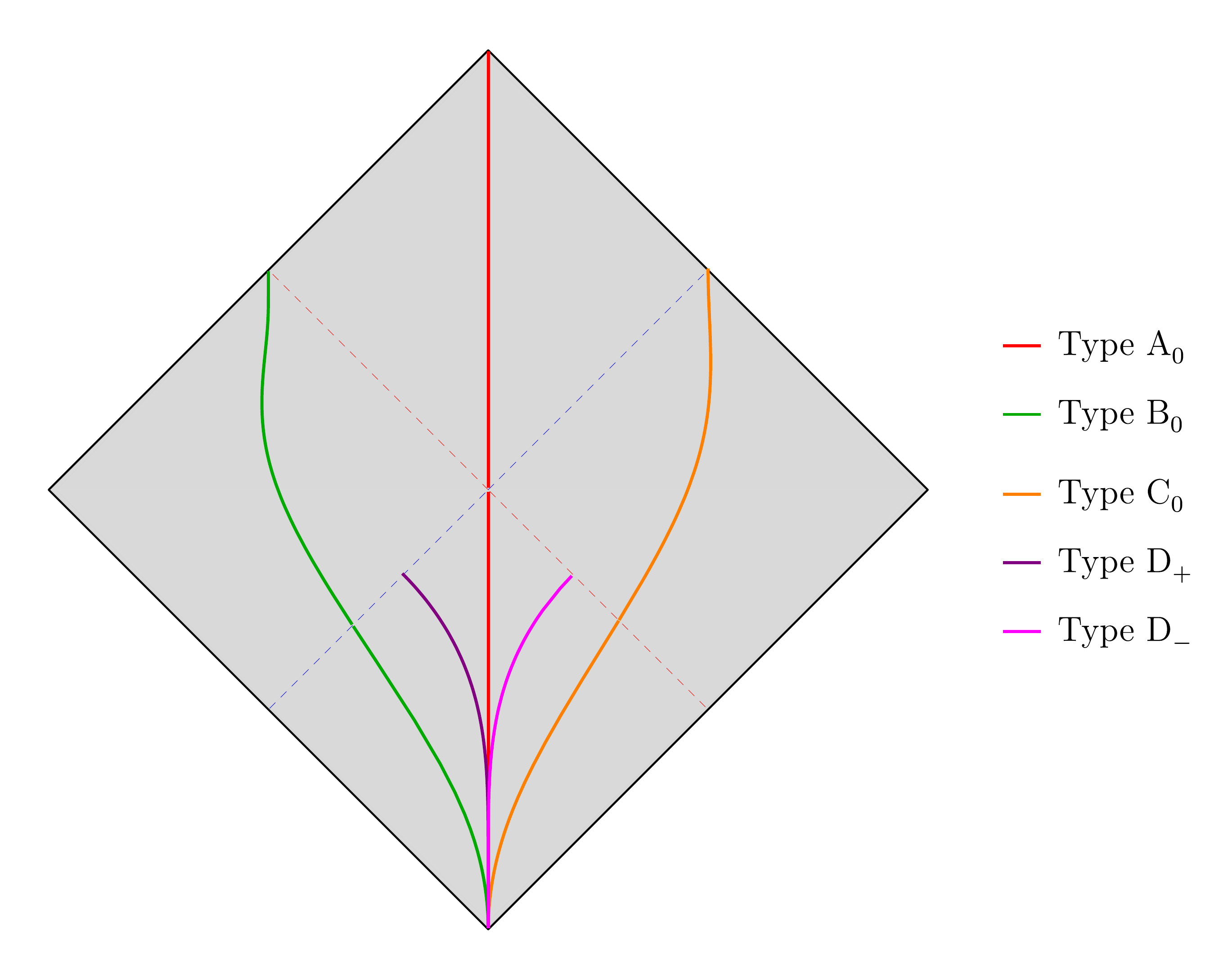}
	\caption{Various trajectories for different types of simple moving mirrors associated with the mapping function $p(u)$.}
	\label{fig:variousmirrors}
\end{figure}

\begin{table}[h!]
	\centering
	\begin{tabular}{ |p{4cm}||p{2cm}|p{3.5cm}|p{3.5cm}|  }
		\hline
		\multicolumn{4}{|c|}{Classes of moving mirrors} \\
		\hline
		Class & Trajectory & 	\centering Criteria & 	Range of $v=p(u)$ \\
		\hline
		Type A \newline (timelike) mirror&  $i^- \to i^+$  & $u \in [-\infty, +\infty]$  & $v\in [-\infty, +\infty]$\\
		\hline
		Type B \newline (escaping) mirror&   $i^- \to \mathcal{I}^+_{\mt{L}}$& $u \in [-\infty,+\infty]$
		\newline  $v_{\rm end}=  \lim\limits_{u \to \infty} p(u)$ 
		\newline  $\to p'(\infty)= 0 $ &  $v \in   [-\infty, \vend]$ \\
		\hline
		Type C \newline (chasing) mirror&  $i^- \to \mathcal{I}^+_{\mt{R}}$   & $u \in [-\infty, u_{\rm end}]$
		\newline $p(u_{\rm end})= + \infty$ 
		\newline $\to p'(u_{\rm end})= + \infty$  &  $v\in [-\infty, +\infty]$ \\
		\hline
		Type D \newline (terminated) mirror&  $i^- \to N$   & $u \in [-\infty, u_{\rm end}]$
		\newline $p(u_{\rm end})=\vend$
		\newline $p'(u_{\rm end})= +\infty / 0$ & $v \in   [-\infty, \vend]$\\
		\hline
		\hline
	\end{tabular}	
	\caption{An overview of key features related to moving mirrors of four distinct types.}
	\label{table:01}
\end{table}

We point out that we need to distinguish the two sides of the moving mirror since we only identify one of them as the physical region. We call the right-hand side of the moving mirror the physical spacetime. In Fig.~\ref{fig:Min}, we show the trajectory of a static mirror in terms of the coordinates $(t,x)$, where the pink region represents the physical spacetime with the static mirror as the boundary. We also depict the static mirror in the Penrose diagram of Minkowski spacetime for later comparison. For asymptotic regions of Minkowski spacetime whose Penrose diagram is shown in Fig.~\ref{fig:Min}, we refer to $i^+, i^-, i^0$ as future timelike infinity, past timelike infinity, spatial infinity, respectively. On the other hand, the null boundaries are labeled by $\mathcal{I}^\pm$ called future/past null infinity, respectively. According to the position of the mirror endpoints, we classify all simple mirrors, which satisfy the assumptions listed before, into four basic types. See Fig.~\ref{fig:variousmirrors} for some characteristic trajectories and table \ref{table:01} for a summary of the main properties. Taking into account the late time behavior, we can further distinguish between three subclasses for each type, as we discuss below.

\subsubsection*{Type A: timelike mirrors}
\begin{figure}[h!]
	\centering		
	\includegraphics[width=3in]{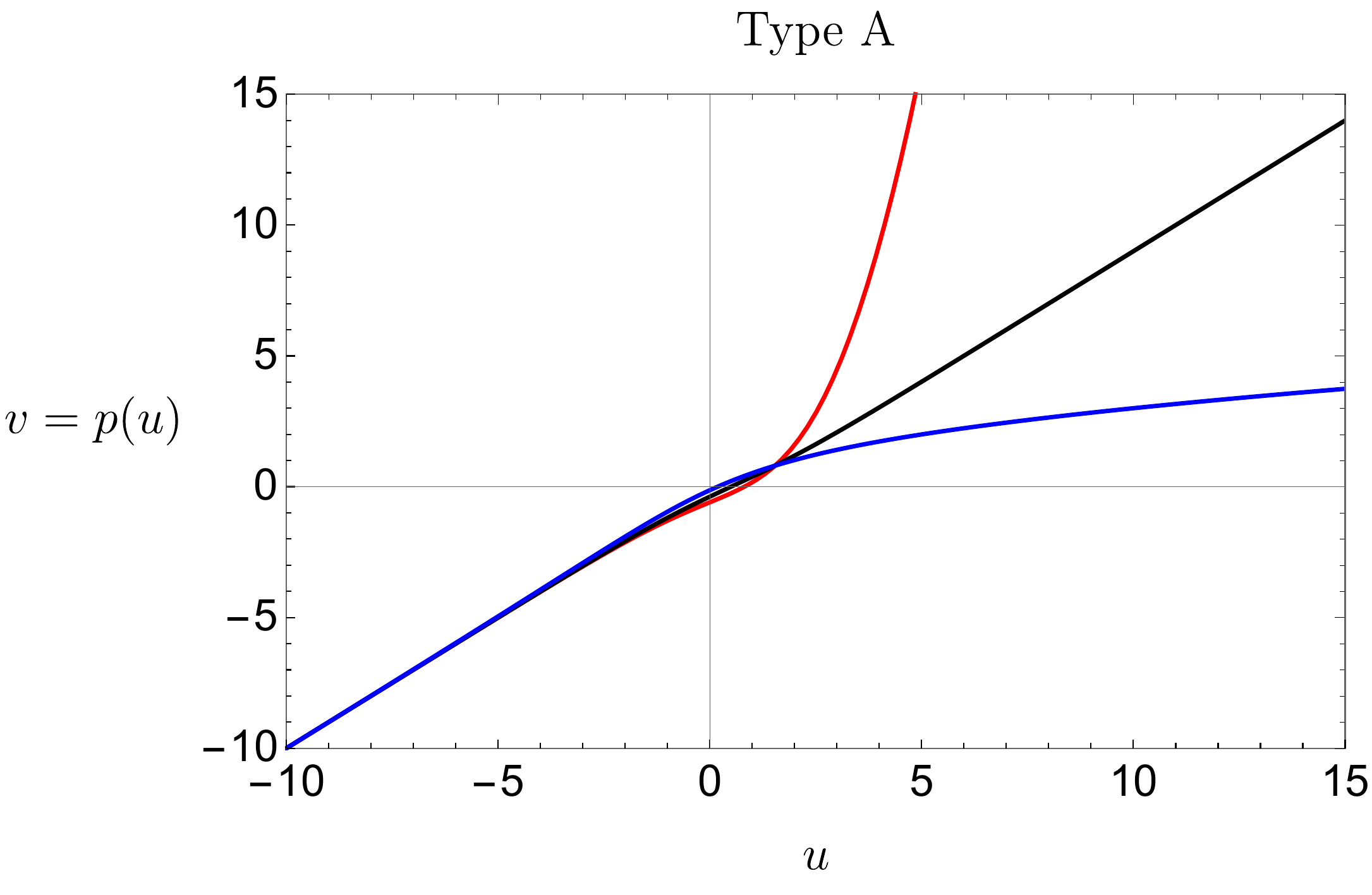}
	\includegraphics[width=3in]{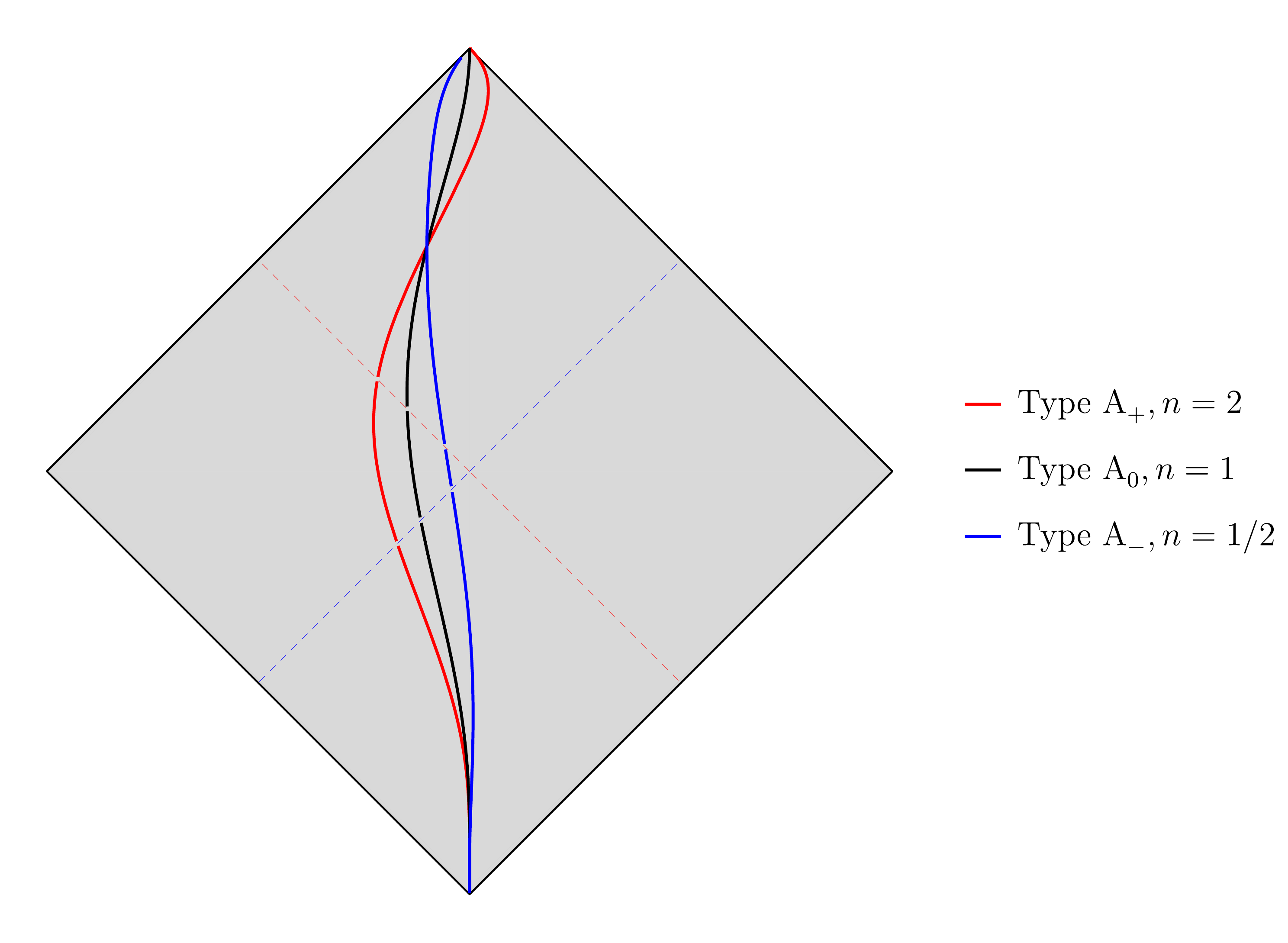}
	\caption{Type A mirrors with differently chosen $p_A(u)$ as introduced in eq.~\eqref{eq:defineTypeA}. The three subclasses type A$_+$, A$_0$, A$_-$ are defined in table \ref{table:AB}. Left: shown are different types of conformal mapping functions with $n=2,1,\frac{1}{2}$, respectively, see also table \ref{table:AB}. Right: Corresponding trajectories of moving mirrors are shown in a Penrose diagram. For both plots, we have chosen $\beta=1=u_0$.}
	\label{fig:TypeA01}
\end{figure}

The first type is the timelike moving mirror, as shown in Fig.~\ref{fig:TypeA01}. As we have explained, in flat two-dimensional spacetime (\ref{eq:define2Dmetric}), moving mirrors can be analyzed by employing a conformal transformation in eq.~\eqref{eq:chiraltrans}, assuming that the state at $t\to -\infty$ is the vacuum state.
The properties of moving mirrors are completely determined by the function $p(u)$. For example, the static mirror is defined by setting
\begin{equation}
p(u)=u\,, \qquad \text{or} \quad Z(t)=0\,.
\end{equation}
For the kink mirror, see \cite{Akal:2021foz} for more details, the mapping function $p(u)$ is given by 
\begin{equation}\label{eq:kink}
p(u)=-\beta\log(1+e^{-u/\beta})+\beta\log(1+e^{(u-u_0)/\beta})\,.
\end{equation}
The static mirror, but also the kink mirror, is always timelike, \ie $p'(u)=1 >0$. Furthermore, its trajectory extends from the past timelike infinity $i^-$ to the future timelike infinity $i^+$. In other words, the trajectories are defined for $u \in [-\infty, +\infty]$. Including the simplest static mirror, all moving mirrors with these features shall be of type A. These trajectories can also be associated with a massive particle moving in Minkowski spacetime. The diagnostic indicators for type A are given by 
\begin{itemize}
	\item $u \in [-\infty, +\infty]$,
	\item $0<p'(u) <\infty$.
\end{itemize}
We do not need to impose the extra constraint $v \in [-\infty, +\infty]$. Combining these two conditions, we automatically arrive at 
\begin{equation}
v \big|_{u =+ \infty}  = \lim\limits_{u \to \infty}  p(u)  = + \infty \,
\end{equation}
due to the absence of a zero point for $p'(u)$.\footnote{This limit cannot have a finite value, because if this limit exists, we can apply L'Hôpital's rule, $ \lim\limits_{u \to \infty} p(u)= \lim\limits_{u \to \infty}  \frac{e^u p(u)}{e^u}=  \lim\limits_{u \to \infty}  (p(u)+ p'(u))$, to arrive at $ \lim\limits_{u \to \infty} p'(u)=0$.}

The endpoints of type A moving mirrors always approach the future timelike infinity with $u=\infty, v=\infty$. 

\begin{figure}[!]
	\centering		
	\includegraphics[width=2.6in]{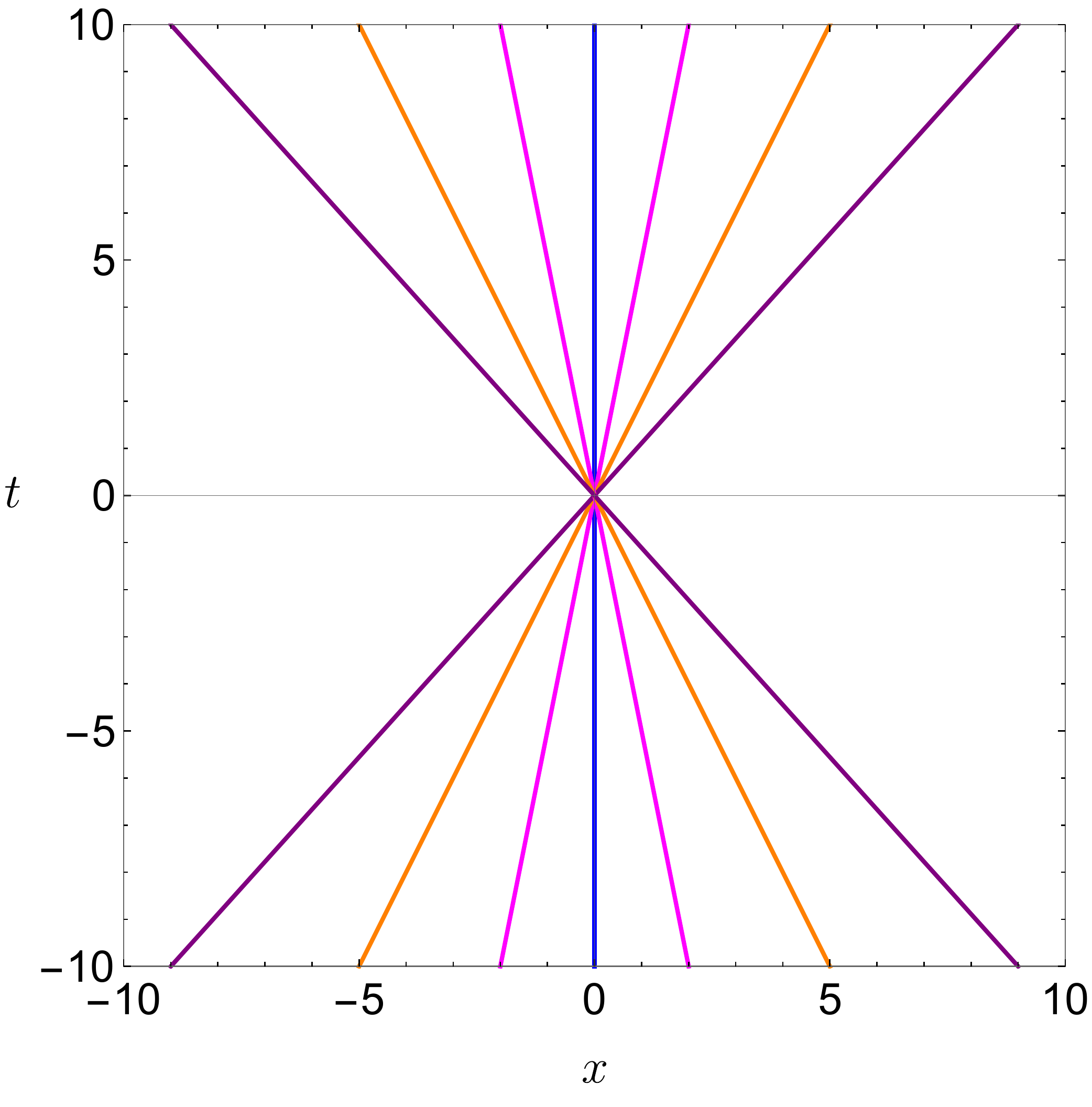}
	\includegraphics[width=3.4in]{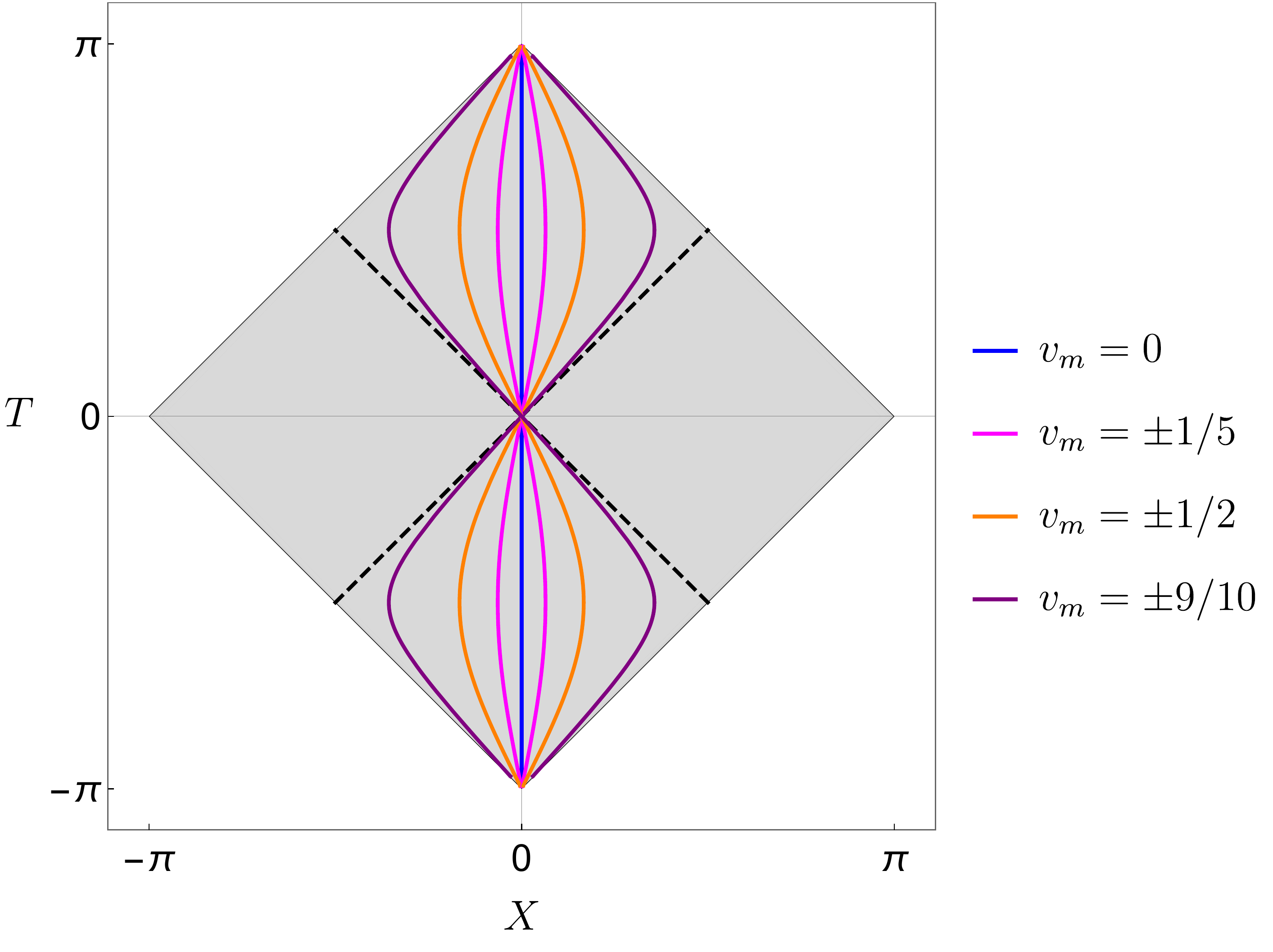}
	\caption{Trajectories of type A mirrors with a constant velocity $v_m$.}
	\label{fig:constantvelocity}
\end{figure}

Several moving mirrors of this type are shown in Fig.~\ref{fig:constantvelocity}, where we have plotted timelike moving mirrors with a constant velocity $v_m$. The corresponding mapping functions are given by 
\begin{equation}
p(u)= \frac{1+v_m}{1-v_m}\, u\,, \quad x=Z(t)= v_m t \,, \quad  \text{with} \quad v_m \in (-1,1)\,. 
\end{equation}
The derivative $p'(u)$ determines the velocity of the mirrors, \ie 
\begin{equation}\label{eq:definevm}
p'(u) = \frac{1+v_m}{1-v_m} \,.
\end{equation}
Most of the cases we explore later do not have a constant velocity. We can define a time dependent velocity
\begin{equation}\label{eq:definevmt}
Z'(t) = v_m(t) \equiv \frac{p'(u)-1}{p'(u)+1} \,, 
\end{equation}
where $Z(t)$ defined in eq.~\eqref{eq:Zt} describes the mirror trajectory.

\subsubsection*{Type B: escaping mirrors}
\begin{figure}[h!]
	\centering		
	\includegraphics[width=3in]{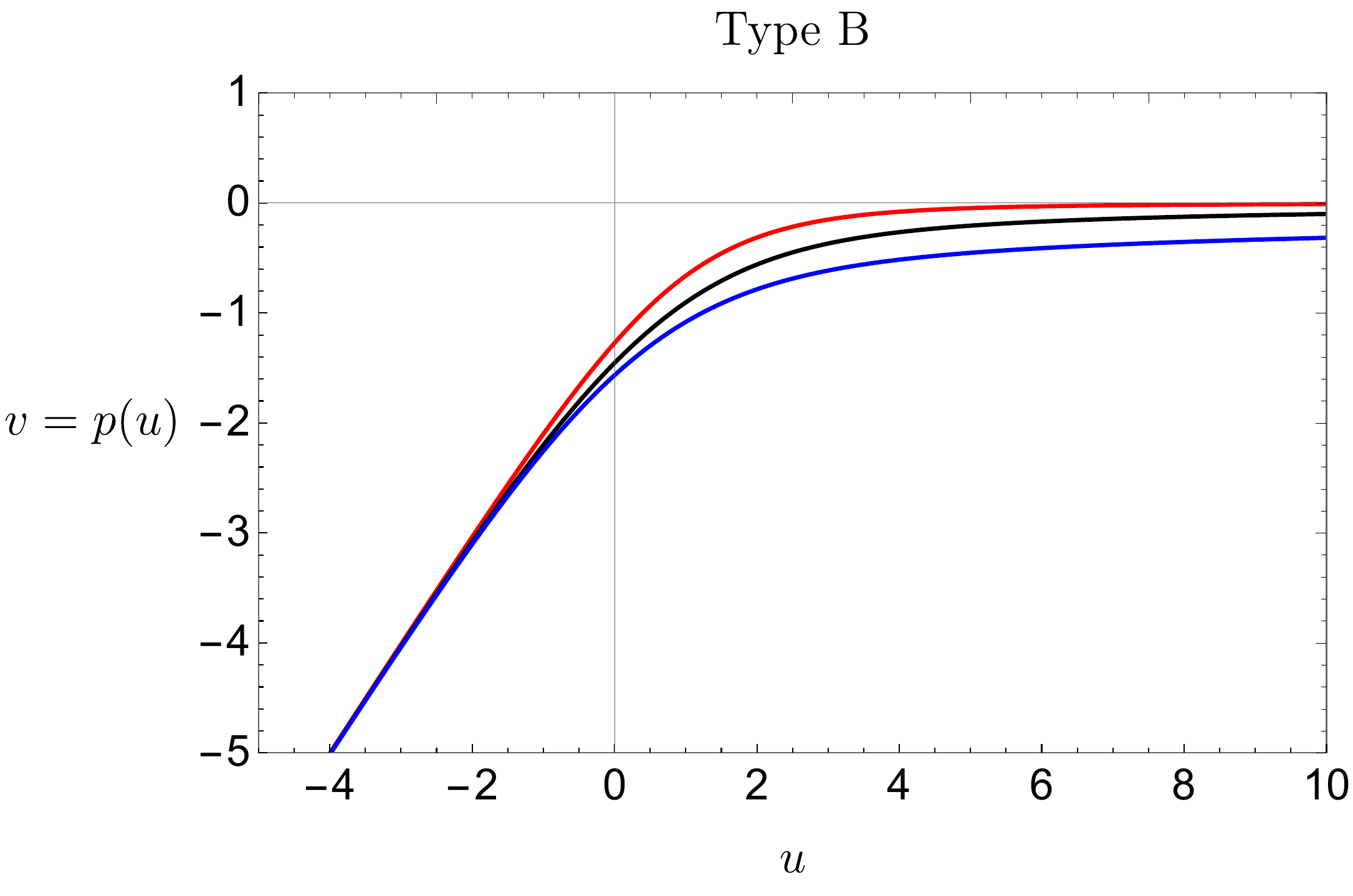}
	\includegraphics[width=3in]{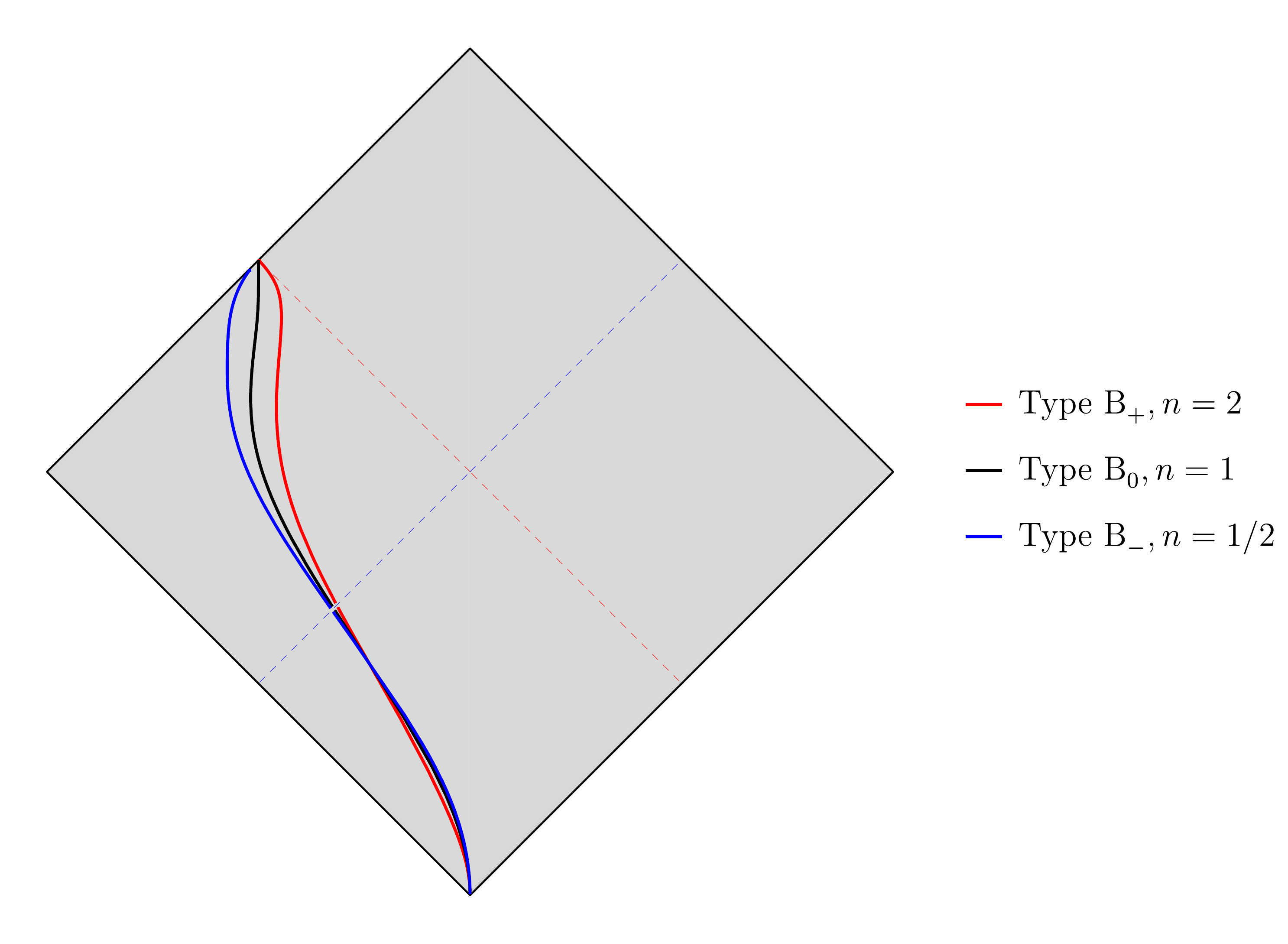}
	\caption{Type B mirrors with $\vend=0$. The three types B$_+$, B$_0$ and B$_-$ are classified in table \ref{table:AB}. Left: different conformal mapping functions $p_{\mt{B}}(u)$ as defined in eq.~\eqref{eq:defineTypeB} with fixing $\beta=1=u_0$ and taking $n=2,1,\frac{1}{2}$, see also table~\ref{table:AB}. Right: the corresponding trajectories of moving mirrors shown in a Penrose diagram.}
	\label{fig:TypeB01}
\end{figure}

Obviously, not all moving mirrors would approach the future timelike infinity, $i^+$. If the mirror can reach the speed of light, it would be able to arrive at the null future infinity. We shall define moving mirrors starting from $i^-$ and ranging to the future null infinity $\mathcal{I}^+_{\mt{L}}, \mathcal{I}^+_{\mt{R}}$ to be of type B, C respectively.  Let us first consider the case with endpoints at  $v=v_{\rm end}=  \lim\limits_{u \to \infty} p(u)$ on $\mathcal{I}^+_{\mt{L}}$. From the viewpoint of a physical observer, the dynamical mirror is moving away from the observer. We, therefore, call these type B escaping mirrors. Another distinguishable feature of type B is that the mirror velocity approaches $\lim\limits_{u \to \infty} v_m \to -1$ due to $\lim\limits_{u \to \infty} p'(u)=0$. As a result, we can summarize the necessary criteria for type B in terms of the mapping function $p(u)$ as follows
\begin{itemize}
	\item $u \in [-\infty,+\infty]\,, \quad  p'(u) \in [0, +\infty )$,
	\item $v_{\rm end}=  \lim\limits_{u \to \infty} p(u) \quad (  \to \lim\limits_{u \to \infty} p'(u)=0)$.
\end{itemize}
Some characteristic escaping mirrors are shown in Fig.~\ref{fig:TypeB01}.

\subsubsection*{Type C: chasing mirrors}
\begin{figure}[h!]
	\centering		
	\includegraphics[width=3in]{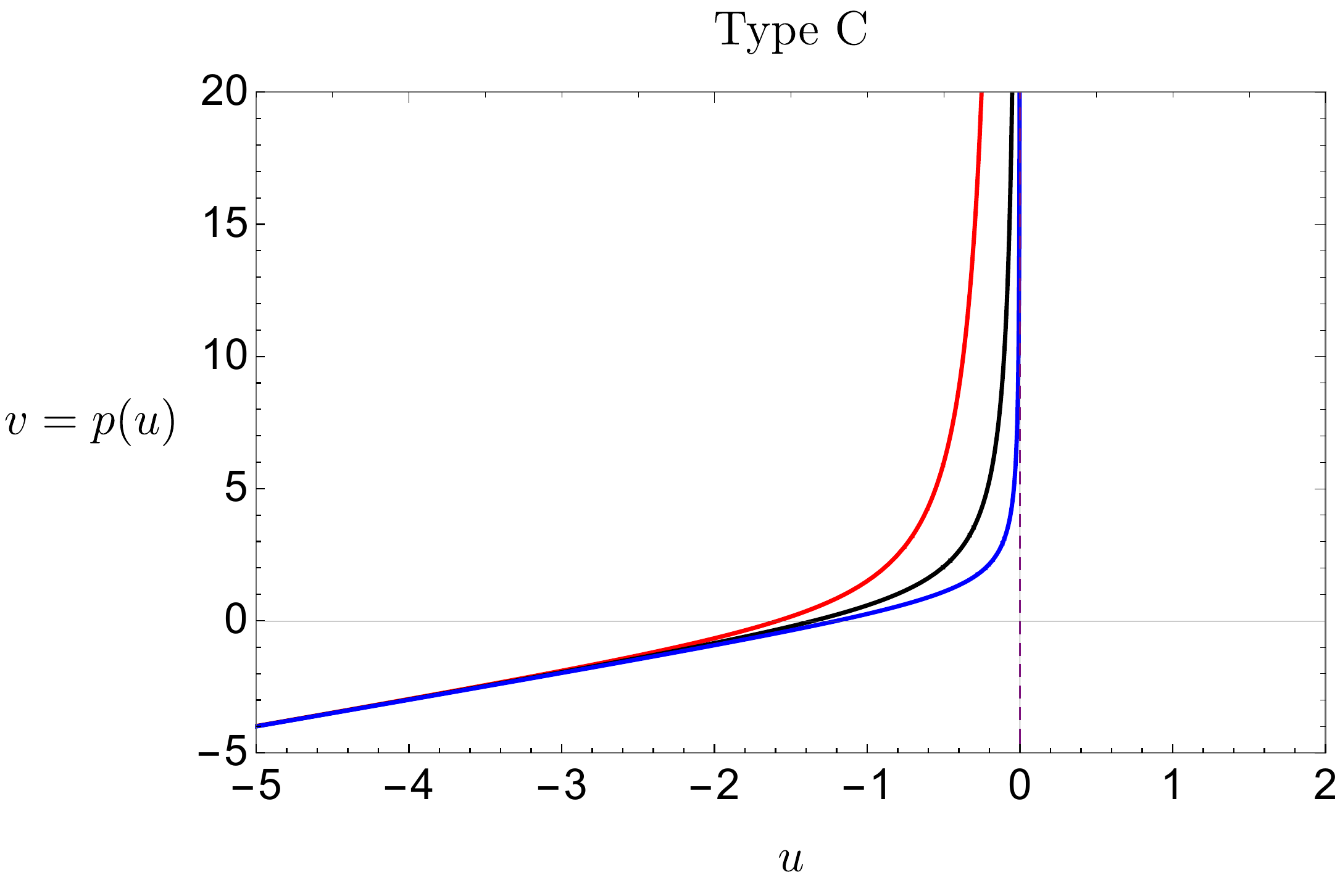}
	\includegraphics[width=3in]{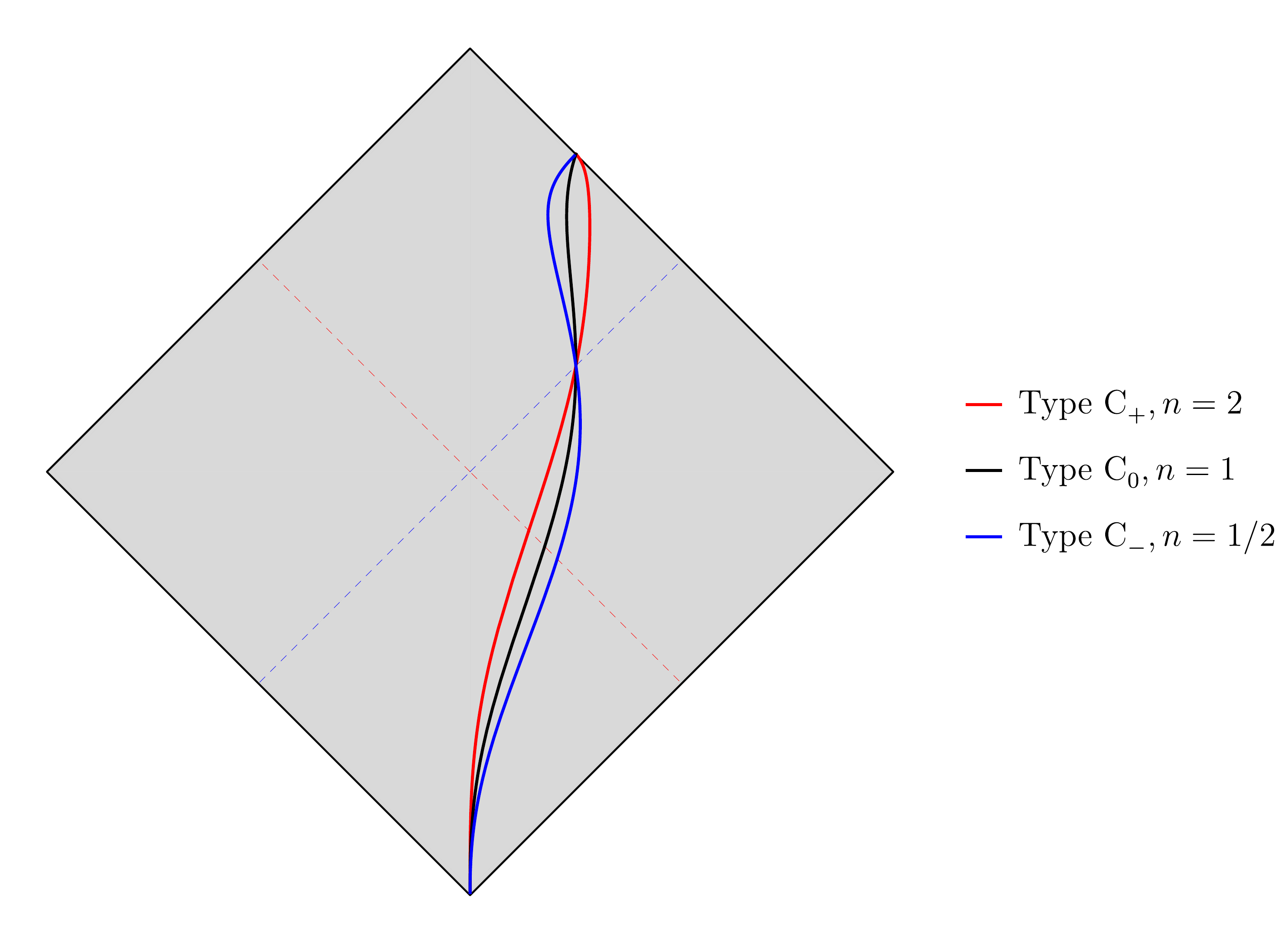}
 	\caption{Type C mirrors with $\uend=0$. Left: Three different types of conformal mapping functions $p_{\mt{C}}(u)$ as defined in eq.~\eqref{eq:defineTypeC} with $n=2,1,\frac{1}{2}$, see also table~\ref{table:CD}. Right: corresponding trajectories for type C$_+$, C$_0$, C$_-$ mirrors. The three subclasses are given in table \ref{table:CD}.}
	\label{fig:TypeC01}
\end{figure}
Similar to type B, the moving mirror could be asymptotically lightlike but arrive on the right future null infinity. We can locate the endpoint of the trajectory at $u=\uend$ so that $p(\uend) \to +\infty$, where the velocity of the moving mirror reaches the speed of light. In contrast to type B, this new type approaches the positive speed of light, \ie $v_m \to +1$ with $ p'(u) \to +\infty$. Since in some sense the mirror is chasing the physical observer, we shall call these moving mirrors, which start from $i^-$ and end at $\mathcal{I}_{\mt{R}}^+$, type C chasing mirrors. Analogous to the escaping mirror, the criteria for type C read\footnote{The condition $ p'(u_{\rm end})= + \infty$ is implied by $p(u_{\rm end})= + \infty$, since we have $p(u_{\rm end}) -p(u_1) =\int^{u_{\rm end}}_{u_1} p'(u) du$.}
\begin{itemize}
	\item $u \in [-\infty, u_{\rm end}]\,, \quad  p'(u)>0$,
	\item $p(u_{\rm end})= + \infty \quad  (  \to p'(u_{\rm end})= + \infty)$.
\end{itemize}
Some characteristic chasing mirrors are shown in Fig.~\ref{fig:TypeC01}.

\subsubsection*{Type D: terminated mirrors}

\begin{figure}[h!]
	\centering		
	\includegraphics[width=3in]{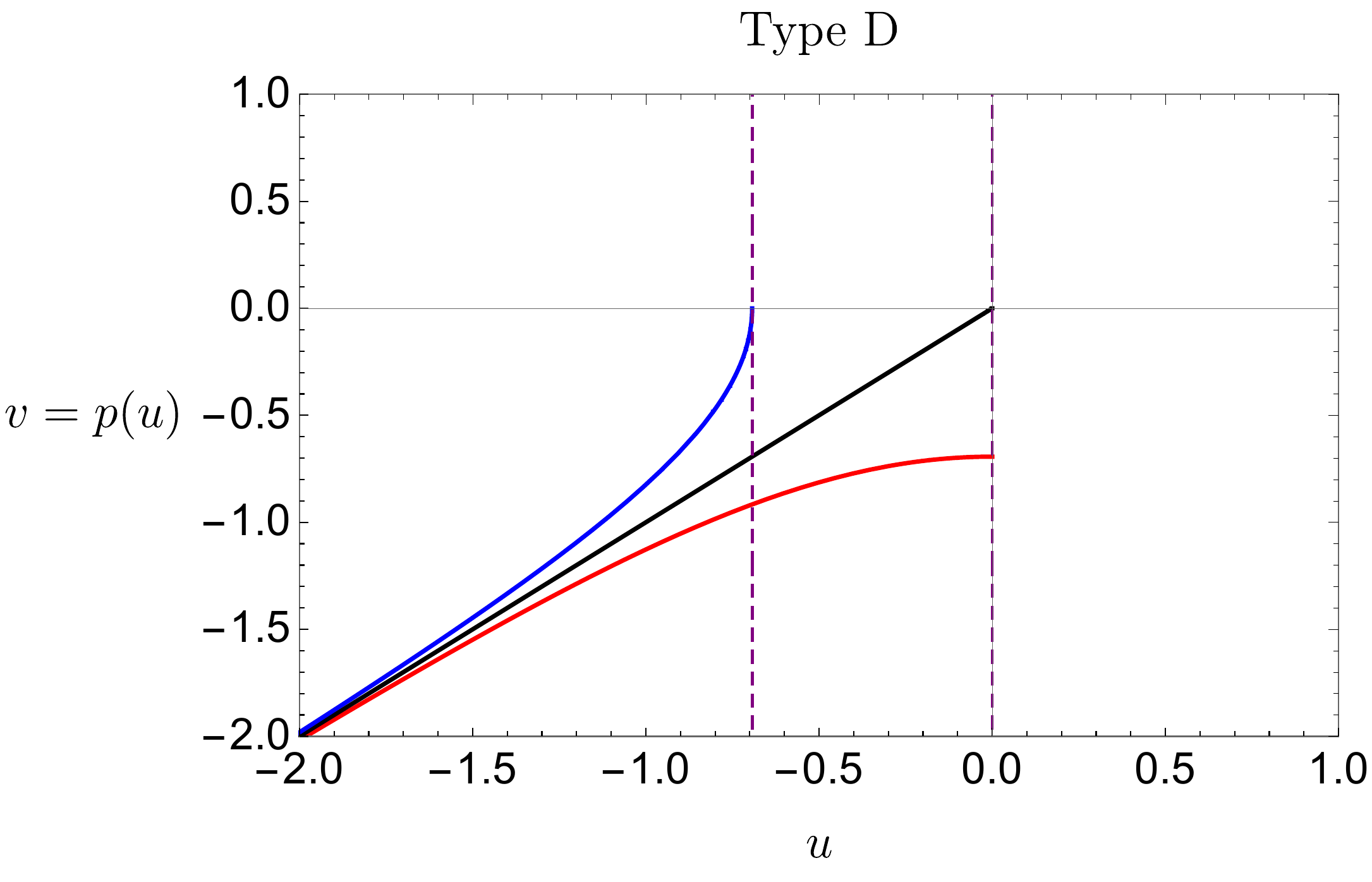}
	\includegraphics[width=3in]{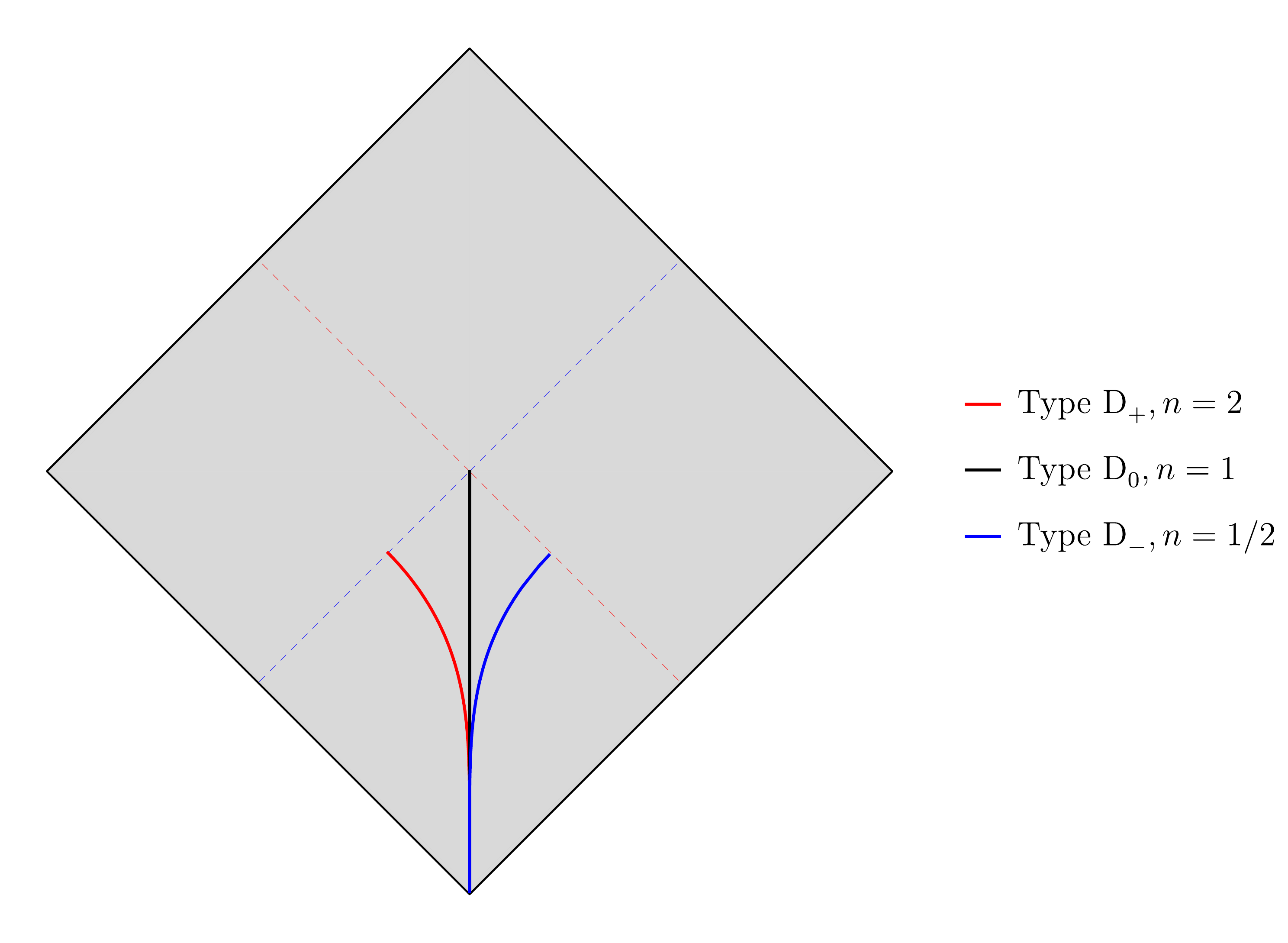}
	\caption{Type D mirrors. The three types D$_+$, D$_0$, and D$_-$ are distinguished in table \ref{table:CD}. Left: corresponding conformal mapping functions $p_{\mt{D}}(u)$. The simple type D$_0$ represents a part of the static case, and the functions $p_{\mt{D}}(u)$ for D$_\pm$ are taken as defined in eqs.~\eqref{eq:defineTypeDp} and \eqref{eq:defineTypeDm} with $n=2$ and $n=\frac{1}{2}$, respectively, see also table~\ref{table:CD}. Right: corresponding trajectories of moving mirrors in a Penrose diagram.}
	\label{fig:TypeD01}
\end{figure}
By now, we have explored all possible mirrors moving from past infinity $i^-$ to future infinity, $i^+$ or $\mathcal{I}^+$. Although we will not discuss piecewise trajectories in this paper, another possibility is that the mirror terminates at a finite time, \ie the endpoint is a finitely valued spacetime point denoted by $u=u_{\rm end}$ and $v=v_{\rm end}=p(\uend)$. The corresponding criteria for type D mirrors thus read 
\begin{itemize}\label{typeD:criteria}
	\item $u \in [-\infty, u_{\rm end}]\,, \quad  p'(u)>0$,
	\item $p(u_{\rm end})=\vend$\,,  \quad $( p'(u_{\rm end})= 0 \quad \text{for} \,\, \text{D}_{+}, \quad  p'(u_{\rm end})= \infty \quad \text{for} \,\, \text{D}_{-})$.
\end{itemize}
Here, we are mainly interested in mirror trajectories truncating at a {\it null point}, \ie a spacetime point where the velocity of the mirror exactly reaches the speed of light. 
Even though a massive object cannot reach the speed of light, this setup is motivated by a process, where a timelike boundary changes into a spacelike boundary in a CFT and the transition point becomes lightlike. Refer to appendix \ref{app:A} for a detailed analysis of such a model. The spacelike boundary is physically meaningful because the future and past spacelike boundaries can be interpreted as a projection onto a product state and preparation of a direct product state, respectively.\footnote{Note that the conformal boundary condition at a given time, \ie the boundary state, can be identified with a state without any real space entanglement \cite{Miyaji:2014mca}.} 

By regarding the spacelike boundary as a final state projection on the black hole singularity and the timelike boundary as a mirror that generates Hawking radiation, we may expect that the black hole evaporation geometry can be modeled by such a BCFT model, thus resembling the black hole final state projection scenario \cite{Horowitz:2003he,Akal:2021dqt}.

From the viewpoint of holography, see section~\ref{sec:gravitytypeD}, we can interpret this as the natural endpoint of the mirror due to the infinite energy excitation. If one considers the termination of the mirror to be happening at a finite point with a velocity less than the speed of light, this can always be obtained by cutting the mirror trajectories in types A, B, C at a finite time, leading to type D$_0$. Similarly, as for the types B and C, we can further distinguish the cases with $v_m = +1$ ($p'(\uend)=\infty$) and $v_m = -1$ ($p'(\uend)=0$) by identifying them as type D$_-$ and type D$_+$, respectively. Trajectories for the three different types of terminated mirrors are shown in Fig.~\ref{fig:TypeD01}.
The null points for the type D$_\pm$ mirror appear in the timelike-spacelike-timelike mirror model extensively discussed in appendix \ref{app:A}.

In summary, we have figured out the properties and criteria of four different types of moving mirrors, as summarized in table \ref{table:01}. The differences between these are reflected in the mapping function $p(u)$ as well as in their trajectories shown in terms of the original spacetime coordinates $(t,x)$, see left panel of Fig.~\ref{fig:variousmirrors}, and also in the global Penrose diagram as depicted in the right panel of Fig.~\ref{fig:variousmirrors}.

\subsection{Subclassification of moving mirrors}
\begin{figure}[h!]
	\centering		
	\includegraphics[width=3in]{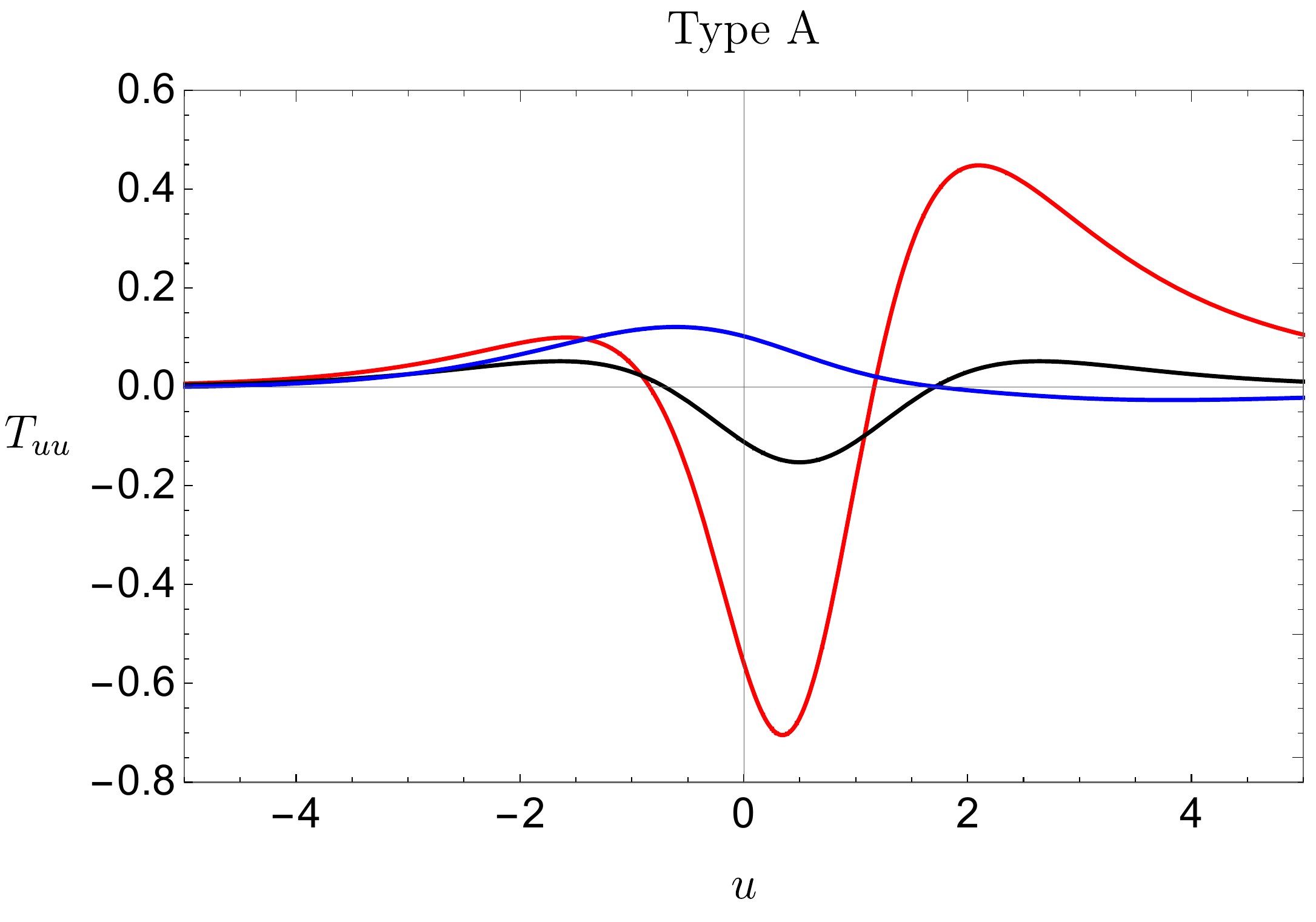}
	\includegraphics[width=3in]{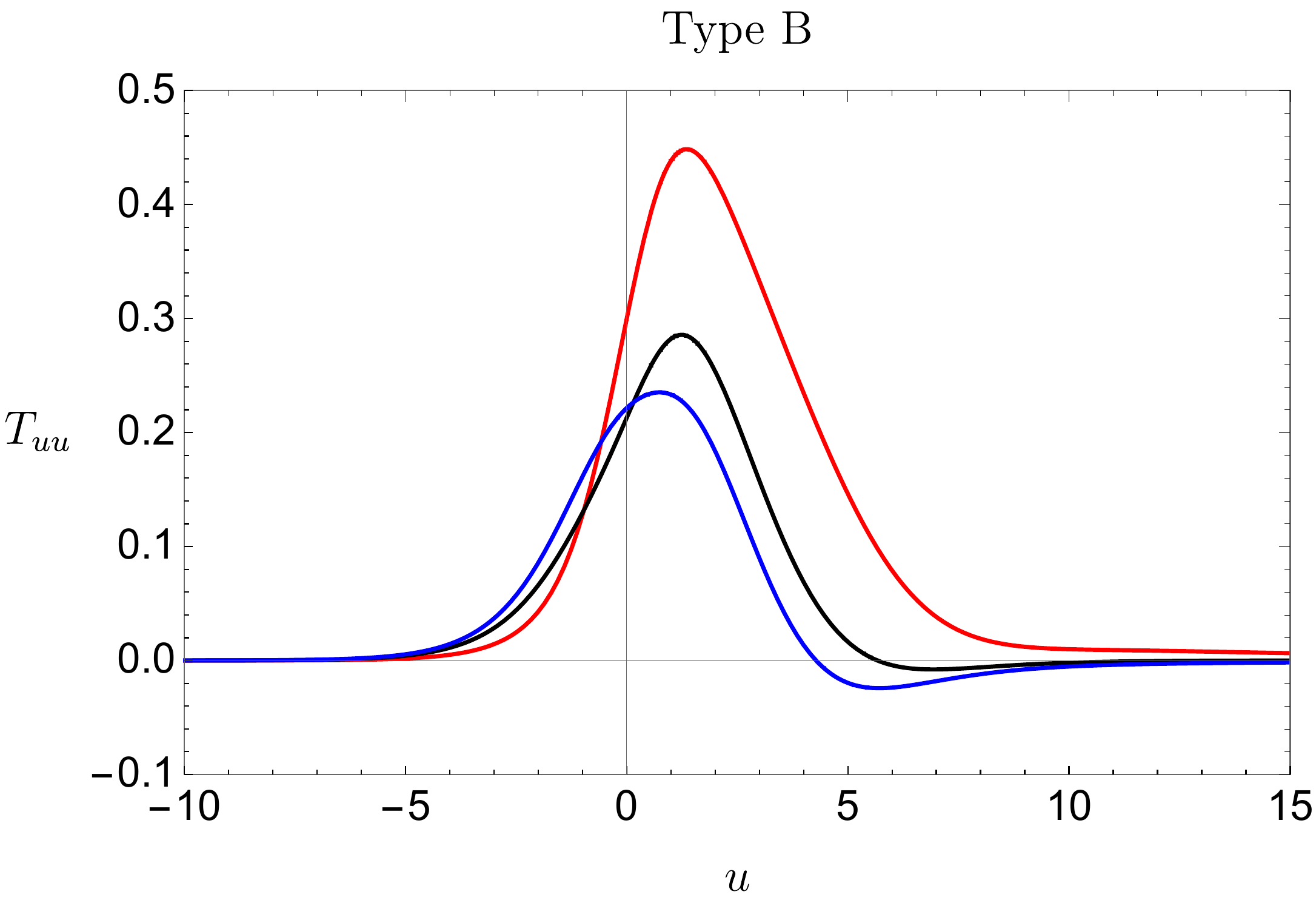}
	\includegraphics[width=3in]{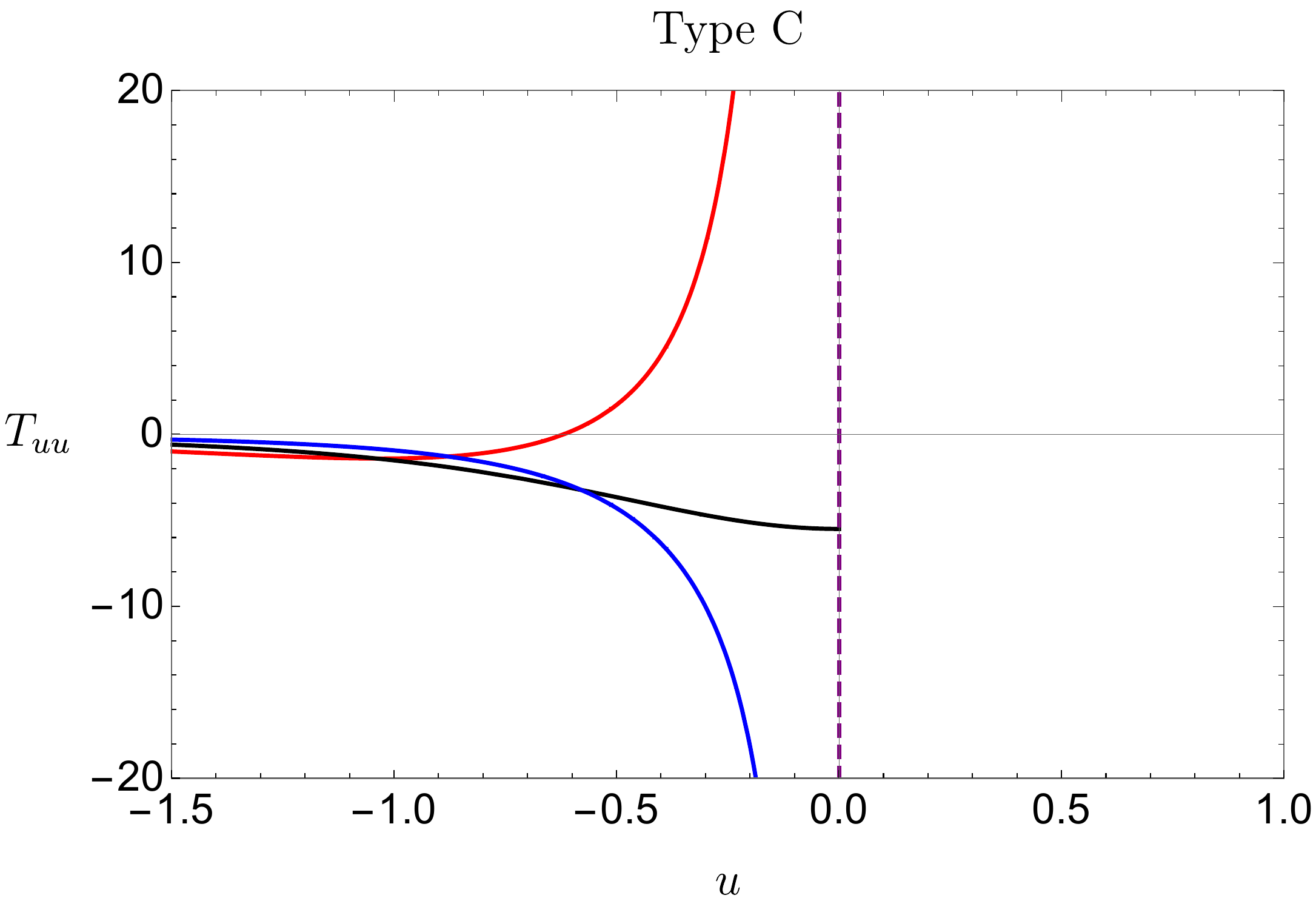}
	\includegraphics[width=3in]{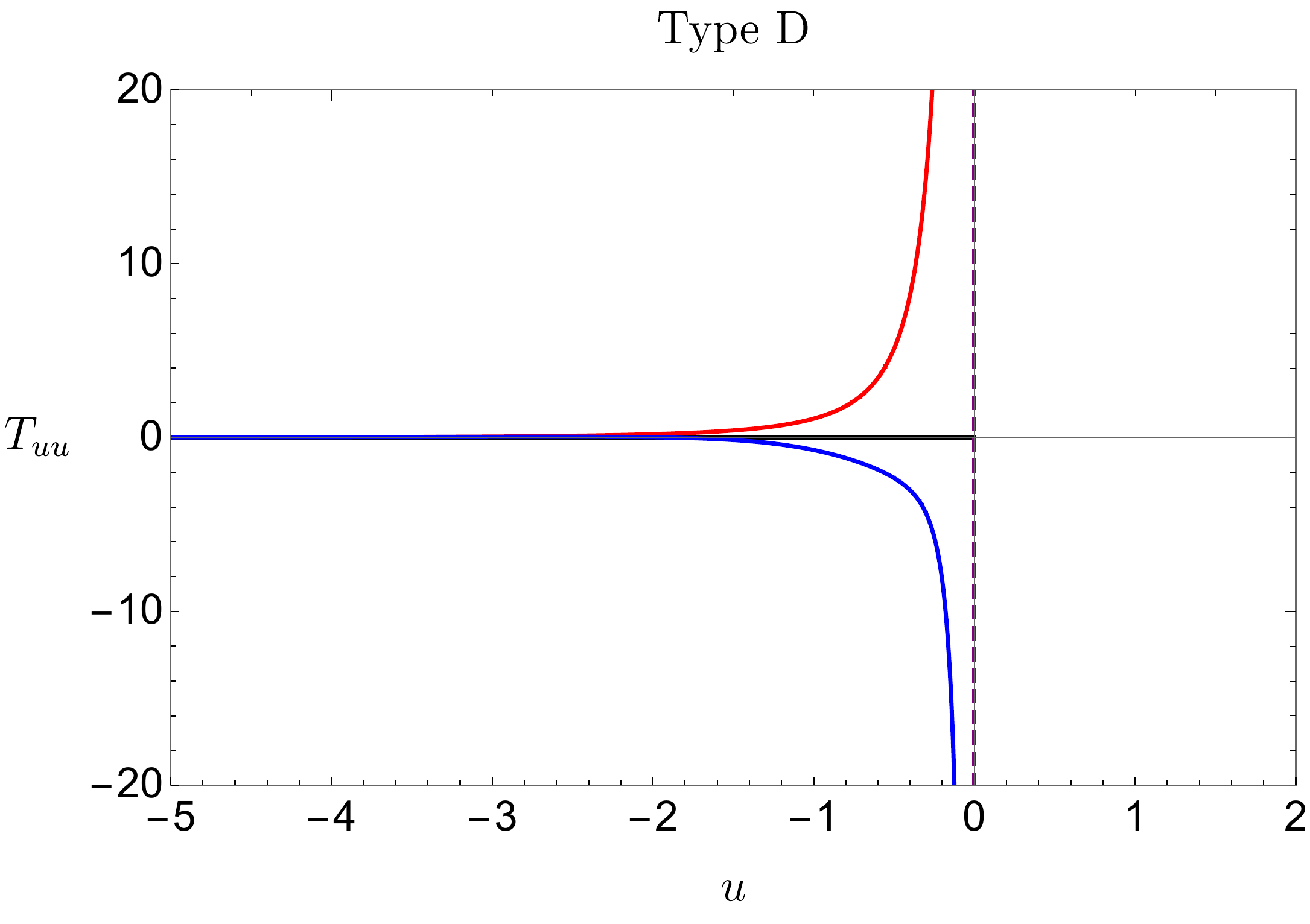}
	\caption{The corresponding stress tensors associated with conformal mapping functions $p_{\mt{A}}(u)$, $p_{\mt{B}}(u)$, $p_{\mt{C}}(u)$, and $p_{\mt{D}}(u)$. The red, black, and blue curves denote the cases with $n=2, n=1$, and $n=\frac{1}{2}$, respectively. The dashed purple line represents the position of $u=\uend$. Here, we have chosen $c=24 \pi$.}
	\label{fig:TuuABCD}
\end{figure}

Because of the nontrivial conformal transformation in eq.~\eqref{eq:chiraltrans}, there is a non-zero energy flux in the original spacetime with a moving mirror. As a result, one can mimic various dynamical spacetime models by using different mirror profiles \cite{Akal:2020twv,Akal:2021foz}. In general, the non-zero stress tensor is given by the Schwarzian derivative, namely,
\begin{equation}\label{eq:stresstensor}
T_{u u}=\left(\frac{d \tilde{u}}{d u}\right)^{2} T_{\tilde{u} \tilde{u}}+\frac{1}{2\pi}\frac{c}{12}\{\tilde{u}, u\}=\frac{c}{24 \pi}\left(\frac{3}{2}\left(\frac{p^{\prime \prime}(u)}{p^{\prime}(u)}\right)^{2}-\frac{p^{\prime \prime \prime}(u)}{p^{\prime}(u)}\right) \,, \quad T_{vv}= T_{uv}=0\,,
\end{equation}
where $c$ denotes the central charge of the CFT we are interested in. It is clear that the non-vanishing energy flux is completely determined by the mapping function $p(u)$, \ie by the trajectory of the mirror. The endpoint of the mirror cannot determine the feature of many physical quantities in the corresponding system, \eg the energy flux and entanglement entropy. In order to describe the underlying dynamics, especially at late times, we shall consider three subclasses for each mirror type, namely, by distinguishing the leading behavior of the mapping function $p(u)$ near the endpoint. We show later that this classification helps us capture the time dependence of physical quantities at late times. 


\subsubsection{Three subclasses}
As a warm-up, let us first consider a simple generalization of the kink mirror by taking a series expansion of a family of mapping functions,
\begin{equation}\label{eq:defineTypeA}
p_{\mt{A}}(u)=-\beta\log(1+e^{-u/\beta})+ \beta \( \log(e^{u_0/\beta}+e^{u/\beta})\)^n,  \qquad \text{with} \qquad n>0,
\end{equation}
which is related to the kink mirror case in eq.~\eqref{eq:kink} if we set $n=1$. Mirror profiles with different values for $n$ are shown in Fig.~\ref{fig:TypeA01}. Of course, $p_{\mt{A}}(u)$ reduces to the static mirror function if $n=1$ and $u_0=0$. The endpoints of moving mirrors associated with eq.~\eqref{eq:defineTypeA}, where $n>0$, are all approaching the future timelike infinity due to $p_{A} \to \infty$ in the late-time limit $u \to +\infty$. However, the asymptotic form of these mapping functions depends on the choice of the parameter $n$, namely
\begin{equation}\label{eq:pAlimit}
\lim\limits_{u \to \infty} p_{\mt{A}}(u) \sim  (u-u_0)^n +\cdots \,.
\end{equation}  
We find that the velocity of the mirrors are different for various $n$. By using eq.~\eqref{eq:definevmt}, one gets 
\begin{equation}
Z'(t) \to 
\begin{cases}
 -1,\quad n>1\\
  v_m \in (-1,1),\quad n=1\\
  +1,\quad n<1\\
\end{cases}.
\end{equation}
The energy flux at late times is related to the value of $n$, because the dominant term is of the form 
\begin{equation}
\lim\limits_{u \to \infty}T_{uu} \approx  \frac{c}{24 \pi} \times \frac{n^2-1}{2u^2}.
\end{equation}
For example, the cases for $T_{uu}$ with $n=2, 1, 1/2$ derived from eq.~\eqref{eq:defineTypeA} are plotted in the left-top panel of Fig.~\ref{fig:TuuABCD}. Later below, we show that entanglement entropy related to these subclasses are distinct. 

Motivated by the simple example in eq.~\eqref{eq:defineTypeA}, we divide all timelike mirrors of type A into three subclasses by taking account of the asymptotic behavior of the mapping function $p(u)$ as in eq.~\eqref{eq:pAlimit}. The three cases with $n>1, n=1$, and $1>n>0$ are referred to as type A$_+$, type A$_0$, and type A$_-$, respectively. According to this subclassification, the static mirror and kink mirror both then become type A$_0$. As a result, we can expect that the late-time behavior for the kink mirror is similar to that of the static mirror, \eg the subsystem entanglement entropy approaches a constant. 

Similarly, we may also define three subclasses for mirrors of other types. Similar to type A in eq.~\eqref{eq:defineTypeA}, a series expansion of a family of mapping functions for type B mirrors is given by 
\begin{equation}\label{eq:defineTypeB}
 p_{\mt{B}}(u)=-\beta\log(1+e^{-u/\beta})+ \beta \( \log(e^{u_0/\beta}+e^{u/\beta}) \)^{-n},  \qquad \text{with} \qquad n\ge 0\,,
\end{equation}
whose mirror trajectories are shown in Fig.~\ref{fig:TypeB01}. In the top-right panel of Fig.~\ref{fig:TuuABCD}, we plot the corresponding stress tensors associated with the mirrors described by $p_{\mt{B}}(u)$ for various values of $n$. Again, we can classify type B mirrors by focusing on the asymptotic behavior of $p(u)$ in the limit $u \to \infty$, which leads to
\begin{equation}\label{eq:typeBlimit}
\lim\limits_{u \to \infty} p_{\mt{B}}(u) \sim   \vend + \frac{c_{-n}}{u^n} +\cdots \,, \qquad \text{with} \qquad c_{-n}< 0\,,
	\end{equation}  
where we denote the finite value $p_{\mt{B}}(+\infty)$ as $\vend$ which parameterizes the endpoint of the mirror in future null infinity $\mathcal{I}^+_L$. The leading term of the non-vanishing stress tensor $T_{uu}$ can also be derived from eq.~\eqref{eq:typeBlimit}. As before, we define the cases with $n>1, n=1$, and $1>n>0$ in eq.~\eqref{eq:typeBlimit} to be of type B$_+$, type B$_0$, and type B$_-$, respectively. 

The requirement that the mirror trajectory $x=Z(t)$ is timelike leads to 
the condition $p'(u)>0$. Under this condition, we can conformally map the trajectories of type A and type B to the static one at $\ti{x}=0$ by using the map in eq.~\eqref{eq:chiraltrans}. Assuming a smooth profile of $Z(t)$, one finds that the energy stress tensor \eqref{eq:stresstensor} is always finite, as shown by the two upper plots in Fig.~\ref{fig:TuuABCD}. 

\begin{table}[]
	\begin{tabular}{|lclcl|}
		\hline
		\multicolumn{5}{|c|}{Energy flux of simple moving mirrors with $u \in [-\infty, +\infty]$}                                                                                                                                                                                                       \\ \hline
		\multicolumn{1}{|c|}{Class}  & \multicolumn{1}{c|}{Trajectory}                                                                                                     & \multicolumn{1}{c|}{Leading term of $ \lim\limits_{u\to +\infty}p(u)$}   & \multicolumn{1}{c|}{Leading term of $\frac{24 \pi }{c}T_{uu}$}               & Limit \\ \hline
		\multicolumn{1}{|l|}{A$_+$} & \multicolumn{1}{c|}{\multirow{3}{*}{\begin{tabular}[c]{@{}c@{}}$i^- \to i^+$\\ timelike\end{tabular}}}                              & \multicolumn{1}{l|}{$c_n u^n + \cdots,\, n>1$} & \multicolumn{1}{c|}{\multirow{3}{*}{$\frac{n^2-1}{2u^2}$}}  &   $+0$               \\ \cline{1-1} \cline{3-3} \cline{5-5} 
		\multicolumn{1}{|l|}{A$_0$} & \multicolumn{1}{c|}{}                                                                                                               & \multicolumn{1}{l|}{$c_n u^n + \cdots,\, n=1$}                           & \multicolumn{1}{c|}{}                                        &   $\sim 0$                    \\ \cline{1-1} \cline{3-3} \cline{5-5} 
		\multicolumn{1}{|l|}{A$_-$} & \multicolumn{1}{c|}{}                                                                                                               & \multicolumn{1}{l|}{$c_n u^n + \cdots,\, 1>n>0$}                           & \multicolumn{1}{c|}{}                                        &       $-0 $                    \\ \hline
		\multicolumn{1}{|l|}{B$_+$} & \multicolumn{1}{c|}{\multirow{3}{*}{\begin{tabular}[c]{@{}c@{}}$i^- \to \mathcal{I}^+_{\mt{L}}$\\ escaping \end{tabular}}} & \multicolumn{1}{l|}{$\vend + \frac{c_{-n}}{u^n}+\cdots  \,, n>1$  }                           & \multicolumn{1}{c|}{\multirow{3}{*}{$  \frac{n^2-1}{2u^2}$}} &          $+0$        \\ \cline{1-1} \cline{3-3} \cline{5-5} 
		\multicolumn{1}{|l|}{B$_0$} & \multicolumn{1}{c|}{}                                                                                                               &        \multicolumn{1}{l|}{$\vend + \frac{c_{-n}}{u^n}+\cdots  \,, n=1$ }                           & \multicolumn{1}{c|}{}                                        &      $\sim 0$            \\ \cline{1-1} \cline{3-3} \cline{5-5} 
		\multicolumn{1}{|l|}{B$_-$} & \multicolumn{1}{c|}{}                                                                                                               & \multicolumn{1}{l|}{$\vend + \frac{c_{-n}}{u^n}+\cdots  \,, 1>n>0$ }                           & \multicolumn{1}{c|}{}                                        &       $-0$                      \\ \hline
	\end{tabular}
	\caption{Classification of type A and type B mirrors based on the leading behavior of the mapping function $p(u)$. We assume that $n$ is finite and nonzero.}
	\label{table:AB}
\end{table}

However, the energy fluxes for type C and type D generically diverge. The divergence is traced back to the fact that the mirrors reach the speed of light at a finite value for $u$, $u=\uend$. 
Similar to the subclassifications of type A and type B, we can further separate the mirrors of type C and type D into three subclasses by focusing on the leading behavior of $p(u)$ around the endpoints. Refer to table~\ref{table:CD} for an overview of the divergent behavior for different moving mirrors. In this paper, we do not pay attention to the type D$_0$, because it represents a portion of the type A. However, we should note that the null endpoint does not always imply a divergence for the stress tensor. For example, we can still get a finite stress tensor for type C mirrors if  
\begin{equation}
\text{type C}_0: \qquad p(\uend) \approx   \frac{c_{-1}}{(\uend-u)}+c_0+ c_1 (\uend-u)\cdots.
\end{equation}
The latter are referred to as type C$_0$ mirrors.\footnote{For type C$_0$ mirrors, the conclusion that $T_{uu}(\uend)$ is a finite constant depends on the subleading term in $p(u)$. If we consider $p(\uend) \sim \frac{c_{-1}}{u-\uend} + \frac{c_m}{(u-\uend)^m} +\cdots$ with $0<m<1$, the stress tensor at the endpoint is still divergent in terms of $T_{uu}(\uend) \sim \(- \frac{c_m}{c_{-1}}\) \frac{1}{(u-\uend)^{m+1}}$. Similarly, for $p(\uend) \sim \frac{c_{-1}}{u-\uend} +c_0+ c_m{(u-\uend)^m} +\cdots $ with $0<m<1$, one has $T_{uu}(\uend) \sim \(\frac{c_m}{c_{-1}}\) \frac{1}{(u-\uend)^{1-m}}$.}

\begin{table}[h!]
	\centering
	\begin{tabular}{ |p{2.5cm}||p{7cm}|p{4.5cm}|}
		\hline
		\multicolumn{3}{|c|}{Divergent energy flux of simple moving mirrors with $u\le \uend$ and $v\le \vend$}\\
		\hline
		Class & Leading term of $ p(u)$ &  Leading term of $\frac{24 \pi }{c}T_{uu}$  \\
		\hline 
		C$_+$: $i^- \to \mathcal{I}^+_{\mt{R}}$  &  $   \frac{c_{-n}}{(\uend-u)^n}+\cdots \,, n>1$   &  $ \frac{n^2-1}{2(u-\uend)^2} \to + \infty $ \\
		\hline
		C$_0$:\, $i^- \to \mathcal{I}^+_{\mt{R}}$  &  $   \frac{c_{-1}}{(\uend-u)}+c_0+ c_1 (\uend-u)\cdots \,,n=1$   &  $ 6 \frac{c_1}{c_{-1}} = \text{finite constant}$ \\
		\hline
		C$_-$:  $i^- \to \mathcal{I}^+_{\mt{R}}$  &  $   \frac{c_{-n}}{(\uend-u)^n}+\cdots \,, 0<n<1$   &  $ \frac{n^2-1}{2(u-\uend)^2} \to - \infty $ \\
		\hline
	D$_+$: $i^- \to N_v$ &  $ \vend + c_n(\uend-u)^n+\cdots \,, n>1$   &  $ \frac{n^2-1}{2(\uend-u)^2} \to +\infty$ \\
		\hline
			D$_0$: \,$i^- \to N_0$ &  $ \vend +c_1(\uend-u)+ c_2(u-\uend)^2+\cdots $   &  $ 6 (\frac{c_2}{c_1})^2 = \text{finite constant}$ \\
		\hline
	   D$_-$: $i^- \to N_u$ & $\vend + c_n(\uend-u)^n+\cdots \,, 1>n>0$  &  $\frac{n^2-1}{2(u-\uend)^2} \to -\infty$ \\
     	\hline
		\hline
	\end{tabular}	
	\caption{The leading behavior of the function $p(u)$ and the stress tensor $T_{uu}$ near the endpoints for various moving mirrors.}
	\label{table:CD}
\end{table}

Before we move on, we would like to note that our classification could also include the cases which cannot be covered by polynomials \eg logarithmic or exponential functions in asymptotic expansion. For example, the special case with $n=0$ in eq.~\eqref{eq:defineTypeB} is given by 
\begin{equation}
p_{\mt{B}}(u)=-\beta\log(1+e^{-u/\beta})\,,  \label{hawkingr}
\end{equation}
which is called escaping mirror in \cite{Akal:2021foz}. Since the late-time limit of the function $p(u)$ is dominated by 
\begin{equation}\label{eq:typeBexp}
 \lim\limits_{u \to \infty} p_{\mt{B}} (u) \approx \vend + c_{\infty} \, e^{-u/\beta}, 
\end{equation}
one finds that the stress tensor approaches a constant,
\begin{equation}
 \lim\limits_{u \to \infty}  T_{uu} =  \frac{c}{48 \pi \beta^2},
 \label{hawkingf}
\end{equation}
and does not decay toward zero. However, we define mirrors for which the mapping function is of exponential form as in eq.~\eqref{eq:typeBexp} to be of type B$_+$, where $n \to \infty$. Furthermore, we define mirrors having dominant logarithmic asymptotic form
\begin{equation}
\lim\limits_{u \to \infty} p_{\mt{B}} \approx \vend + \frac{c_0}{\log \(u/\beta\) } \,,
\end{equation}
to be of type B$_-$ with $n \to 0$, since the corresponding energy flux becomes $ T_{uu} \approx -\frac{c}{48 \pi u^2}$. Similar exponential/logarithmic behavior also exists for other types. We summarize the subclassifications of the four mirror types in tables~\ref{table:AB} and \ref{table:CD}, and the special cases in table~\ref{table:ABCD}.

\begin{table}[h!]
	\centering
	\begin{tabular}{ |p{1cm}|p{3cm}|p{1.5cm}||p{1cm}|p{4cm}|p{2cm}|}
		\hline
		\multicolumn{6}{|c|}{Special cases} \\
		\hline
		Class & Leading $ p(u)$ &  Leading $\frac{24 \pi }{c}T_{uu}$ 	& Class & Leading  $ p(u)$ &  Leading $\frac{24 \pi }{c}T_{uu}$ \\
		\hline 
		A$_+$ & $  c_{\infty} e^{u/\beta}$   &   $ \frac{1}{2\beta^2} $ & C$_+ $ &  $  c_{\infty} \exp \( \frac{\beta}{\uend -u}\) $    &  $  \frac{\beta^2}{2(\uend-u)^4}$\\
		\hline
		A$_-$  & $  c_{0} \log \( u/\beta\)$   &  $ \frac{-1}{2u^2}$   
		& C$_-$ & $  c_{0} \log \( \frac{\beta}{\uend -u} \) $   &  $\frac{-1}{2(\uend-u)^2}$ \\
		\hline
		B$_+$  &  $ \vend+ c_{\infty} e^{-u/\beta}$  &  $ \frac{1}{2\beta^2} $ 
	&	D$_+$ &  $ \vend+ \frac{c_{\infty}}{\exp \( {\beta}/(\uend -u)\) }$    &  $ \frac{\beta^2}{2(\uend-u)^4}$ \\
		\hline
	B$_-$ &  $  \vend+ \frac{c_0}{\log \( u/\beta\)} $   &  $ \frac{-1}{2u^2}$ 
	& D$_-$ & $  \vend+ \frac{c_{\infty}}{\log \( {\beta}/(\uend -u)\) }$   &  $\frac{-1}{2(\uend-u)^2}$ \\
		\hline
	\end{tabular}	
	\caption{Special cases corresponding to $n \to \infty$ and $n \to 0$ for each type.}
	\label{table:ABCD}
\end{table}

\section{Entanglement entropy in moving mirrors}
\label{sec:3}
In this section, we study the behavior of the time evolution of entanglement entropy in moving mirror models. We start with a brief review of how we construct the gravity duals of moving mirrors and of how we calculate the entanglement entropy, mainly following \cite{Akal:2020twv,Akal:2021foz}. Then we move on with our analysis of entanglement entropy in various types of moving mirror models.

\subsection{AdS/BCFT and holographic entanglement entropy}
\begin{figure}[h!]
	\centering		
	\includegraphics[width=3in]{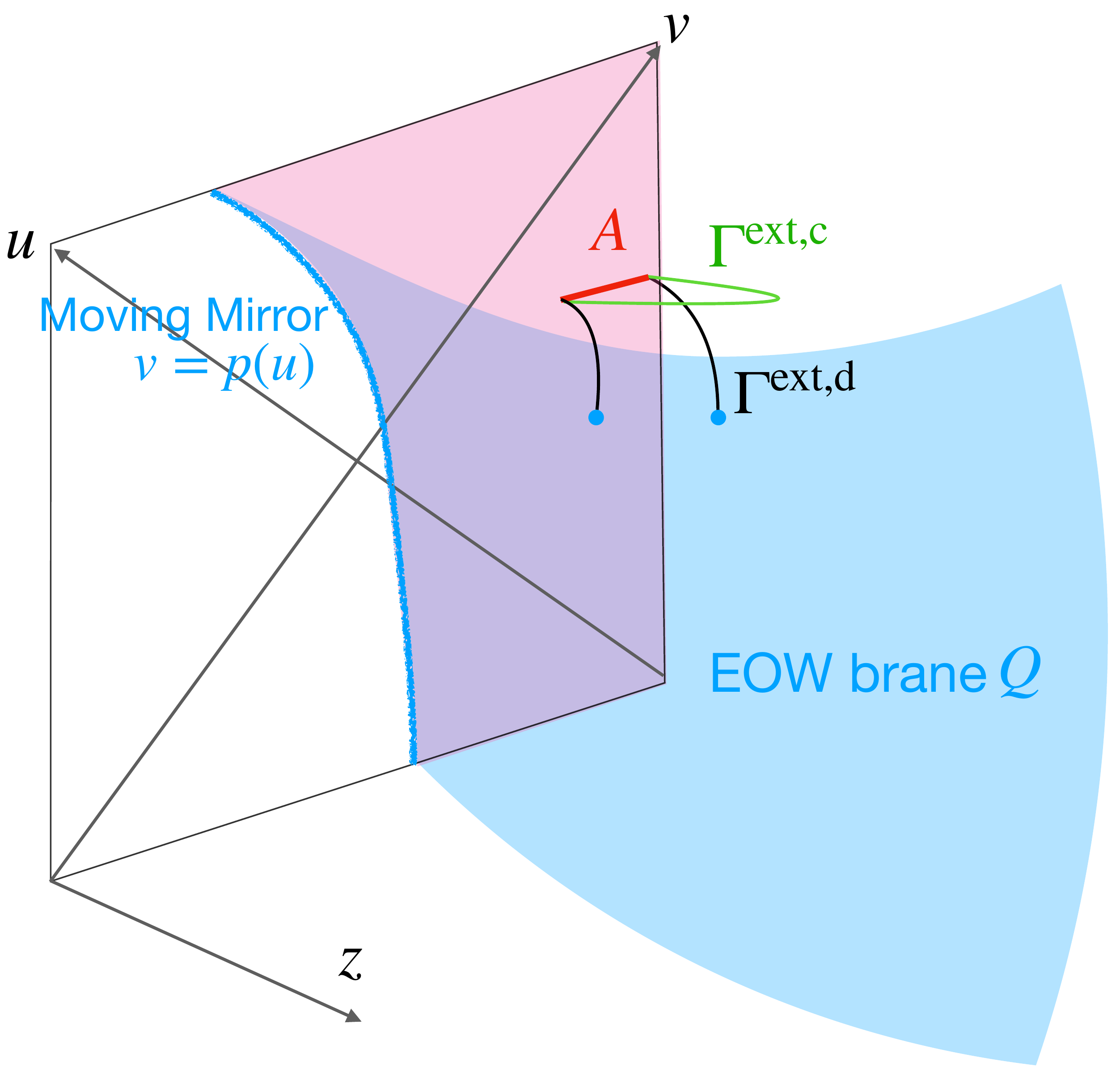}
	\caption{A sketch of the AdS/BCFT construction and computation of holographic entanglement entropy.}
	\label{fig:AdSBCFTHEEHEE}
\end{figure}
An important class of gravity duals of CFTs with conformal boundaries, so-called BCFTs, is provided by the AdS/BCFT construction \cite{Takayanagi:2011zk,Fujita:2011fp}. 
In AdS/BCFT, the gravity dual of a BCFT on a manifold $\Sigma$ with its boundary $\de\Sigma$, 
is given by a region $M$, surrounded by an asymptotic AdS boundary region $\Sigma$ and EOW brane $Q$ such that $\de M=\Sigma\cup Q$. 
In order to preserve the boundary conformal invariance on $\de\Sigma$, the EOW brane satisfies the Neumann boundary condition,
\ba
K_{ab}-h_{ab}K=-{\cal T}h_{ab}.
\label{ndycon}
\ea
The metric of the gravity dual $M$ is determined by solving the Einstein equations with a negative cosmological constant, where the Neumann boundary condition eq.~\eqref{ndycon} is imposed on $Q$, and the standard Dirichlet boundary condition is imposed on $\Sigma$.

The holographic entanglement entropy (HEE) \cite{Ryu:2006bv,Ryu:2006ef,Hubeny:2007xt} in AdS/BCFT setups can be computed according to the following rule as worked out in \cite{Takayanagi:2011zk,Fujita:2011fp}. Consider a subsystem $A$ on a constant time slice and introduce the reduced density matrix $\rho_A$ by tracing out its complement $\bar{A}$, leading to $\rho_A=\mbox{Tr}_{\bar{A}}[|\Psi\lb\la\Psi|]$, where $|\Psi\lb$ is the quantum state at the specified time. The entanglement entropy $S_A$ is defined as the von Neumann entropy $S_A=-\mbox{Tr}[\rho_A\log\rho_A]$. In AdS/BCFT, we can calculate the holographic counterpart of this entanglement entropy as the area of a codimension two extremal surface $\Gamma^{\rm ext}_A$ which ends on the boundary $\de A$ on $\Sigma$. The important point is that the extremal surface can also end on the EOW brane $Q$.
Typically, there are thus two candidates: one is the connected extremal surface, $\Gamma^{\rm ext,c}_A$, and the other one is the disconnected surface, $\Gamma^{\rm ext,d}_A$. The HEE is given by
the smaller one \cite{Takayanagi:2011zk,Fujita:2011fp},
\ba
S_A=\mbox{min}\left\{S^{\text{con}}_A,S^{\text{dis}}_A\right\},
\label{eq:HEE}
\ea
where we have introduced the connected HEE, $S^{\text{con}}_A$, and the disconnected one, $S^{\text{dis}}_A$, defined by 
\ba
S^{\rm con}_A=\frac{\mbox{Area}(\Gamma^{\rm ext,c}_A)}{4G_N},\ \ \ \ 
S^{\rm dis}_A=\frac{\mbox{Area}(\Gamma^{\rm ext,d}_A)}{4G_N}.  \label{EXTS}
\ea
Refer to Fig.~\ref{fig:AdSBCFTHEEHEE} for a sketch of this calculation.

In this paper, we focus on AdS$_3/$BCFT$_2$, where the spacetime $M$ is three-dimensional, and work out the gravity duals of moving mirrors. When we choose the subsystem $A$ to be an interval, the surface $\Gamma^{\rm ext,c}_A$ is the geodesic connecting the two endpoints of $A$. On the other hand, $\Gamma^{\rm ext,d}_A$ is the union of the two disconnected geodesics connecting each of the two endpoints of $A$ with a point on the surface $Q$. The latter point is determined by minimizing the geodesic length.

\subsection{AdS/BCFT with moving mirrors}\label{sec:AdSBCFT}

As seen in the previous section, we can describe moving mirrors by the conformal map (\ref{eq:chiraltrans}) from a half plane. The gravity dual in the AdS/BCFT can be found by lifting this two-dimensional conformal map to the three-dimensional AdS geometry. First, the gravity dual of a half plane $X>0$ 
is given by the right half of Poincar\'e AdS$_3$,
\ba
&& ds^2=\frac{-dUdV+d\eta^2}{\eta^2},  \label{Poin} \no
&&  V-U+2\lambda \eta>0 \,, \label{EOWla}
\ea
where we have introduced the light cone coordinates $U=T-X$ and $V=T+X$.  The EOW brane is located at
$X+\lambda \eta=0$, and this solves the Neumann boundary condition (\ref{ndycon}) with tension 
\begin{equation}\label{relavqz}
    \lambda=\frac{{\cal T}}{\s{1-{\cal T}^2}}\,.  
\end{equation} 
We note that the brane tension for {\it timelike} brane $Q$ satisfies $|\mathcal{T}|<1$, \ie
\begin{equation}\label{eq:contrainT}
-1\le \mathcal{T} = \frac{\lambda}{\sqrt{1+\lambda^2}} \le  1\,, 
\end{equation}
with $\lambda$ being a real dimensionless parameter. Then, we perform the coordinate transformation dual to the conformal map (\ref{eq:chiraltrans}),
\begin{equation}\label{cordtr}
\begin{split}
U&=p(u)\,,\\
V&=v+\frac{p''(u)}{2p'(u)}z^2\,,\\
\eta&=z \s{p'(u)}\,.  
\end{split}
\end{equation}
Eventually, the gravity dual of the moving mirror whose trajectory is parameterized by $v=p(u)$, can be derived from Poincar\'e coordinates. With the transformation 
eq.~\eqref{cordtr}, one obtains
\ba
ds^2=\frac{dz^2}{z^2}+\frac{12\pi}{c}T_{uu}(u)(du)^2-\frac{1}{z^2}dudv \,,
\label{metkads}
\ea
with
\ba
T_{uu}(u)=\frac{c}{12\pi} \times \frac{3(p'')^2-2p'p'''}{4p'^2} \,,
\ea
as the non-vanishing energy stress tensor.

\subsection{Entanglement entropy}

In gravity duals of moving mirrors, we can calculate the HEE by following the formulae (\ref{eq:HEE}) and (\ref{EXTS}), and using the transformation into Poincar\'e AdS (\ref{cordtr}), where the calculation of HEE is straightforward.\footnote{It is, in fact, nontrivial, why such a computation can successfully reproduce entanglement entropies in moving mirror setups. See appendix \ref{sec:LIandDM}.} 

For example, let us take subsystem $A$ as a finite interval $A=[x_1,x_2]$. There are both connected and disconnected geodesics, and we need to take the smaller one as in (\ref{eq:HEE}). Each of them at the time $t$ explicitly reads
\ba
&& S^\text{con}_A=\frac{c}{6}\log\frac{(v_2-v_1)(p(u_2)-p(u_1))}{\ep^2\s{p'(u_1)p'(u_2)}}\,,
\label{conee2} \no
&& S^\text{dis}_A=\frac{c}{6}\log\frac{v_1-p(u_1)}{\ep\s{p'(u_1)}}+\frac{c}{6}\log\frac{v_2-p(u_2)}{\ep\s{p'(u_2)}} + 2 S_{\rm bdy}\,,
\label{disee2} 
\ea
where $u_{1,2}=t-x_{1,2}$, $v_{1,2}=t+x_{1,2}$, and 
\begin{equation}\label{eq:bdry entropy}
S_{\rm bdy}=\frac{c}{6}\log\s{\frac{1+{\cal{T}}}{1-{\cal{T}}}}= \frac{c}{6}  \log \( \sqrt{\lambda^2 + 1}+\lambda  \),
\end{equation}
here denotes the boundary entropy \cite{Affleck:1991tk}. 
Notice that the entropy result above can only be applicable to holographic CFTs as the two point function of twist operators in the presence of a boundary is not universal.  

A special case is the choice when subsystem $A$ is taken to be a semi-infinite line $A=[x_0,\infty)$. In this case, it is clear that only the disconnected HEE is available.
Therefore we find
\begin{equation}\label{eq:defineSEEA}
S_{A} =\frac{c}{6} \log \frac{V_{0}-U_{0}}{\epsilon \sqrt{p^{\prime}\left(u_{0}\right)}}+S_{\mathrm{bdy}}=\frac{c}{6} \log \frac{v_{0}-p\left(u_{0}\right)}{\epsilon \sqrt{p^{\prime}\left(u_{0}\right)}}+S_{\mathrm{bdy}} \,, 
\end{equation}
where we have set $u_0=t-x_0$ and $v_0=t+x_0$. It is important to note that this result \eqref{eq:defineSEEA} is true for any BCFT including those not holographic. This is because the entanglement entropy involves the one-point function of twist operator in the replica method calculation \cite{Calabrese:2004eu,Calabrese:2009qy}, which is universal up to the value of the boundary entropy.

\subsection{Entanglement entropy for static, semi-infinite interval}
\label{sec:EEa}

\begin{figure}[h!]
	\centering		
	\includegraphics[width=3in]{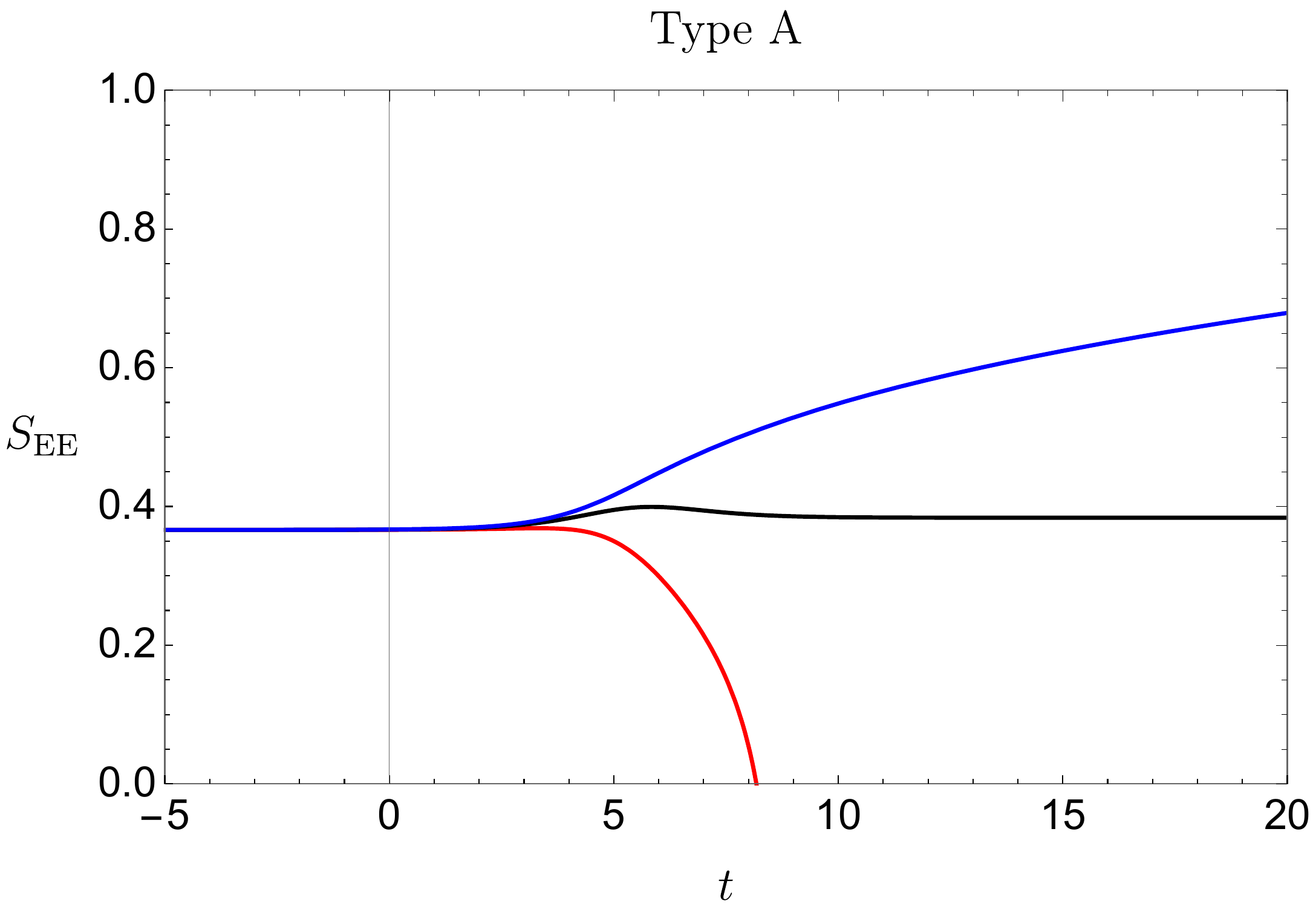}
	\includegraphics[width=3.0in]{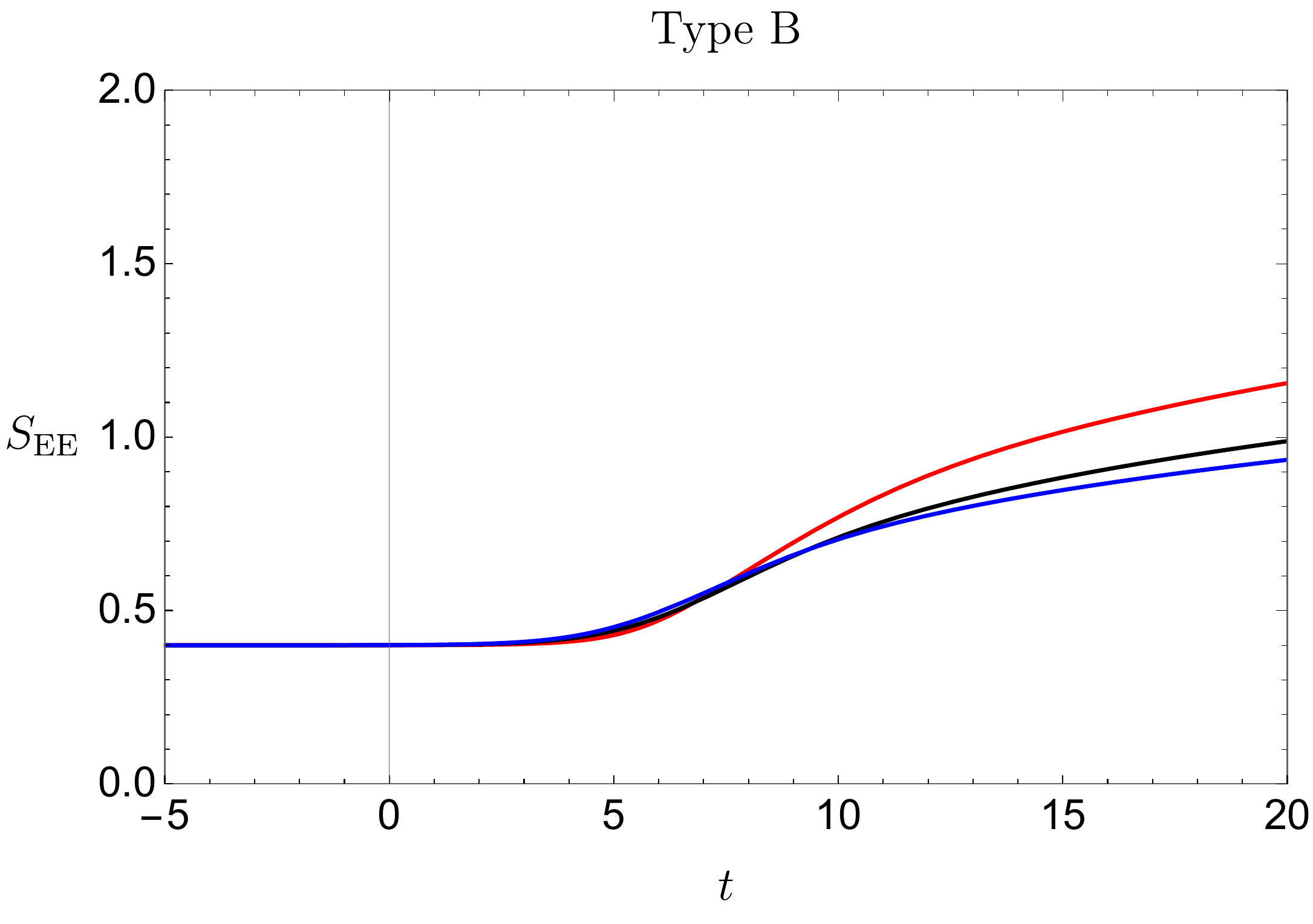}
		\centering	
	\includegraphics[width=3in]{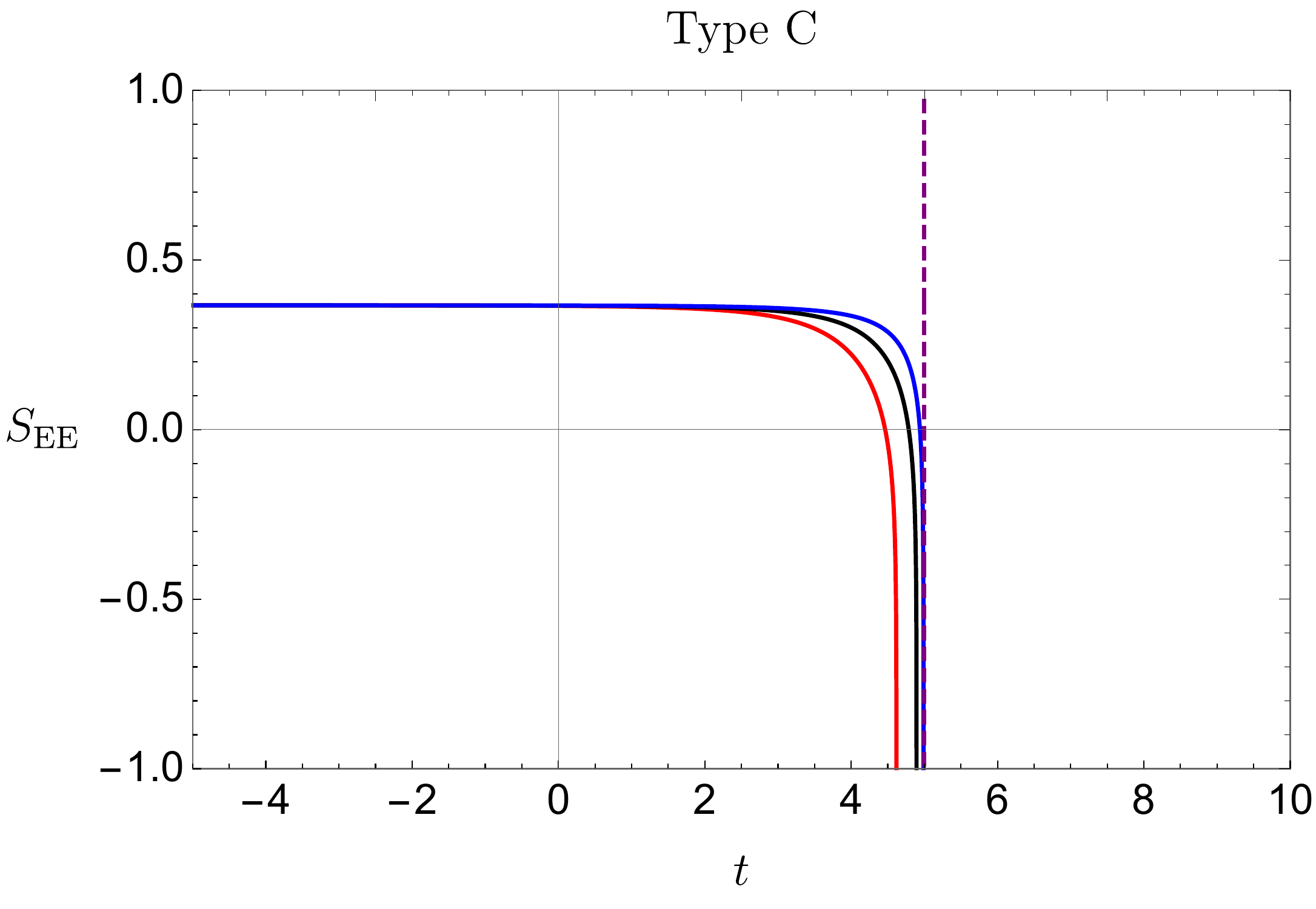}
	\includegraphics[width=3in]{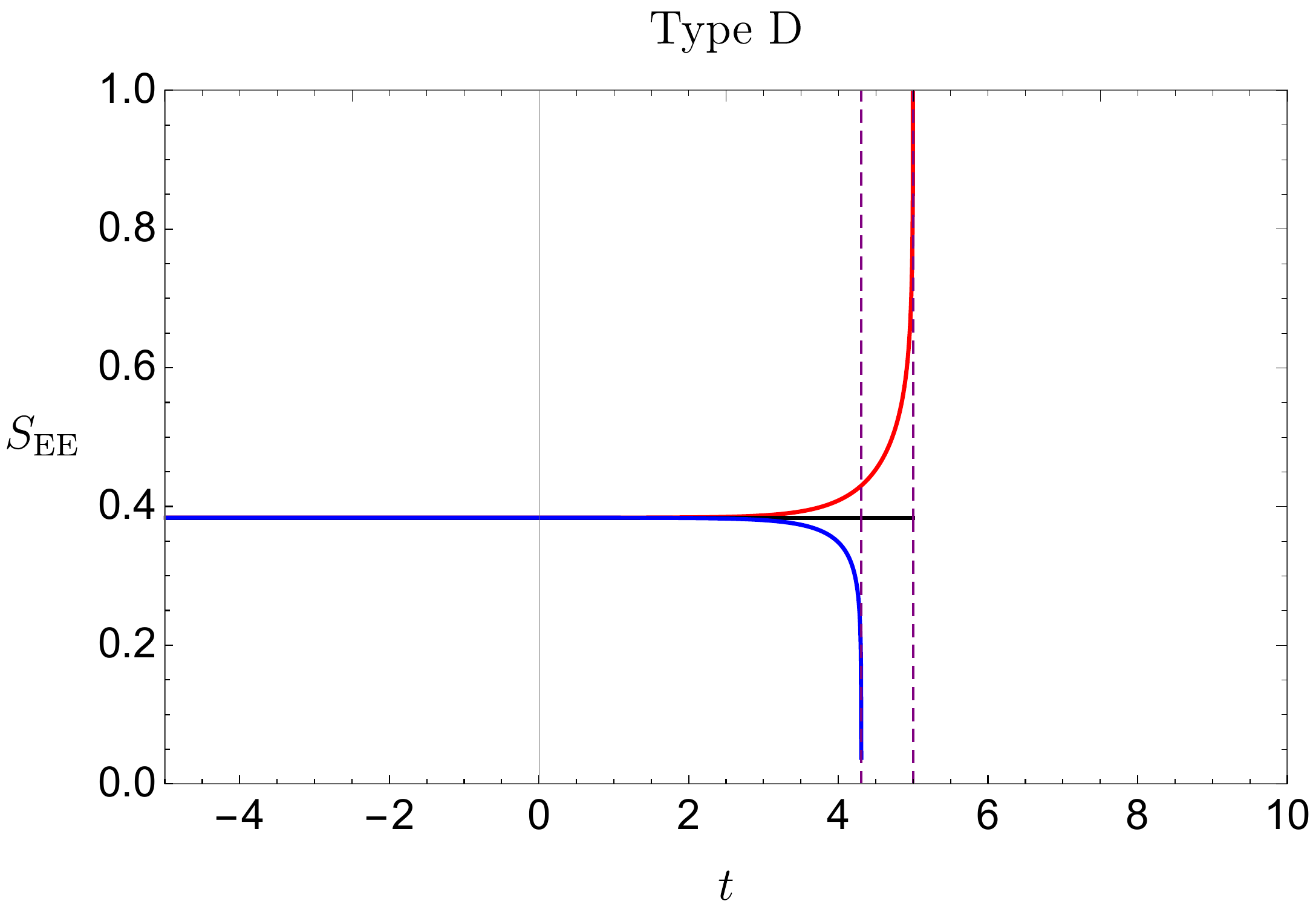}
	\caption{Entanglement entropy for a static, semi-infinite interval $A \in [x_0=5,+\infty]$. The mapping functions $p(u)$ for mirrors of type A, B, C, and D are respectively defined in eq.~\eqref{eq:defineTypeA}, eq.~\eqref{eq:defineTypeB}, eq.~\eqref{eq:defineTypeC}, and eq.~\eqref{eq:defineTypeDp}, eq.~\eqref{eq:defineTypeDm}. The red, black, and blue curves correspond to subclasses $+, 0, -$, for which we have chosen $n=2, n=1$, and $n=\frac{1}{2}$ in each plot.}
	\label{fig:SEEABCD}
\end{figure}
First, we examine the case where subsystem $A$ is a static, semi-infinite line, so we fix $x_0$ as a time-independent constant. The entanglement entropy $S_A$ associated with various moving mirrors follows from eq.~\eqref{eq:defineSEEA} by using the corresponding mapping function $p(u)$. In Fig.~\ref{fig:SEEABCD}, we show the results for the previously identified four mirror types and their subclasses. Because the initial state at past infinity is the ground state due to $p(u) \sim u$ around $u \to - \infty$, the entropies $S_A$ for all four types turn out to be constant at early times, \ie
\begin{equation}
S_A \approx \frac{c}{6} \log \frac{2x_0}{\epsilon} +S_{\rm bdy}\,.
\end{equation}

In order to consider the time-dependence of the entanglement entropy eq.~\eqref{eq:defineSEEA} for the static, semi-infinite interval $A$, one can find that the time derivative is given by 
\begin{equation}\label{eq:timeSEE}
\frac{\partial S_A}{\partial  t} =   \frac{c}{6} \(   \frac{1- p'(u_0)}{v_0 - p(u_0)}  - \frac{p''(u_0)}{2p'(u_0)} \).
\end{equation}
With this decomposition, we may understand the physical origin of the time evolution of the entanglement entropy as follows. The first term in eq.~\eqref{eq:timeSEE} goes back to the relative motion between the moving mirror and the interval $A$. The numerator $1-p'(u_0)$ captures the relative velocity of the mirror $ v_m = \frac{p'(u)-1}{p'(u)+1}$, see eq.~\eqref{eq:definevm}, where the denominator $v_0- p(u_0)$ is the distance between the interval $A$ and the mirror. The second term in eq.~\eqref{eq:timeSEE} captures the influence of the non-vanishing energy flux $T_{uu}(u)$. 

In what follows, we explain the features of the time evolution at late times for various mirror types shown in Fig.~\ref{fig:SEEABCD}. First, we note that both type A and type B mirrors extend to future infinity at $u \to \infty$, where the energy flux decays to zero, because $\frac{p''(u)}{2p'(u)}$ vanishes in the late-time limit. Note that the escaping mirror defined by eq.~\eqref{hawkingf} is an exception (with $n\to\infty$) because it leads to a constant and non-vanishing flux (\ref{hawkingf}). One should notice that type A$_+$ mirrors are moving closer to the interval $A$ and finally collide with it when $v_0= p(u_0)$. As a result, one always gets 
\begin{equation}
S_A \to 0
\end{equation}
at a finite time for type A$_+$ mirrors. On the contrary, mirrors of type A$_-$ move away from interval $A$. Correspondingly, one can consider the late-time behavior of $S_A$ by taking the limit $t \to \infty$. Noting that the late-time behavior for type A$_-$ mirrors is dominated by $p(u) \sim c_n u^n \sim c_n t^n$ with $0<n<1$, one obtains 
\begin{equation}
\frac{\partial S_A}{\partial  t} \bigg|_{t \to \infty} \approx  \frac{c}{6} \(  \frac{1}{v_0 -c_n t^n }    -   \frac{n-1}{2t}  \)\approx  \frac{c}{6} \frac{3-n}{2t} \,,  
\end{equation}
which indicates the logarithmic growth in the entropy, namely
\begin{equation}
S_A \sim \frac{c}{6} \log \frac{(t)^{(3-n)/2}}{\epsilon} +\cdots.
\end{equation}

Similarly, we can find the logarithmic growth, shown in the top-right panel of Fig.~\ref{fig:SEEABCD}, for type B$_+$, type B$_0$ as well as type B$_-$ mirrors. More explicitly, one gets
\begin{equation}
S_A \sim \frac{c}{6} \log \frac{(t)^{(3+n)/2}}{\epsilon} +\cdots \,,
\end{equation}
whose growth rate depends on the explicit subclass in type B mirrors. Note that the escaping mirror (\ref{hawkingr}), corresponding to the limit $n\to\infty$, is an exception and its entanglement entropy at late times follows a linear growth, $S_A\propto t$, \cite{Akal:2020twv,Akal:2021foz}.

The time evolution of $S_A$ for the three subclasses in type C is similar to the one in type A$_+$, because type C mirrors keep moving towards the static interval $A$ and finally intersect with it at a finite time when $v_0= p(u_0)$. 

Finally, we find that the entanglement entropy for type D$_\pm$ obtains a singular behavior at 
\begin{equation}
 \uend =u_0 \equiv  t-x_0\,,
\end{equation}
as shown in the bottom-right panel of Fig.~\ref{fig:SEEABCD}. This is traced back to the fact that 
\begin{equation}
\begin{split}
p'(\uend)  = \begin{cases}
0 & \text{for} \qquad \text{D}_+ \,,\\
+\infty & \text{for} \qquad \text{D}_- \,.\\
\end{cases}
\end{split}
\end{equation}
However, we would like to address that this is different from the decreasing behavior for type C or type A$_+$. At the critical time $t= \uend +x_0$, subsystem $A$ hits a null shockwave of infinite energy, which origins from the endpoint of the type D$_\pm$ mirror. Namely, the endpoint of type D$_\pm$ reaches the (positive/negative) speed of light, the null shockwave at $u=\uend$ carries an infinite amount of energy and causes the divergence of $S_A$ when the subsystem passes through. Around the mentioned critical time, one finds 
\begin{equation}
S_A \approx  \begin{cases}
 \frac{c}{6} \log  \frac{1}{\epsilon (\uend +x_0 -t )^{(n-1)/2}}  \to +\infty & \text{for} \qquad \text{D}_+ ,\\
 \frac{c}{6} \log  \frac{ (\uend +x_0 -t )^{(1-n)/2}}{\epsilon} \to 0  & \text{for} \qquad \text{D}_- .\\
\end{cases}\label{EEDpm}
\end{equation}
Hence, this result motivates to interpret the null shockwave with infinite energy as the `boundary' of the physical spacetime. In the next section, we further explore this understanding from the viewpoint of the holographic dual spacetime. 

\subsubsection{Entanglement entropy and energy flux}
We here would like to comment on the close connection between energy flux and entanglement entropy for a static subsystem. Considering the non-vanishing energy stress tensor defined in eq.~\eqref{eq:stresstensor} and integrating by parts, one finds the total amount of energy that can be recast as 
\begin{equation}\label{eq:decomE01}
E_{\rm{st}}\equiv \int_{u_{\rm{min}}}^{u_{\rm{max}}} T_{u u} \,du = \frac{c}{48 \pi} \int \( \frac{p''(u)}{p'(u)} \)^2 du  -\( \frac{c}{24 \pi} \frac{p''(u)}{p'(u)} \)\bigg|_{u_{\rm{min}}}^{u_{\rm{max}}}  ,
\end{equation}
where the observer stays on future null infinity $\mathcal{I}_{R}^+$ along $u \in [u_{\rm min}, u_{\rm max}]$. The second term is a surface term that would not contribute for mirrors with vanishing $p''/p'$ as $u \to u_{\rm{min}}, u_{\rm{max}}$. As shown in Fig.~\ref{fig:TuuABCD}, the stress tensor $T_{uu}$ at a finite time $u$ can be even negative. However, ignoring the surface term in the integrated energy flux, we can find the effective energy flux given by the density 
\begin{equation}\label{eq:Teff}
T_{\rm{eff}}(u)  \sim    \frac{c}{48 \pi}  \( \frac{p''(u)}{p'(u)} \)^2  \ge 0 \,,
\end{equation}
which is positive by definition. 

For mirrors of type A and type B, we can evaluate the total energy flux in the physical spacetime by taking $u_{\rm min}, u_{\rm max} \to \pm \infty$. Because of 
\begin{equation}
\lim\limits_{u \to \pm \infty}\frac{p''(u)}{p'(u)} =  0 \,,
\end{equation}
we conclude that the total energy excited by mirrors of type A or type B is always positive, namely 
\begin{equation}
E_{\rm{st}} \ge 0 \,,
\end{equation}
This is nothing but the averaged null energy condition (ANEC), $\int T_{uu}du \ge 0$. However, the surface term,  
\begin{equation}
\partial E_{\rm{st}}=-  \frac{c}{24 \pi}\frac{p''(u_{\rm{end}})}{p'(u_{\rm{end}})}\,,
\end{equation}
still non-trivially contributes in the case of type C and type D mirrors, where the integral is performed along a null geodesic with $u \in (-\infty, \uend]$. In particular, from the time derivative of $S_A$ in eq.~\eqref{eq:timeSEE}, we find that the surface term $\partial E_{\rm st}$ influences the evolution of entanglement entropy at late times. For example, using the asymptotic analysis in table \ref{table:CD}, we get $\partial E_{\rm{st}} \sim \frac{n-1}{\uend -u}>0$ for type D$_+$ mirrors, $\partial E_{\rm{st}} < 0$ for type D$_-$, and $\partial E_{\rm{st}} \sim -\frac{n+1}{\uend -u}< 0$ for all type C mirrors. These results are consistent with the time evolution of entanglement entropy for type C and type D mirrors, as shown in Fig.~\ref{fig:SEEABCD}.

Of course, the decomposition in eq.~\eqref{eq:decomE01} is not unique. Considering the HEE determined by the disconnected geodesic, one finds the following decomposition of the non-vanishing stress tensor 
\begin{equation}
2 \pi T_{uu} = \partial_u \partial_u S_{A} + \frac{6}{c} \( \partial_u S_{A} \)^2 \,.
\end{equation}
The integrated energy flux can then be rewritten as 
\begin{equation}\label{eq:decomE02}
2 \pi E_{\rm{st}}= \int  \frac{6}{c} \( \partial_u S_{A}  \)^2 \, du +\(  \partial_u S_{A}     \)\bigg|_{u_{\rm{min}}}^{u_{\rm{max}}}  \,.
\end{equation}
Taking $u_{\rm min } \to -\infty$, we find that the time derivative of entanglement entropy at $u_\ast$ is related to the total energy integrated over $u \in (-\infty, u_\ast)$, 
\begin{equation}
\partial_u S_A (u_\ast) = 2 \pi E_{\rm{st}} -  \int^{u_\ast}_{-\infty}  \frac{6}{c} \( \partial_u S_{A}  \)^2 \, du\,,
\end{equation}
where we have assumed that the state at past infinity is a vacuum state. Hence, a negative total energy results in a monotonic decrease of entanglement entropy.

\subsection{Entanglement entropy for co-moving, semi-infinite interval}\label{sec:EEb}

\begin{figure}[h!]
	\centering		
	\includegraphics[width=3.0in]{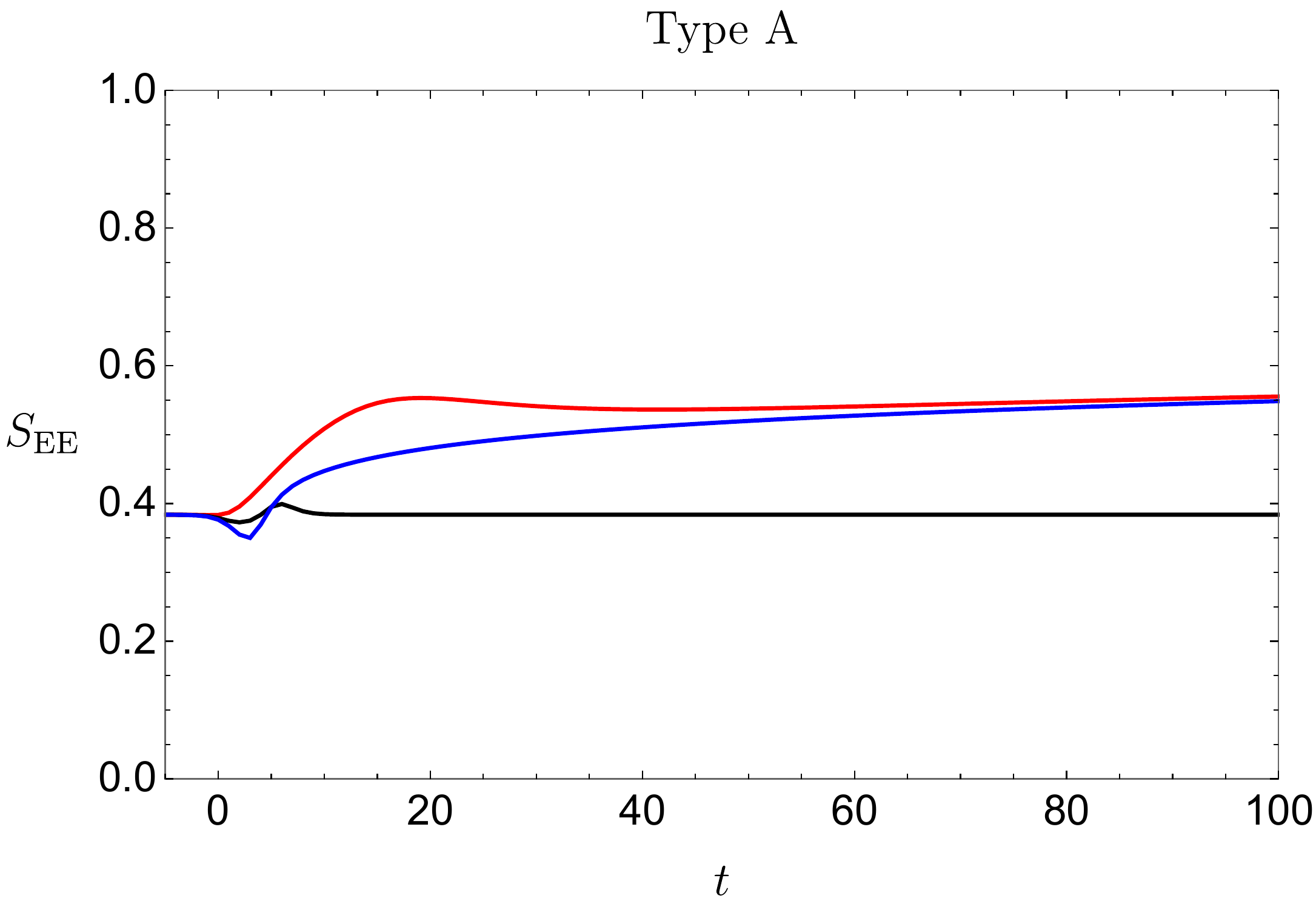}
	\includegraphics[width=3.0in]{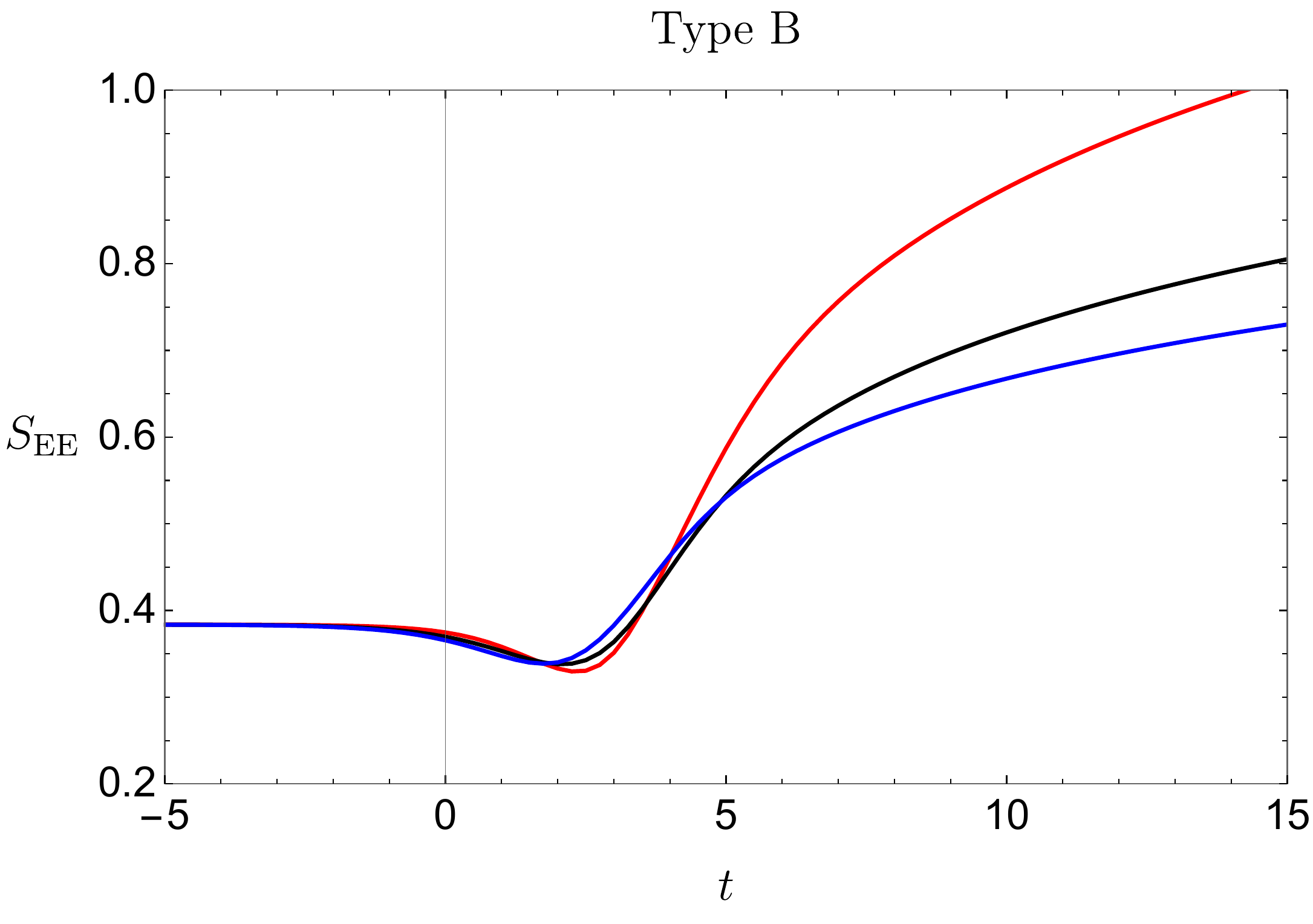}
	\includegraphics[width=3.0in]{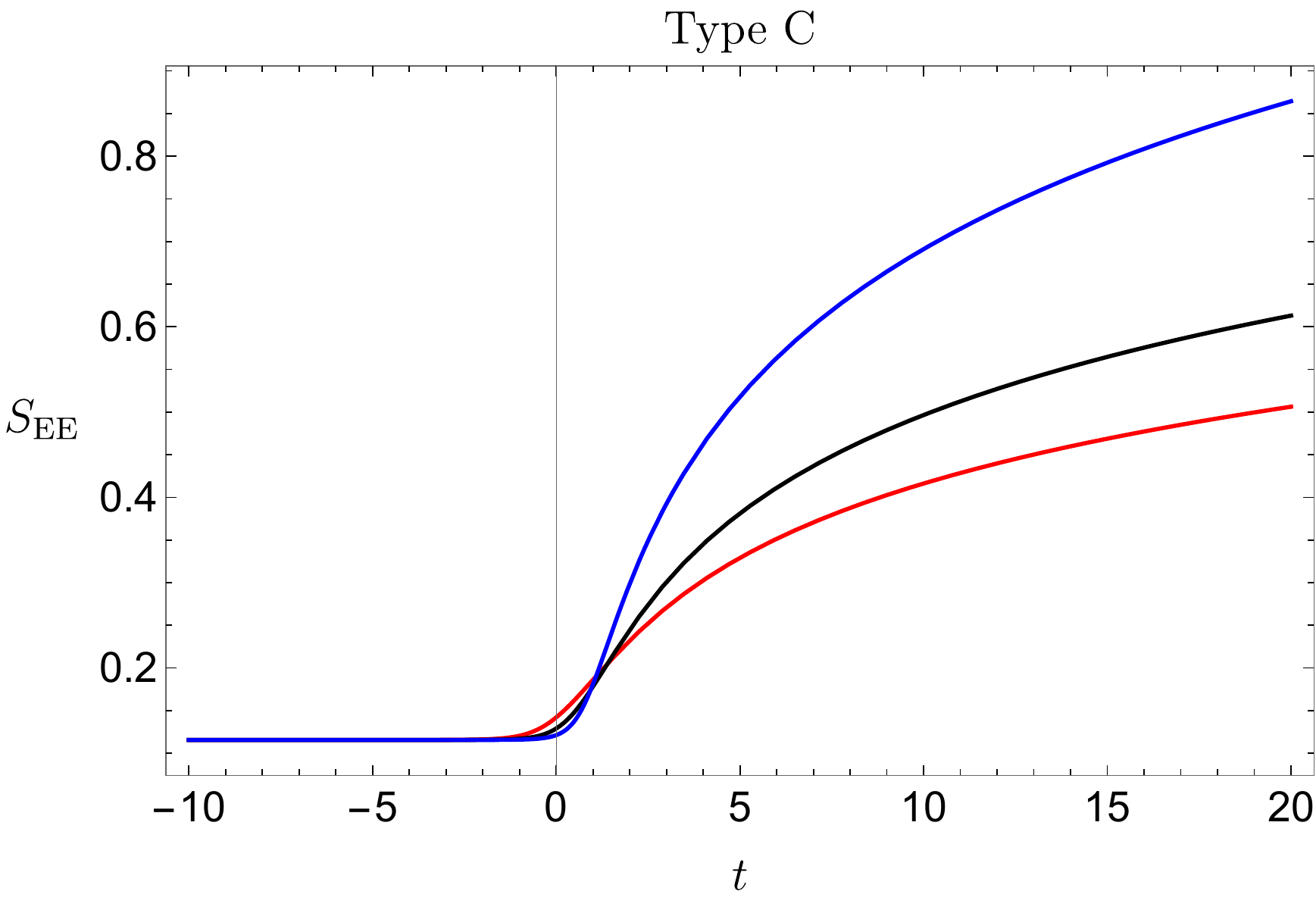}
	
	\centering	
	\caption{Entanglement entropy for a co-moving, semi-infinite interval with fixed endpoint $x_a=Z(t)+\Delta x$, where $\Delta x=5$ for all three plots. The red, black, and blue curves correspond to subclasses $+, 0, -$ with $n=2,1$, and $\frac{1}{2}$, respectively. The mapping functions $p(u)$ are chosen as in Fig.~\ref{fig:SEEABCD}.}
	\label{fig:SEEmovingABCD}
\end{figure}

\begin{figure}
    \centering
    \includegraphics[width=5in]{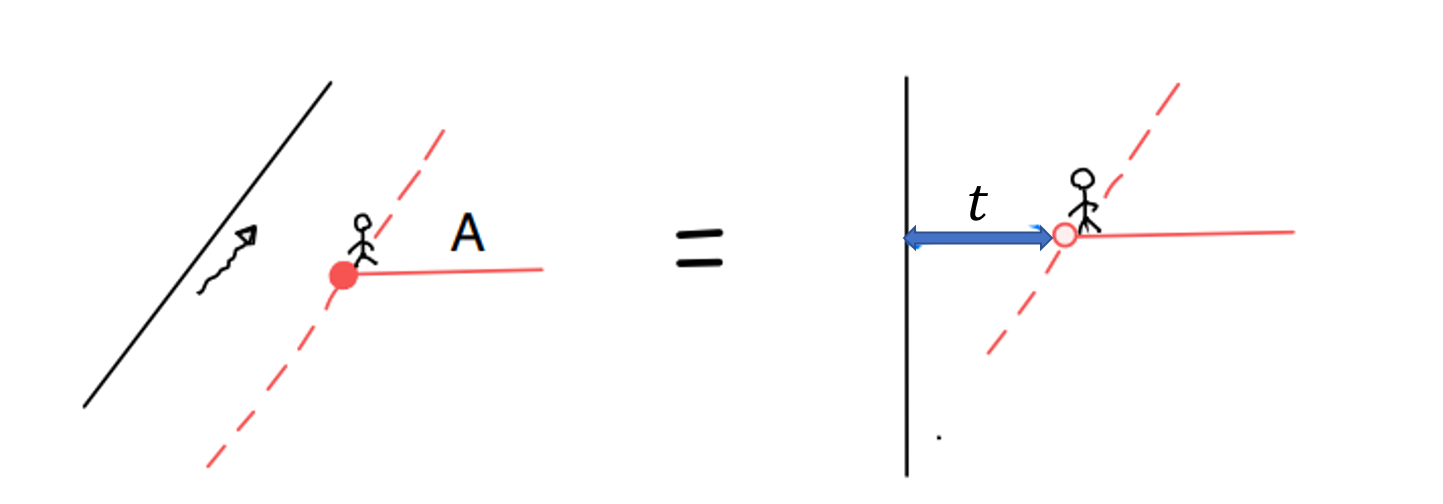}
    \caption{The physical interpretation for the universal behavior of entanglement entropy for a co-moving interval in the case of type C mirrors.}
    \label{fig:cute}
\end{figure}

As we have explained before, the intersection between mirror and a static interval arises in the case of type A$_-$ and type C, since the mirror moves toward the specified subsystem. This could be avoided by making the interval move as well. We, therefore, would like to examine the time evolution of entanglement entropy with a moving subsystem endpoint $x_0(t)$. In particular, we take the distance between $x_0$ and the mirror trajectory to be a constant $\Delta x$,
\begin{equation}
x_0 (t)= Z(t)+ \Delta x \,.
\end{equation}
The time evolution of entanglement entropy of a comoving interval for various mirror types\footnote{Since moving mirrors of type D terminate at a finite time, we ignore those here.} are shown in Fig.~\ref{fig:SEEmovingABCD}. Due to the co-moving subsystem $A$, the time evolution of entanglement entropy deviates from the one in the static interval case. In terms of the mirror trajectory function, $x= Z(t)$, we may write
\begin{equation}\label{eq:timeSee02}
\frac{\partial S_A}{\partial  t} =   \frac{c}{6} \(   \frac{1+Z'(t)- p'(u_0)(1-Z'(t))}{v_0 - p(u_0)}  -  \frac{p''(u_0)}{2p'(u_0)} (1-Z'(t)) \) \,,
\end{equation}
where $Z'(t) = \frac{p'(u)-1}{p'(u)+1}$ is referred to as the velocity of the mirror at time $t$. Considering the late-time limit $t \to \infty$, one finds that $\partial_t S_A \to 0$ for type A. Thus, the entanglement entropy for all three subclasses of type A always approaches a constant value. The entanglement entropy for type B mirrors, however, grows logarithmically as
\begin{equation}
    S_A\sim \frac{c}{6}\log\frac{t^{\frac{n+1}{2}}}{\ep}\,. 
\end{equation}
Similar logarithmic growth also applies to the late-time entanglement entropy of a co-moving, semi-infinite subsystem for type C mirrors,
\begin{equation}
S_A\sim \frac{c}{6}\log\frac{t}{\ep}\,.  
\label{EEloghl}
\end{equation}
A physical interpretation of this universal behavior independent of $n$ for type C mirrors can be given as follows. The velocity of the mirrors approaches the speed of light at late times. Therefore, an observer located on the endpoint of the co-moving interval also approaches the speed of light, and they cannot receive any information generated by the mirror trajectory. Since the mirror is initially static, this experience of the observer should be able to be explained by an (imaginary) always-static mirror. In this case, the distance between the observer and the always-static mirror grows as $t$ at late times. The behavior in (\ref{EEloghl}) can merely be predicted from the vacuum entanglement entropy in a two-dimensional CFT \cite{Holzhey:1994we}. In fact, we can explicitly show that the behavior of the entanglement entropy for such a half-infinite interval, \ie the one whose end-point moves to the right at the speed of light, is universally independent from the mirror profile. For such a subsystem, we have the endpoint given by $x_0(t)= t+\epsilon + \delta(t)$, where $\delta(t)\rightarrow0$ as $t\rightarrow\infty $, and hence
\begin{equation}
     v_0(t)= t+ x_0(t) = 2 t  +\epsilon + \delta(t),\quad u_0(t)= t- x_0(t) = -\epsilon -\delta(t)\,.
\end{equation}
We may consider an expansion of the entanglement entropy formula around $u_0 = -\epsilon$ at late times. From (\ref{eq:defineSEEA}), it follows that
\begin{equation}
    S_A(t) = \frac{c}{6}\log{\frac{2t + \epsilon +\delta(t)-(p(-\epsilon)-\delta(t)p'(-\epsilon)+\cdots)}{\epsilon\sqrt{p'(-\epsilon)}(1-\frac{1}{2}\delta(t)\frac{p''(-\epsilon)}{p'(-\epsilon)}+\cdots)}} \xrightarrow{t\rightarrow\infty} \frac{c}{6}\log\frac{t}{\epsilon} \,.
\end{equation}
We refer to Fig.~\ref{fig:cute} for an intuitive sketch.

\section{Gravity dual of moving mirrors}
\label{sec:4}

In this section, we explore the gravity duals of various moving mirrors by studying the EOW brane in AdS$_3$ bulk spacetime. In particular, we briefly review the results associated with type A and type B mirrors which have been studied before \eg in \cite{Akal:2020twv,Akal:2021foz}, and then focus on type C and type D mirrors, for which we find that the null surface related to the infinite energy flux at $u=\uend$ plays an important role. 

\subsection{End-of-the-world brane in AdS/BCFT}
Using the position of the brane $Q$ in Poincar\'e coordinates, we find that the induced geometry on the EOW brane $Q$ 
reads 
\begin{equation}\label{eq:braneQAdS2}
d s^{2}=\frac{\left(1+\lambda^{2}\right) d \eta^{2}-d T^{2}}{\eta^{2}}\,,
\end{equation}
which is nothing but an AdS$_2$ geometry. Obviously, the intersection between the EOW brane and the asymptotic boundary located at $\eta \to 0$ is the mirror at $\tilde{x}=0=X$. We note here that this has implied that the intersection, \ie the mirror, is timelike owing to $ds^2 \propto- dT^2 <0 $.
On the other hand, we can start from a holographic AdS$_3$ spacetime using the Ba\~nados map and get the corresponding brane geometry 
\begin{equation}\label{eq:braneQAdS}
 d s^{2}=\frac{d z^{2}}{z^{2}}+\left(\frac{p^{\prime \prime}}{z p^{\prime}}+\frac{2 \lambda \sqrt{p^{\prime}}}{z^{2}}\right) d u d z+\left(\frac{p^{\prime \prime 2}}{4 p^{2}}-\frac{p^{\prime}}{z^{2}}+\frac{\lambda p^{\prime \prime}}{z \sqrt{p^{\prime}}}\right) d u^{2} \,,
\end{equation}
by fixing the position of the brane as 
\begin{equation}\label{eq:EOWbraneeq}
v_{\rm{brane}}=-\frac{p^{\prime \prime}z^{2} }{2 p^{\prime}}  + p(u) -  2\lambda z \sqrt{p^{\prime}} \,.
\end{equation}
Of course, the two induced metrics on the brane, \ie eq.~\eqref{eq:braneQAdS} and eq.~\eqref{eq:braneQAdS2}, are equivalent according to the coordinate transformation 
\begin{equation}\label{eq:coordinatetrans}
\begin{split}
\eta &= z \sqrt{p'(u)} \,,\\
T&= p(u)- \lambda z \sqrt{p'(u)}\,.
\end{split}
\end{equation}
We remind that the physical spacetime is given by the right region of the EOW brane $Q$.

\subsection{Gravity duals of type A and type B mirrors}
\begin{figure}[!]
	\centering		
	\includegraphics[width=3in]{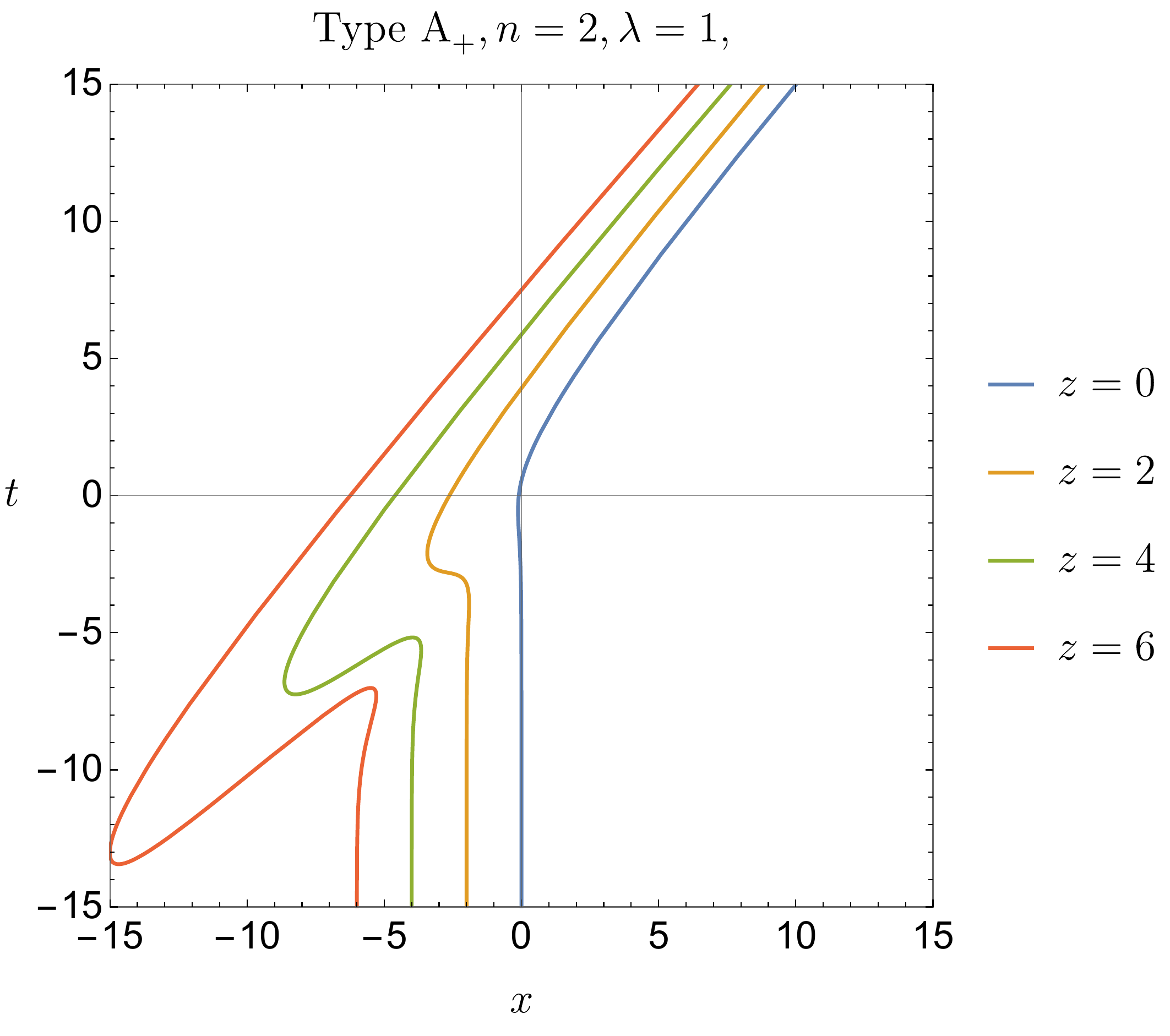}
	\includegraphics[width=3.0in]{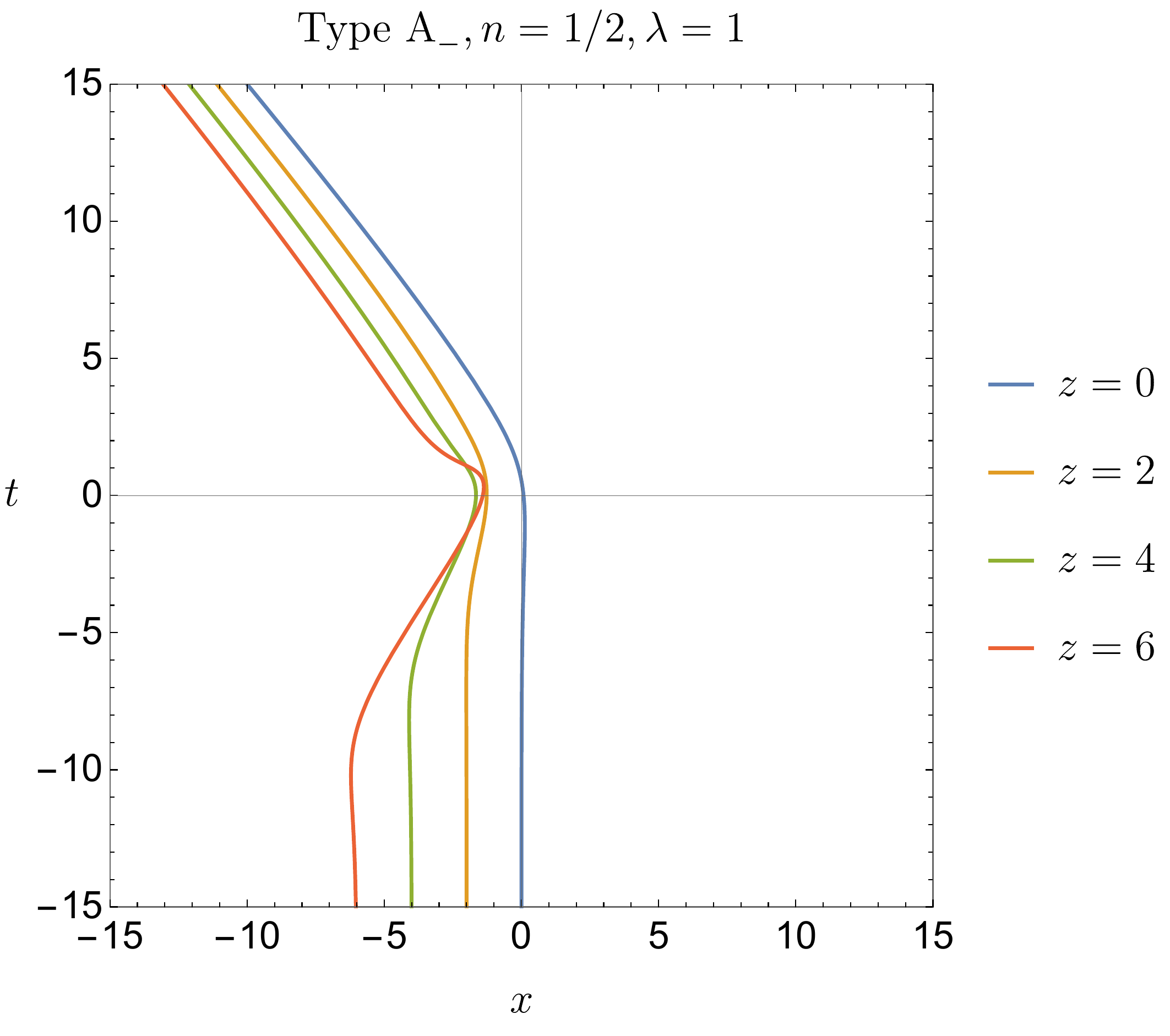}
	\includegraphics[width=3.0in]{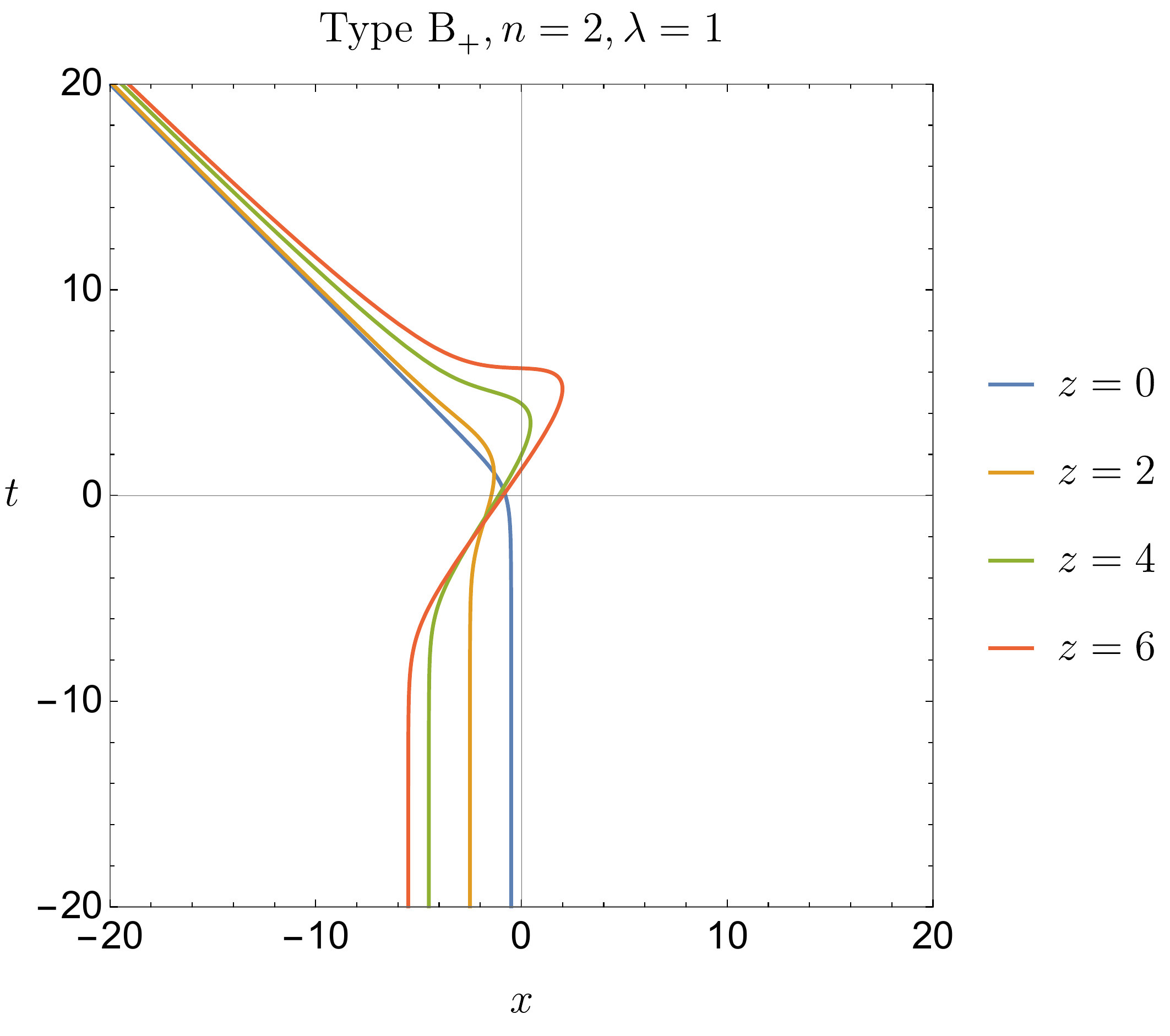}
	\includegraphics[width=3.0in]{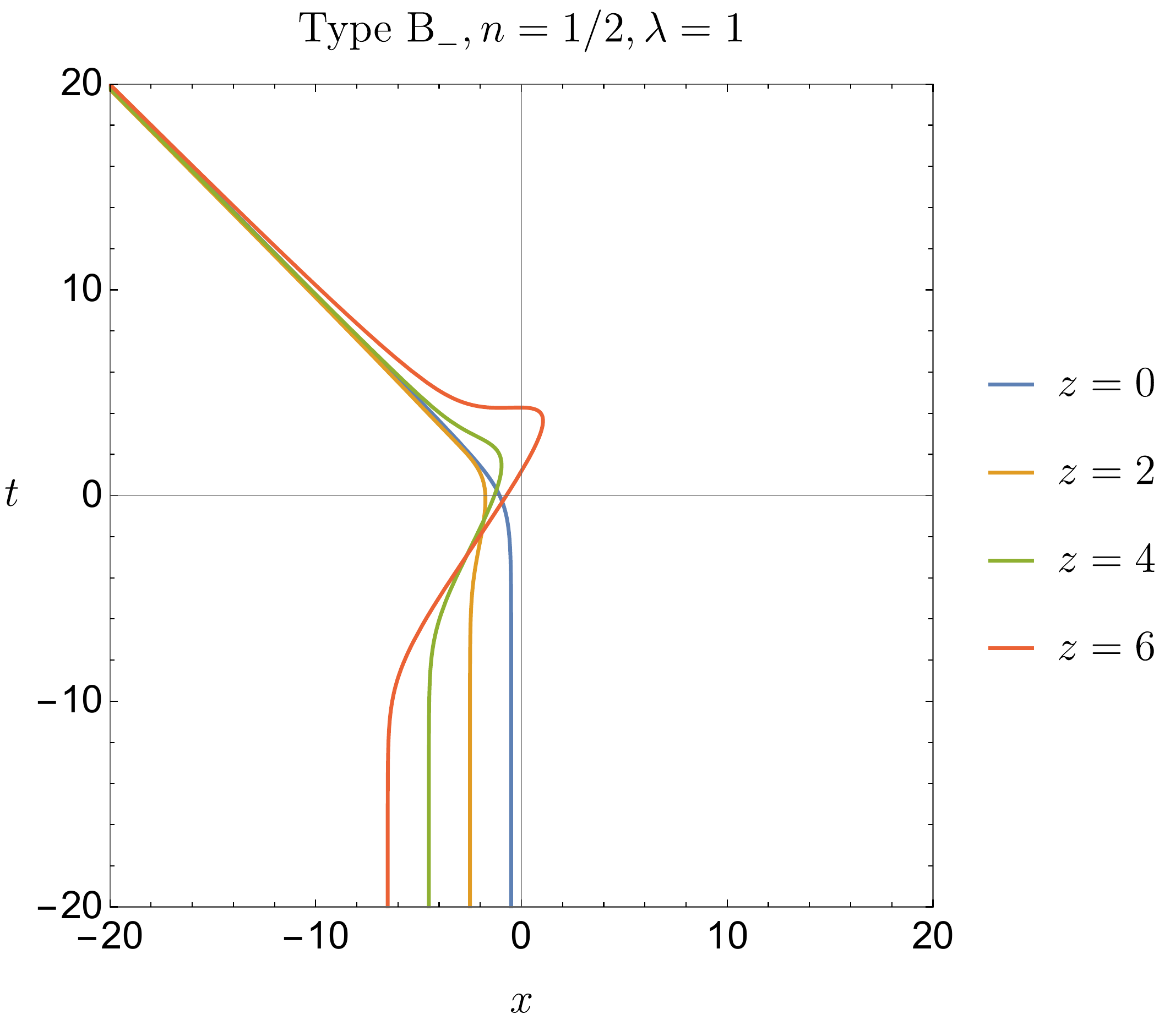}
	\caption{Profiles of brane $Q$ associated with type A and type B mirrors in terms of $(t,x)$ coordinates and for various $z$. Top: the mapping functions for type A are given in eq.~\eqref{eq:defineTypeA}. Here we have chosen $\beta=1$ and $u_0=0$ as previously. Bottom: the mapping functions for type B are given in eq.~\eqref{eq:defineTypeB}. We have set $\beta=1=u_0$.}
	\label{fig:BraneAB}
\end{figure}

 The profiles of EOW branes described by $v_{\rm{brane}}$ for type A mirrors are depicted in Fig.~\ref{fig:BraneAB}. As we may already expect, the brane profiles for type A$_+$ and type A$_-$ show a different behavior, because mirrors of type A$_-$ move away from the physical system. Correspondingly, the holographic duals for the escaping mirrors of type B$_\pm$ and type B$_0$ are similar. These have analogous features with mirrors of type A$_-$, as shown in  Fig.~\ref{fig:BraneAB}. We remark that the profiles of the EOW brane for type A and type B mirrors are always regular, since the mirrors keep being timelike at any finite time. We refer interested readers to \cite{Akal:2021foz} for more detailed studies about holographic duals of moving mirrors of type A and type B. We discuss moving mirrors of type C and type D in more detail in the following,

\subsection{Gravity dual of type C mirrors}
In contrast to the type A and type B cases, moving mirrors of type C are timelike at early times but become lightlike at a finite $u$, $u=u_\text{end}$. In other words, the mirror accelerates in the $x$ direction and finally reaches the speed of light at $\uend$. They are moving in the opposite direction to mirrors of type B, so we call mirrors of type C chasing mirrors.

In order to show the differences between the type A and type B cases, we consider a family of mapping functions describing type C mirrors,
\begin{equation}\label{eq:defineTypeC}
\text{type C}: \qquad p_{\mt{C}}(u)=u + \frac{\beta}{ \(e^{\uend/\beta}-e^{u/\beta}\)^n} \,,  \qquad \text{with} \qquad n>0\,.
\end{equation}
The asymptotic behavior around the endpoint of the trajectory at $u=\uend$ is of the form 
\begin{equation}\label{eq:type C}
\lim\limits_{u \to \uend}  p_{\mt{C}}(u) \approx  \frac{\beta}{ \( e^{\uend/\beta} (\uend -u )\)^n} \,.
\end{equation} 
This implies that the mapping function eq.~\eqref{eq:defineTypeC} is of type C$_+$, C$_0$ and C$_-$ by taking $n>1, n=1$, and $n<1$, respectively.
By taking $u_{\rm{end}}=0$ we show the corresponding mirror trajectories in Fig.~\ref{fig:TypeC01} as well as the non-vanishing stress tensor in Fig.~\ref{fig:TuuABCD}.

\subsubsection{Type C$_+$}
\begin{figure}[h]
	\centering		
	\includegraphics[width=3.0in]{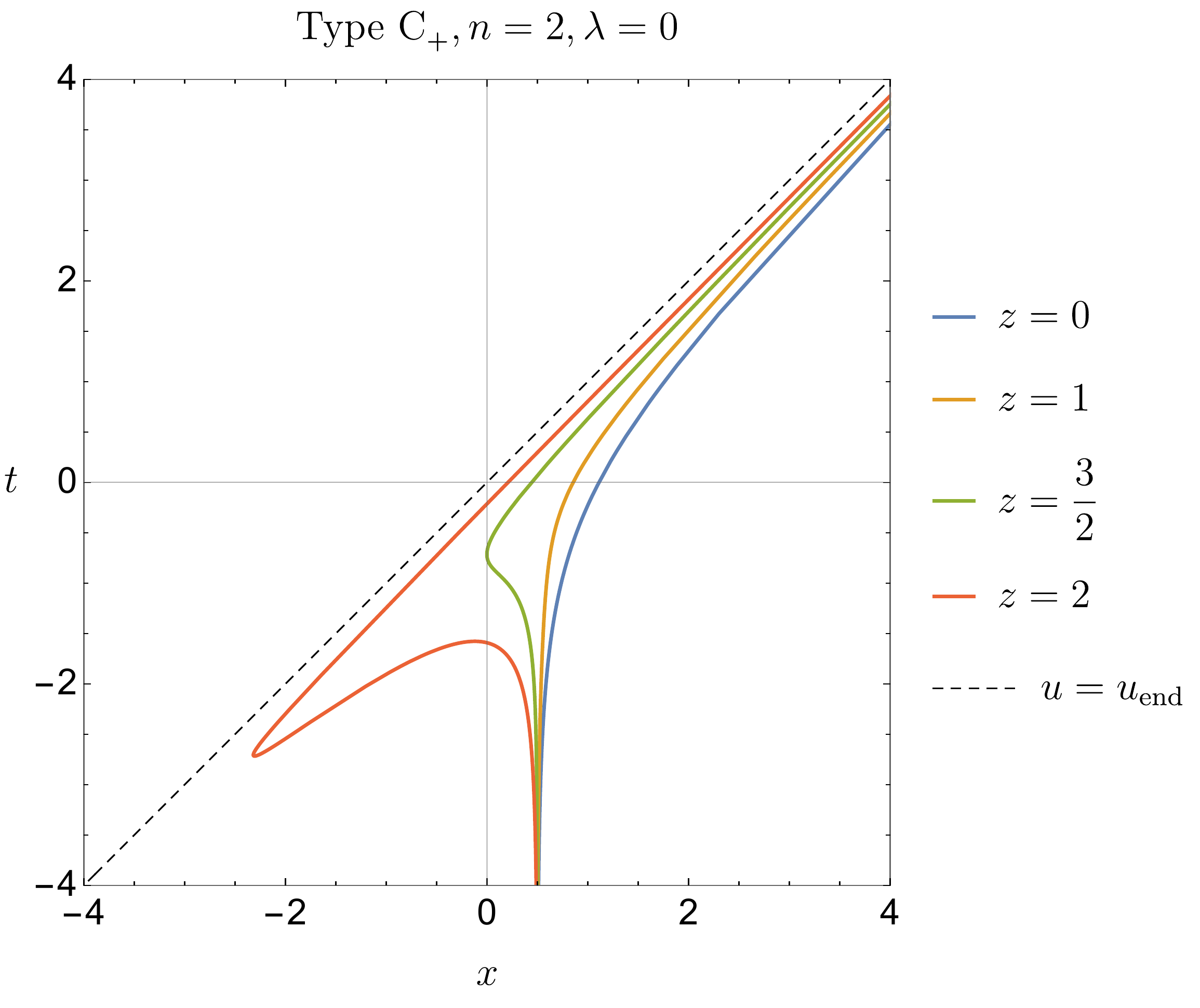}
	\includegraphics[width=3.0in]{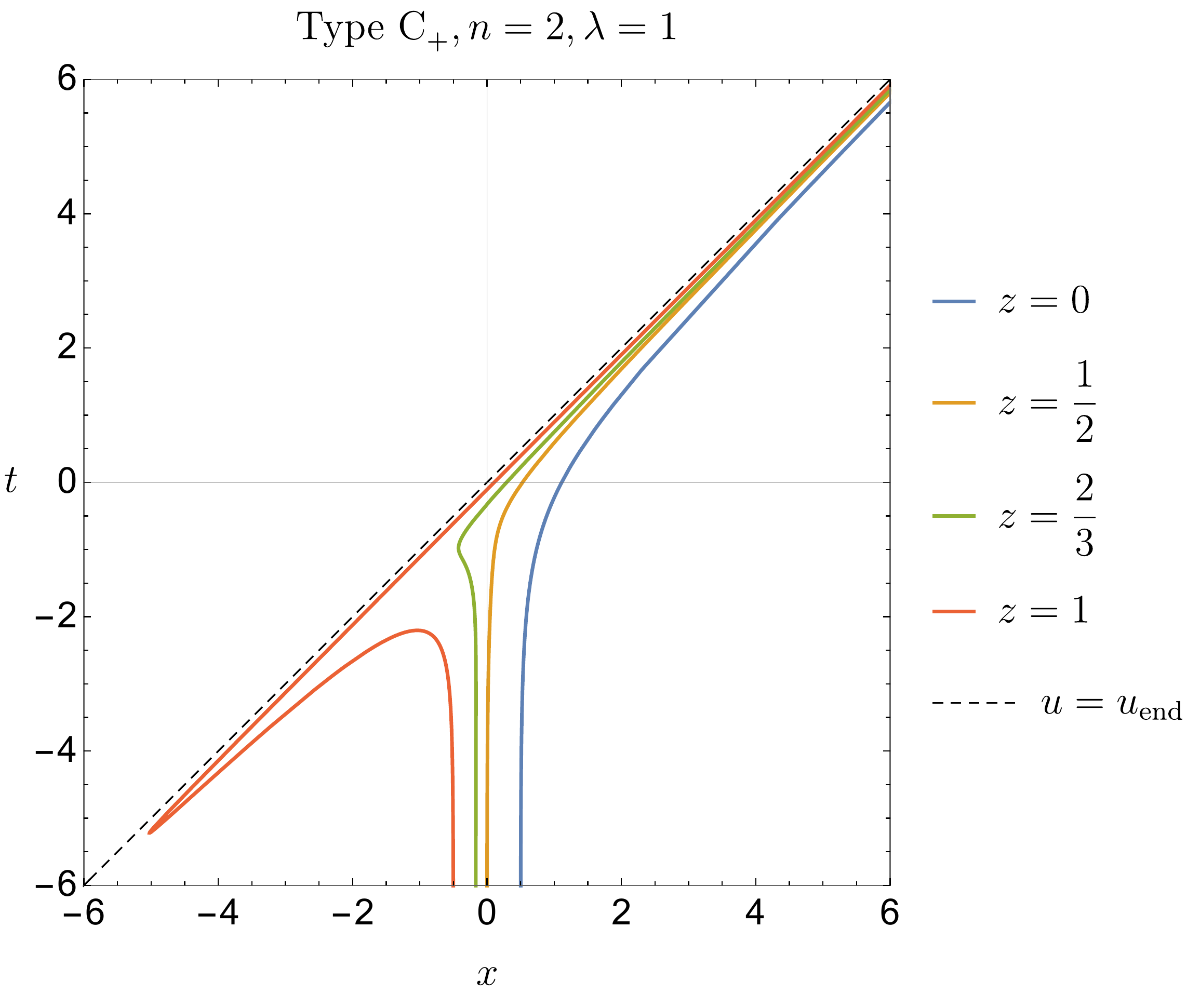}
	\caption{The profiles of brane $Q$ associated with moving mirrors of type C$_+$. The mapping functions are defined in eq.~\eqref{eq:defineTypeC}. We have chosen $\beta=1, \uend=0$, and $n=2$ for both plots.}
	\label{fig:BraneCplus}
\end{figure}
Applying the formula eq.~\eqref{eq:EOWbraneeq}, we can study the holographic dual of type C mirrors by focusing on the profiles of the EOW brane $Q$. Substituting the mapping function eq.~\eqref{eq:defineTypeC} with $n=2$, we have plotted the EOW brane profiles for type C$_+$ in Fig.~\ref{fig:BraneCplus}. One finds that the brane profiles for type C$_+$ are similar to those for type A$_+$ \eg see the top-left plot in Fig.~\ref{fig:BraneAB}. However, it is worth noting that the EOW brane for type C$_+$ is always bounded by the null surface located at $u=\uend$. The region lying to the right of the EOW brane is physical, since the brane naturally plays the role of a spacetime boundary. 
In particular, the constant $z$ slice of the EOW brane for type C$_+$ is smoothly connected. In the following, we show that the brane profiles drastically change for type C$_0$ and type C$_-$ mirrors. 

\subsubsection{Type C$_-$}
\begin{figure}[h!]
	\centering		
	\includegraphics[width=3.0in]{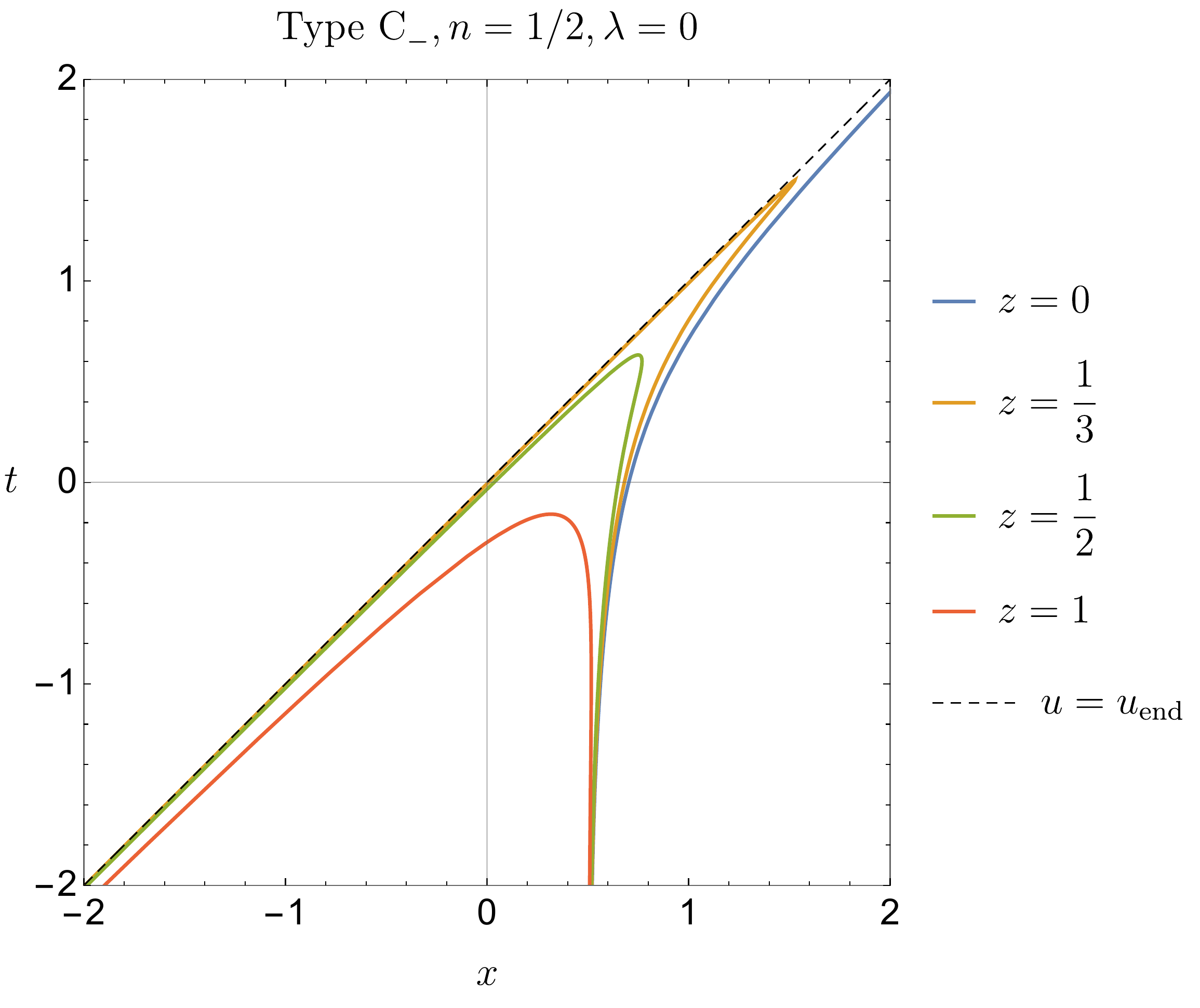}
	\includegraphics[width=3.0in]{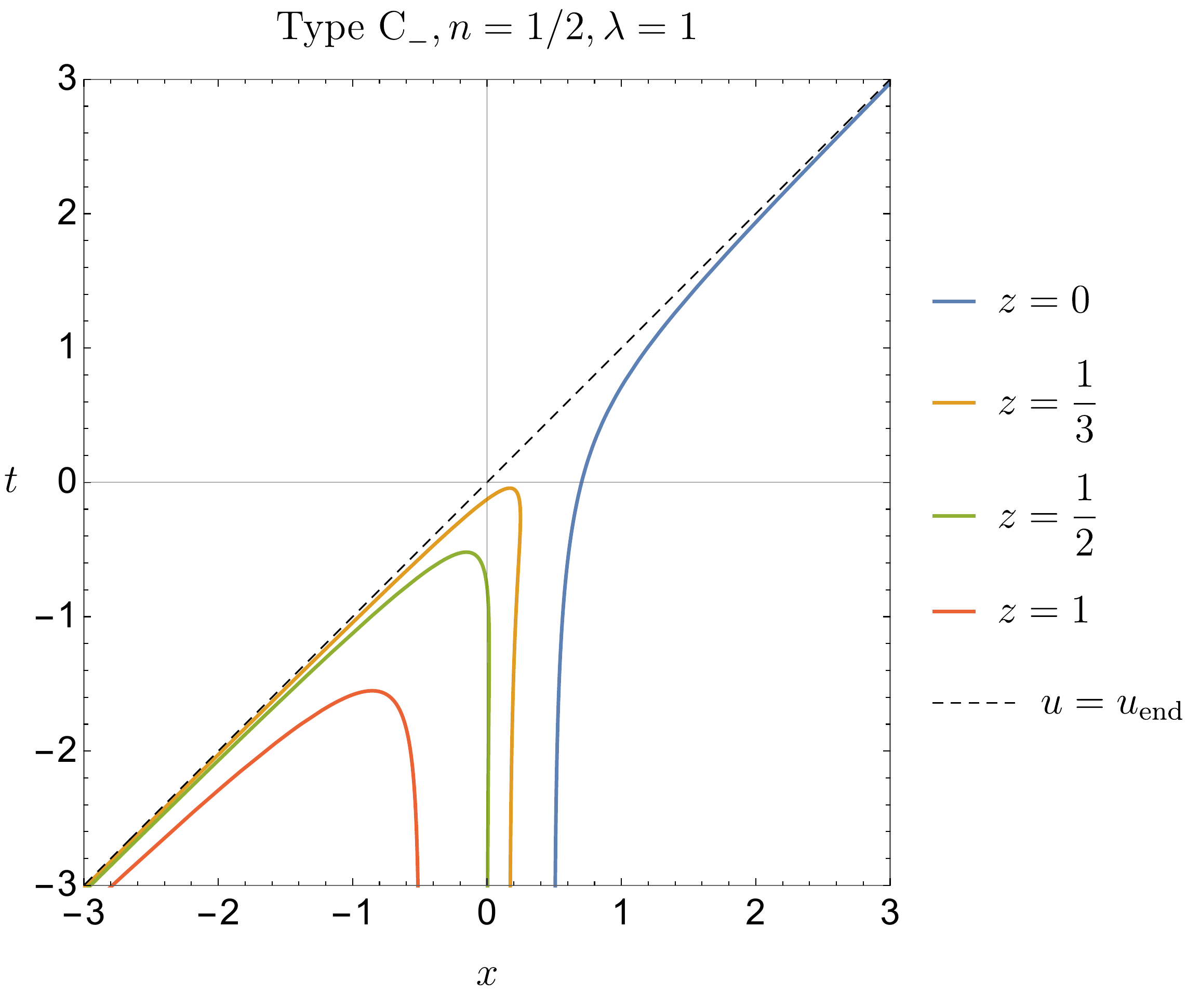}
	\caption{The profiles of the brane $Q$ associated with moving mirrors of type C$_-$. The mapping functions are defined in eq.~\eqref{eq:defineTypeC}. We have set $\beta=1, \uend=0$, and $n=1/2$ for both plots.}
	\label{fig:BraneCminus}
\end{figure}

In order to show the singular behavior for type C$_0$ and type C$_-$ mirrors, let us first examine the case for type C$_-$. As shown in Fig.~\ref{fig:BraneCminus}, the profiles of the EOW brane for type C$_-$ have some rather exotic behavior. The mirror trajectory (\ie $z=0$ slice) is not connected to the EOW brane profiles at $z>0$. In fact, $z=0$ is special because it eliminates the divergent terms appearing in $v_{\rm brane}$, \ie $-\frac{p^{\prime \prime}z^{2} }{2 p^{\prime}} -  \lambda z \sqrt{p^{\prime}} $ in eq.~\eqref{eq:EOWbraneeq}.
This discontinuity looks like a serious problem for applying the AdS/BCFT correspondence because the bulk brane does not extend to the boundary. However, we note that considering a cut-off surface located at $z=\epsilon \to 0$ automatically avoids this problem. On the other hand, a natural question is what is the physical region of the gravity dual. Of course, we should always consider the EOW brane as the boundary of the physical spacetime. As shown in Fig.~\ref{fig:BraneCminus}, the spacetime bounded by the EOW brane contains a large portion beyond $u=\uend$. However, we note that not only the mirror trajectory for type C but also the conformal map eq.~\eqref{eq:chiraltrans} is defined for $u\le \uend$. As a result, the coordinate transformation in eq.~\eqref{eq:coordinatetrans} identifies the bulk spacetime in Poincar\'e coordinates with the physical spacetime being defined by 
\begin{equation}
 v > v_{\rm brane}\,, \qquad  u < \uend \,.
\end{equation}
Especially, the null surface $u=\uend$ is just mapped to the infinite line at $U = p(\uend) \to +\infty, \eta = z \sqrt{p'(\uend)} \to +\infty$ via the coordinate transformation defined in eq.~\eqref{cordtr}. Physical quantities like geodesics do not go beyond $u > \uend$. 
In other words, we conclude that the boundary of the dual spacetime related to the type C mirror is given by the EOW brane at $v=v_{\rm brane}$ as well as the null surface at $u=\uend$. Physically, we interpret this as the consequence of the negative divergence for the energy flux $T_{uu}$ at $u=\uend$. This is why we have not seen this exotic behavior for type C$_+$, where the energy flux approaches positive infinity rather than negative infinity. 

\subsubsection{Type C$_0$}
\begin{figure}[h!]
	\centering		
	\includegraphics[width=3.0in]{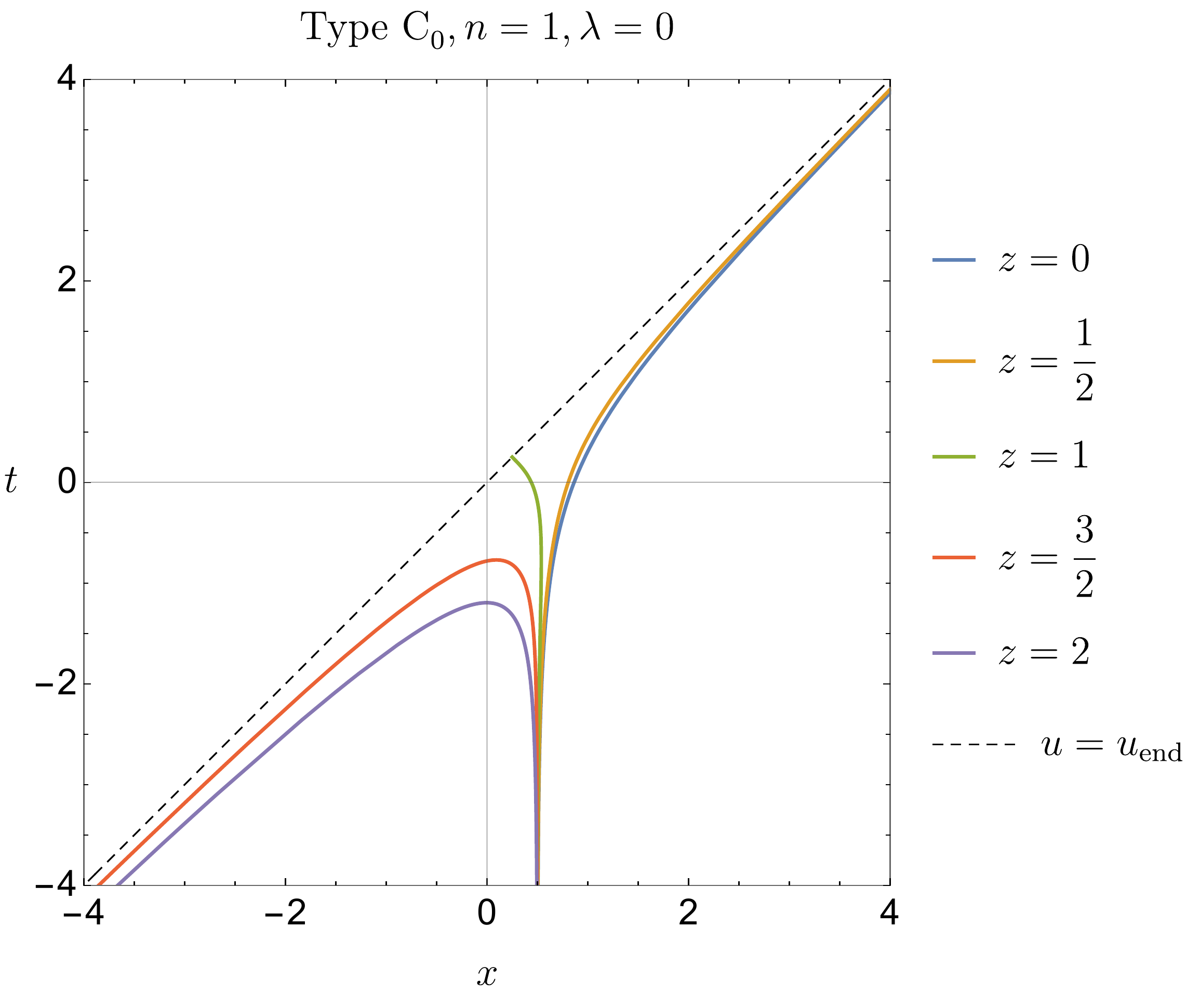}
	\includegraphics[width=3.0in]{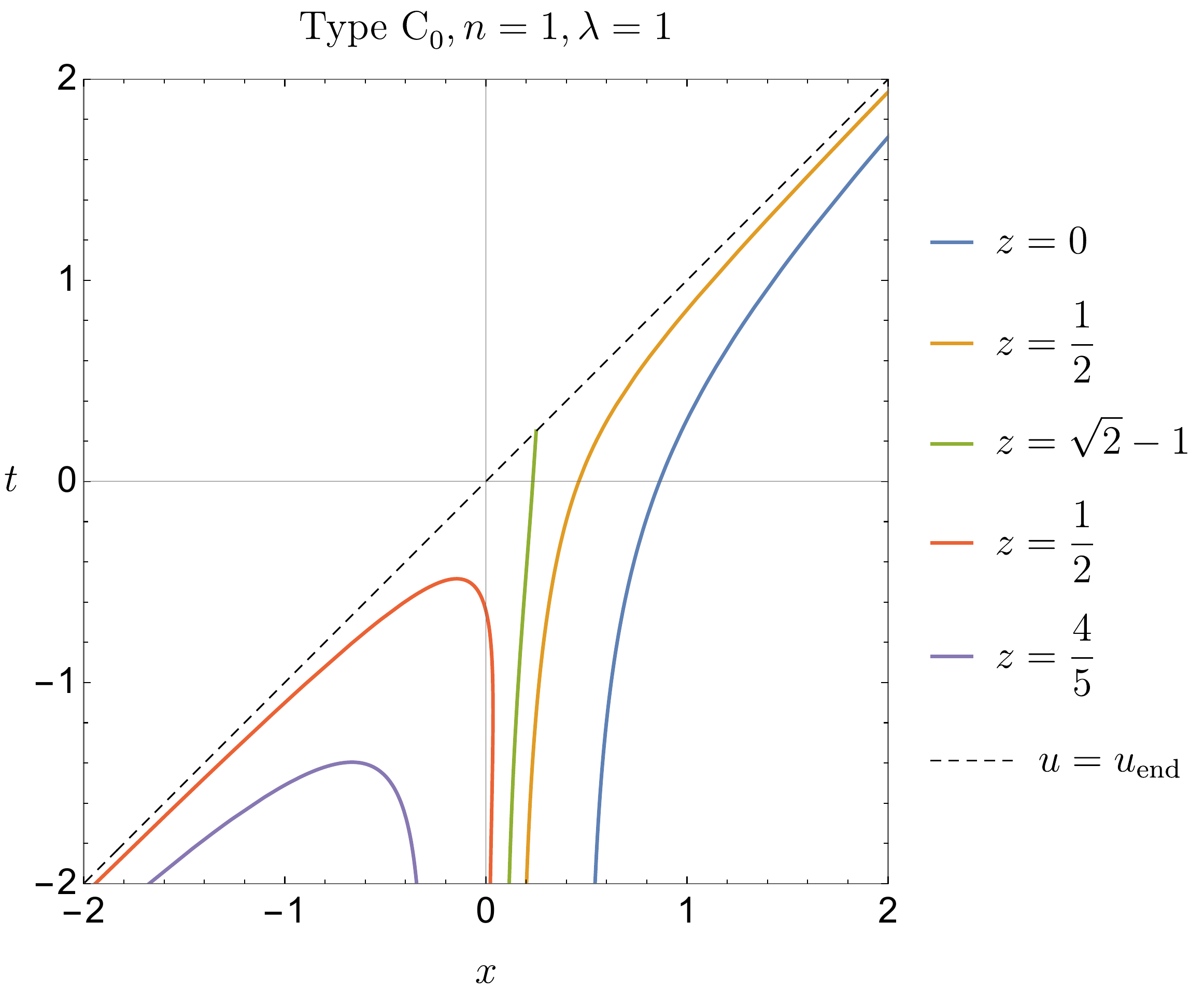}
	\caption{The profiles of the brane $Q$ associated with mirrors of type C$_0$. The mapping functions are defined in eq.~\eqref{eq:defineTypeC}. We have set $\beta=1, \uend=0$, and $n=1$ for both plots.}
	\label{fig:BraneCzero}
\end{figure}

The profiles of EOW branes for type C$_0$ mirrors share certain features with those for type C$_\pm$ mirrors. In Fig.~\ref{fig:BraneCzero}, we show the corresponding profiles of EOW branes for type $C_0$ mirrors defined by 
 \begin{equation}\label{eq:typeCzero}
      p_{\mt{C}_0}(u) = u + \frac{\beta}{1-e^{u/\beta}},\quad u < 0\,,
 \end{equation}
 corresponding to the mapping function in eq.~\eqref{eq:defineTypeC} with $n=1$ and $\uend=0$. More explicitly, we find that the profile of the EOW brane is similar to the one for type C$_+$ in the case of small $z$. However, the similarity is shared with the one for type C$_-$ in the case of large $z$. 
 In particular, there is a transition point at 
 \begin{equation}\label{eq:solution}
 z=z_c=\beta  \left(\sqrt{\lambda ^2+1}-\lambda \right) \,.
 \end{equation}
We derive the critical value $z_c$ by taking the expansion of $v=v_{\rm brane}$ around the endpoint $u=\uend$. Substituting eq.~\eqref{eq:typeCzero} into eq.~\eqref{eq:EOWbraneeq}, one obtains the following expansion 
\begin{equation}
v= v_{\rm brane} \approx \frac{-\beta ^2+z^2+2 \beta  \lambda  z}{u} +\frac{\beta }{2}+ \mathcal{O}(u) \,.
\end{equation} 
It is clear that we have $v_{\rm brane} \to + \infty$ for $z\ll \beta$ as in the case of type C$_+$, and $v_{\rm brane} \to - \infty$ for $z\gg \beta$ as in the case of type C$_+$. Instead, the transition point with finite $v_{\rm brane} (z=z_c, u=\uend)$  
is given by the solution of $-\beta ^2+z^2+2 \beta =0$, \ie eq.~\eqref{eq:solution}. The analysis above can be generalized to any case with type C mirrors, which we briefly touch upon in the next subsection.

\subsubsection{Asymptotic behavior of EOW brane}
Following our classification of moving mirrors into four different types, the mapping functions for type C mirrors turn out to have a universal divergence around $u \approx \uend$,
 \begin{equation}\label{eq:typeCexpansion}
 \lim\limits_{u \to \uend}p(u) = \frac{c_{-n}}{(u_{\text{end}}-u)^n} + \cdots \,.
\end{equation}
Here, the constant $c_{-n}$ is always positive because the mirror trajectories are timelike for $u<u_{\text{end}}$. As a reminder, the position of the brane is determined by the mapping function via 
\begin{equation}\label{eq:brane}
v_{\rm{brane}}=-\frac{p^{\prime \prime}z^{2} }{2 p^{\prime}}  + p(u) -  \lambda z \sqrt{p^{\prime}} \,. 
\end{equation}
Using the universal expansion formula eq.~\eqref{eq:typeCexpansion} for type C mirrors, we find that the brane position $v_{\rm brane}$ is dominated by 
\begin{equation}
v_{\text{brane}} \approx  -  \frac{n+1}{2 n } \frac{z^2}{(u_\text{end} - u)}+\frac{c_{-n}}{(u_\text{end}-u)^n} - 2\lambda z \frac{\sqrt{c_{-n} n}}{(u_\text{end}-u)^{\frac{n+1}{2}}}+ \cdots\,,
\end{equation}
with a leading term depending on the value of $n$. 

Considering the moving mirror of type C$_+$ with $n>1$, $v_{\rm brane}$ is dominated by $p(u)$,
\begin{equation}
v_{\text{brane}} \approx   \frac{c_{-n}}{(u_\text{end}-u)^n} \rightarrow +\infty\,,
\end{equation}
for any $z$. This indicates the universal behavior of EOW brane profiles for type C$_+$ mirrors at any constant $z$ slice, see Fig.~\ref{fig:BraneCplus}. On the contrary, the brane profile for type C$_-$ mirrors satisfies 
\begin{equation}
   v_{\text{brane}}  \approx   -\frac{n+1}{2n}\frac{z^2}{u_\text{end}-u} \rightarrow -\infty \,,
\end{equation}
due to $1>\frac{n+1}{2}>n$. As a result, the profiles of $v_{\rm brane}$ develop a zigzag shape as shown in Fig.~\ref{fig:BraneCminus}. The asymptotic behavior of $v_{\rm brane}$ in the special type C$_0$ case with $n= \frac{n+1}{2}=1$ is given by  
\begin{equation}
v_{\text{brane}} \approx   \frac{-z^2 + c_{-1}-2\lambda z \sqrt{c_{-1}}}{u_\text{end}-u} + \mathcal{O}((\uend -u)^0) \,. 
\end{equation}
Here, we have assumed that the subleading term is not divergent. The sign of the numerator depends on the value of $z$.  It is positive for $z \ll 1$, but negative for $z\gg 1$. The transition happens at $z=z_c$, where $v_{\rm brane}$ is located at a finite value and does not diverge. We then get
\begin{equation}
z_c =\sqrt{c_{-1}}( \sqrt{\lambda^2 +1}-\lambda ) \,.
\end{equation}
Recalling the relation between the boundary entropy $S_{\rm bdy}$ and the parameter $\lambda$, see eq.~\eqref{eq:bdry entropy}, we rewrite the critical value as 
\begin{equation}
z_c = \sqrt{c_{-1}} e^{-\frac{6}{c}S_{\text{bdy}}}\,.
\end{equation}
We conclude that the EOW brane profiles are similar to the mirror trajectories for $z<z_c$ and evolve in the opposite direction for $z>z_c$. We shall note that we have assumed that the mapping function $p(u)$ around $\uend$ is of some polynomial asymptotic form, \eg
$p(u) \sim \frac{c_{-1}}{u_\text{end}-u} + c_0 +c_1 (u-u_\text{end})+\cdots$. Correspondingly, we have obtained a finite value for $v_\text{brane}$ at $z=z_c $. If there should appear some logarithmic contributions in subleading terms, \eg $p(u) \sim \frac{c_{-1}}{u_\text{end}-u} + c_{\text{log}} \log{(u_\text{end}-u)}+\cdots$, $v_\text{brane}$ may also approach $\pm \infty$ at $z=z_c $.

\subsubsection{Global structure of EOW brane for type C$_{-}$ and type C$_0$}
\begin{figure}[h]
	\centering		
	\includegraphics[width=3in]{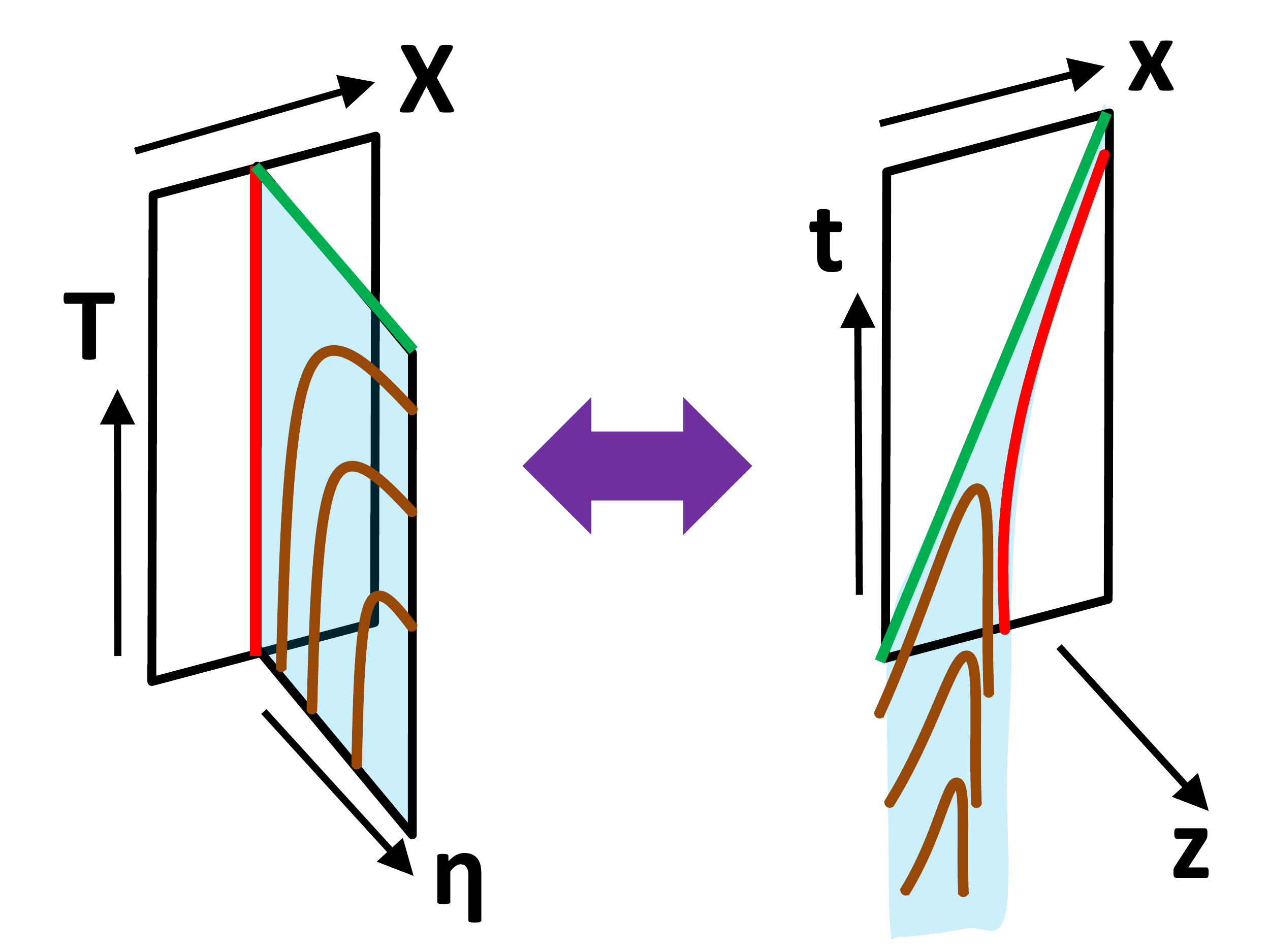}
	\caption{Sketch of the transformation of EOW brane from Poincar\'e AdS$_3$ (left) to the gravity dual of type C$_{-}$ and C$_0$ moving mirror (right). The red curve (right) is the mirror trajectory mapped to the red, straight line in the left figure. The EOW branes correspond to the blue-colored regions.}
	\label{fig:TypeCp}
\end{figure}
 For type C$_0$ and type C$_-$ mirrors, the EOW branes turn over as $z$ gets larger. So one might wonder whether or not the physical spacetime extends to the other region, $u>u_\text{end}$. To better understand this point, let us examine the global structure of the gravity dual. This becomes clear if we transform the EOW brane described by $(u,v,z)$ to the one described by Poincar\'e coordinates $(U,V,\eta)$. This is sketched in Fig.~\ref{fig:TypeCp}. In terms of Poincar\'e coordinates, one should pick up only the region associated with $u<u_\text{end}$, since the coordinate transformation $U=p(u)$ in eq.~(\ref{cordtr}) maps the $u<\uend$ region to the whole region assigned to the static mirror (\ie $V-U>0$). Especially, the $u=\uend$ line is mapped to $U = + \infty$, and any physical probe cannot reach $u\geq \uend$. This may also be understood as an effect caused by the singular metric on $u =\uend$ with $z >0$ that arises due to the divergence of the energy flux. In summary, in addition to type C$_{+}$ mirrors, the AdS/BCFT setups in the presence of EOW branes dual to type C$_{-}$ and type C$_0$ mirrors turn out to be physically sensible and are defined in a region with $u<u_\text{end}$.


\subsection{Gravity dual of type D mirrors}
\label{sec:gravitytypeD}
\begin{figure}[h]
	\centering		
	\includegraphics[width=3in]{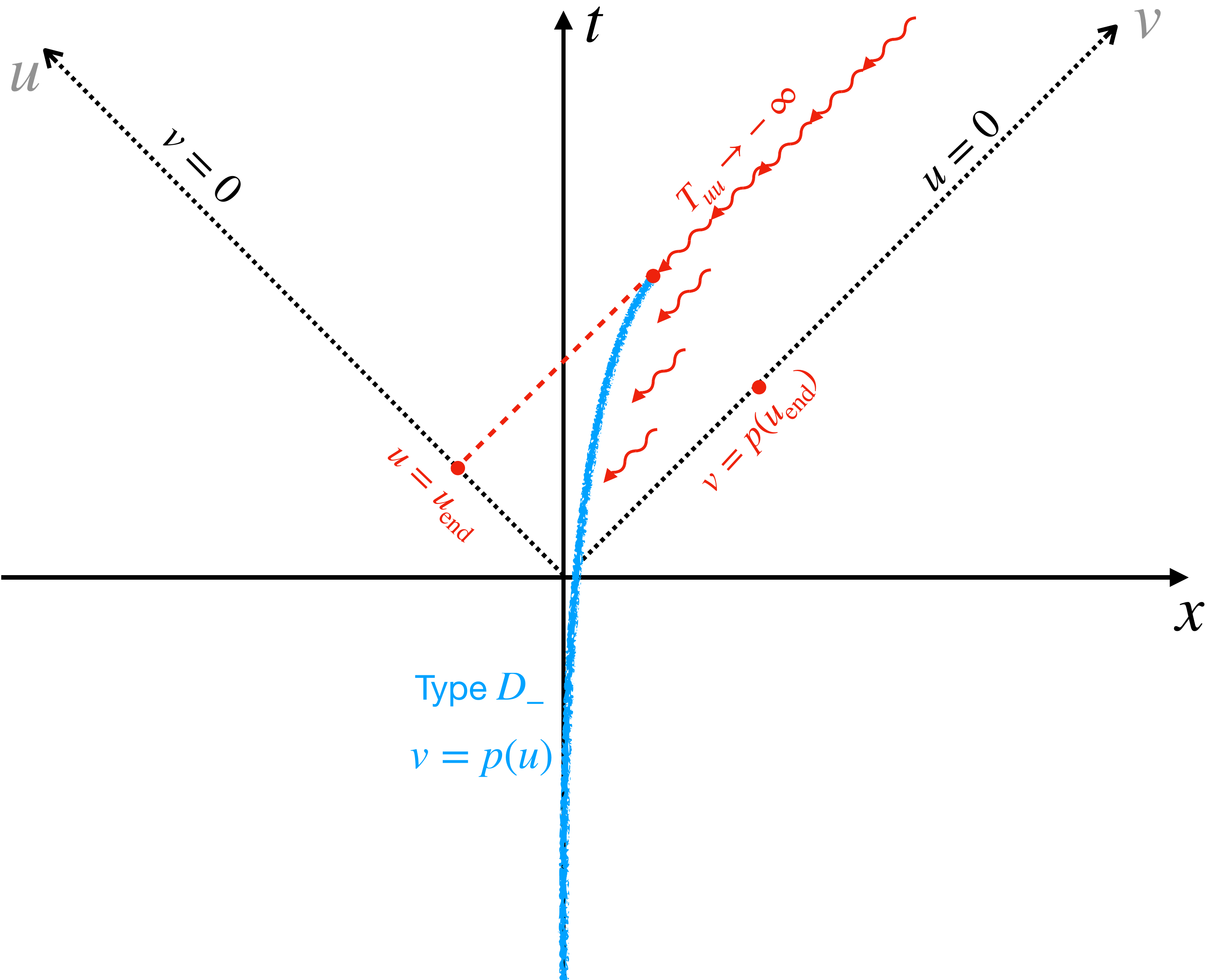}
	\includegraphics[width=3in]{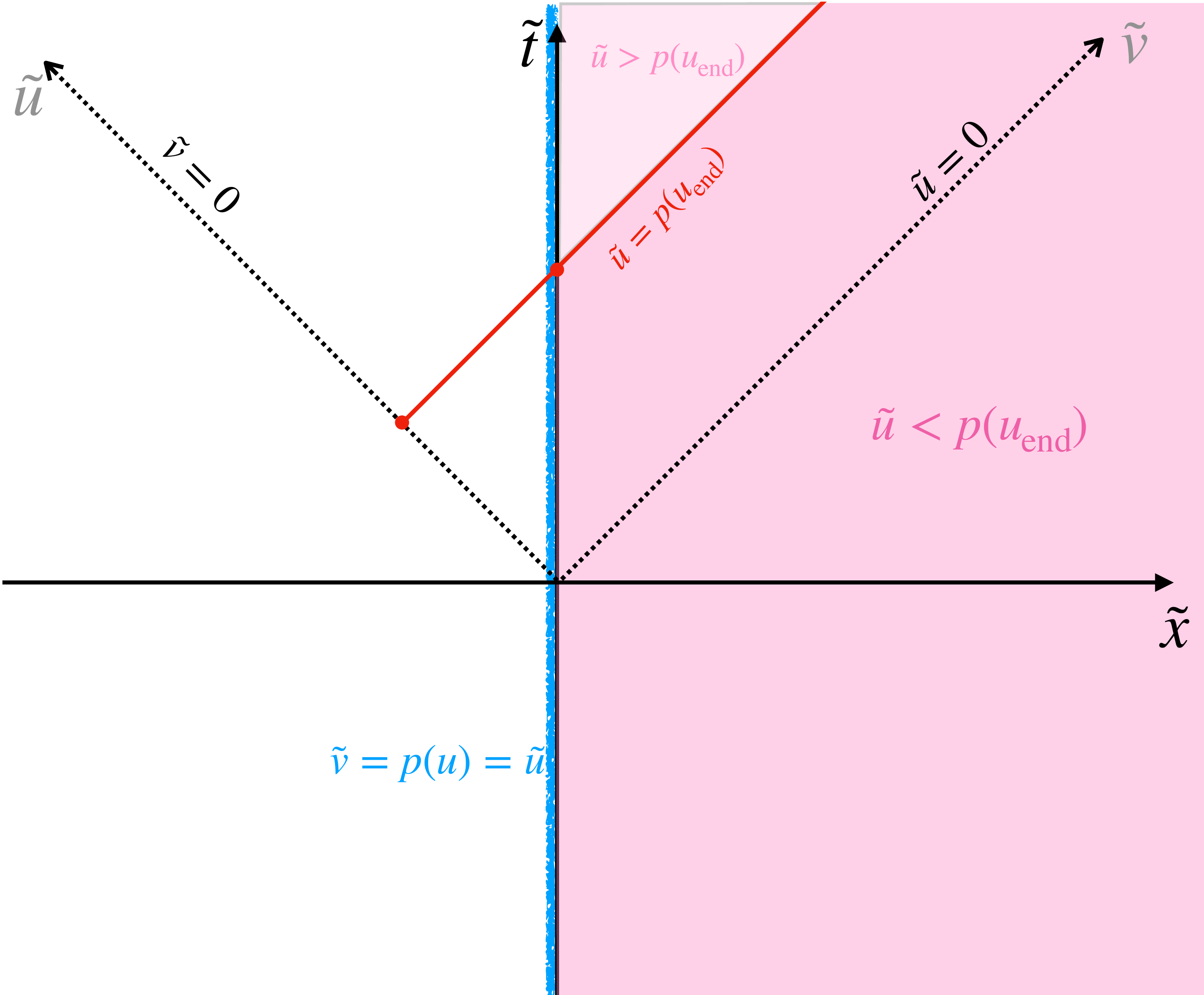}
	\caption{Physical spacetime (pink shaded area) is divided by a null surface into two parts. Left: the null surface is located at $u=\uend$ in the original spacetime with a moving mirror of type D$_-$ which generates infinite-energy flux along the null surface at its endpoint. Right: the physical region in $(\tilde{t}, \tilde{x})$ coordinates is the pink shaded region divided by the null surface at $\tilde{u}= p(\uend)=\vend$ into two parts.}
	\label{fig:cutspacetime}
\end{figure}

In the previous sections, we have seen that the singular (or divergent) behavior in the stress tensor, entanglement entropy, and holographic dual can be traced back to the appearance of $p'(u)= 0 $ or $p'(u)= \infty$, as the mirror moves at the speed of light. 
The unique feature associated with type D mirrors is that the endpoint of the mirror is located in the bulk spacetime rather than at the infinitely far boundaries. In terms of lightlike coordinates, $(u,v)$, moving mirrors of type D$_\pm$ are constrained by
\begin{equation}
 u \in (- \infty, \uend]\,, \qquad v \in (- \infty, \vend]\,, 
 \end{equation}
 with $\vend = p(\uend)$.

Because type D$_+$/D$_-$ mirrors reach the speed of light at their endpoint, 
the corresponding stress tensor $T_{uu}$ does not vanish and is given by 
\begin{equation}
\lim\limits_{u \to \uend} T_{uu} \approx \frac{c}{48 \pi} \frac{n^2-1}{(\uend-u)^2}\,,
\end{equation}
indicating a positive/negative divergence along the null surface $u=\uend$. For later discussion, we remark that the divergence in the stress tensor is associated with the divergent term $p''/p'$, since 
\begin{equation}
T_{uu} = \frac{c}{24 \pi }  \(  \frac{1}{2}\( \frac{p''}{p'}\)^2 - \( \frac{p''}{p'}  \)' \) \,,
\end{equation}
where 
\begin{equation}
\lim\limits_{u \to \uend} \frac{p''(u)}{p'(u)} \sim \frac{1-n}{\uend - u}\,,
\end{equation}
holds for moving mirrors of type D$_\pm$. For example, in Fig.~\ref{fig:TuuABCD}, we show the time evolution of the non-vanishing stress tensor, where we use the conformal mapping function $p(u)$ for mirrors described by eq.~\eqref{eq:defineTypeDp} and eq.~\eqref{eq:defineTypeDm}. The divergent energy flux also results in a divergence for the entropy $S_A$. Taking the subsystem $A$ to be a static, semi-infinite interval, the universal divergent behavior has been seen in eq.~\eqref{EEDpm}. This can be seen as a consequence of the infinite-energy flux inserted along $u_0=\uend$, which leads to 
\begin{equation}
\lim\limits_{u \to \uend}  \frac{\partial S_A}{\partial t} \approx  -\frac{c}{12} \frac{p''(u_0)}{p'(u_0)} \sim   \frac{c}{12} \frac{n-1}{\uend -u} \,.
\end{equation}

As illustrated in Fig.~\ref{fig:cutspacetime}, the lightlike mirror at $(\uend, \vend)$ induces an infinite-energy shockwave moving along the null surface
\begin{equation}
 u =\uend\,.
\end{equation}
The null surface separates the spacetime into two parts, such that only the right part becomes accessible to an observer living there. 

Our entanglement entropy calculations are based on standard BCFT techniques in the presence of a static mirror at $\tilde{x}=0$. Since the mapping function $p(u)$ for type D$_\pm$ mirrors is truncated at $u=\uend$, one can only obtain part of the right half-plane, 
\begin{equation}
 \tilde{u} \in (-\infty, p(\uend)]\,,
\end{equation}
via the map $\tilde{u}=p(u)$ and $\tilde{v}=v$. One may imagine that there is another mirror that connects the truncated mirror of type D and extends to infinity. As a result, the joint mirror, similar to the types A, B, or C, would be dual to the complete static mirror with the physical spacetime covered by the right half-plane. However, one can find infinitely many such smooth mirrors, where the first half is the same as for type D$_\pm$ mirrors. This is not a contradiction because the null surface at $u=\uend$ or $\tilde{u}=p(\uend)$ divides the spacetime into two parts. Physical quantities like entanglement entropy for a subsystem can therefore be determined on one side before reaching the null line. This also motivates us to consider type D$_\pm$ mirrors as a particular case. 

Finally, let us consider the dual gravitational spacetime associated with type D$_\pm$ mirrors, again following the AdS/BCFT dictionary. As introduced before, the dual spacetime is nothing but AdS$_3$ with an EOW brane extending into the bulk. In the following, we analyze the corresponding EOW branes for type D$_\pm$ and show that their features differ from each other.

\subsubsection{Type D$_+$}
\begin{figure}[h]
	\centering		
	\includegraphics[width=3in]{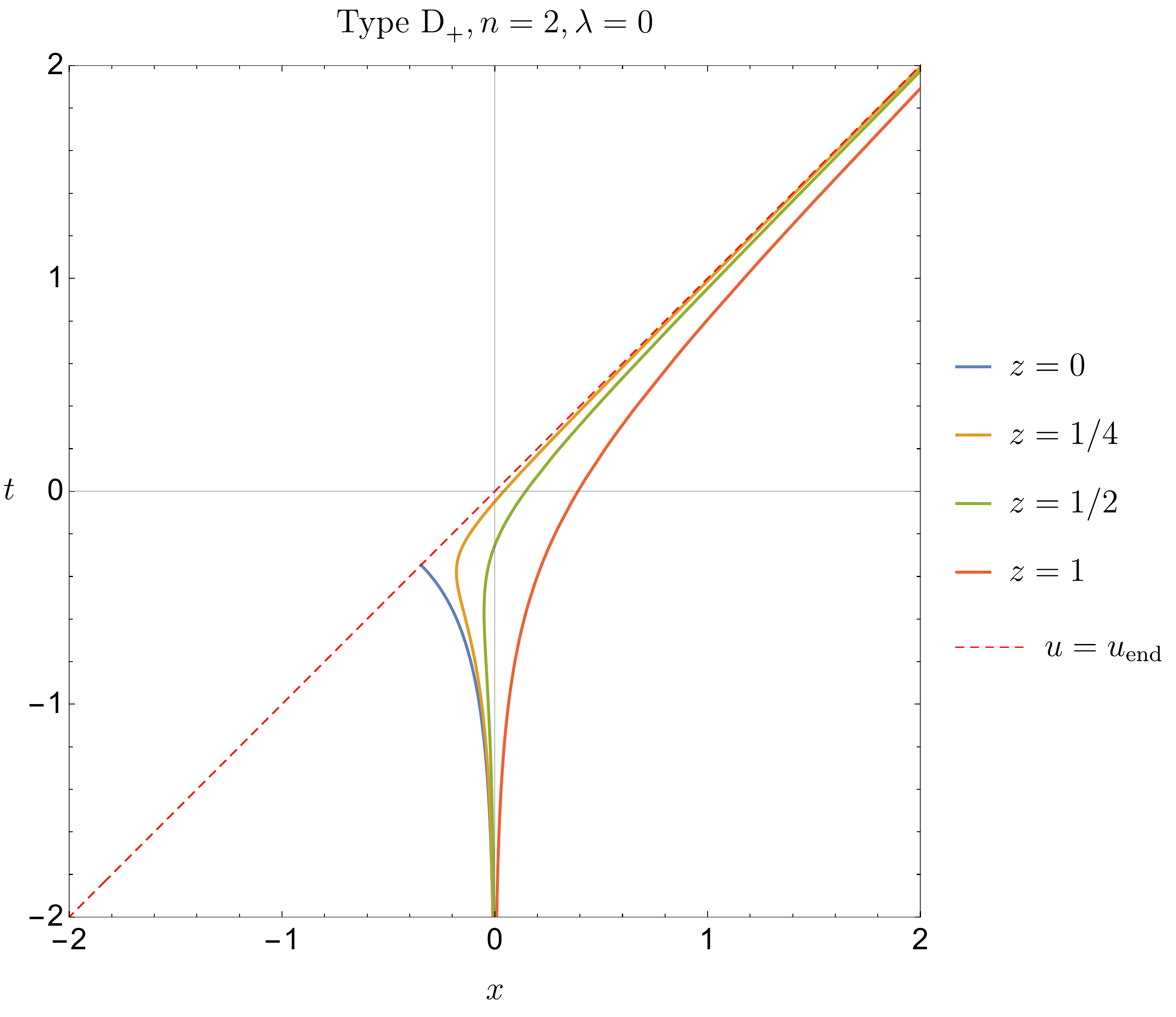}
	\includegraphics[width=3in]{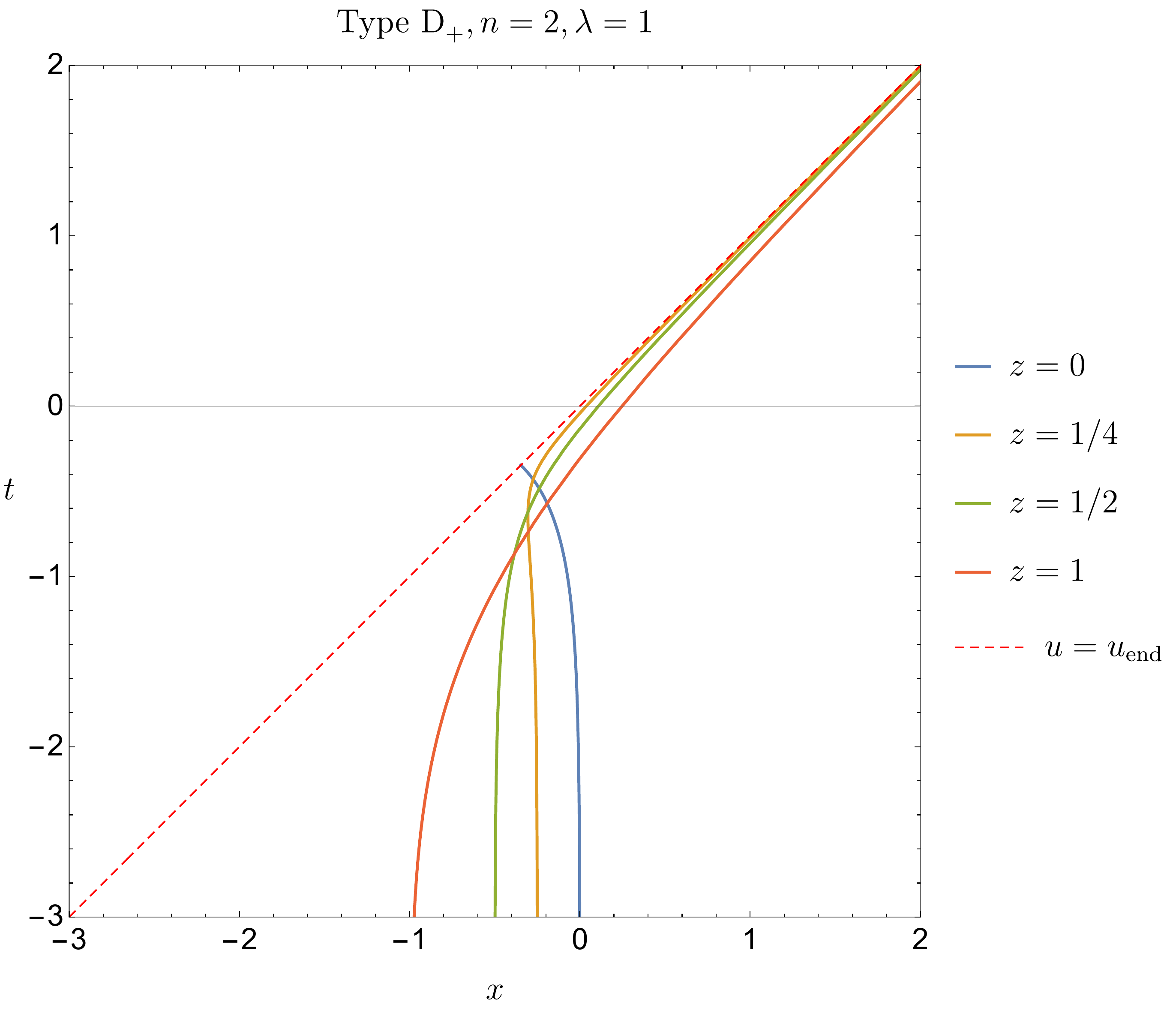}
	\caption{Profiles of the brane $Q$ associated with mirrors of type D$_+$ in coordinates $(t,x)$ for various $z$. The mapping functions are given in eq.~\eqref{eq:defineTypeDp}. Here, we have chosen $\beta=1$ for both plots.}
	\label{fig:TypeDbrane}
\end{figure}
Let us consider the following mapping function for type D$_+$ mirrors
\begin{equation}\label{eq:defineTypeDp}
\begin{split}
 p_{\mt{D}_+}(u)&=  -\beta \log ( e^{-u/\beta}+e^{u/\beta}) \,,  \qquad \text{with} \qquad n=2 \,,\\
\end{split}
\end{equation}
where the endpoints are fixed at $u=\uend=0$ and $\vend=-\beta \log 2$. The generalization to an arbitrary $\uend$ can be made by shifting $u$ as $u \to u-\uend$. Various mapping functions and trajectories for mirrors of type D are plotted in Fig.~\ref{fig:TypeD01}. The trajectory of the mirror explicitly reads
\begin{equation}\label{eq:trajectoryDp}
x=Z(t) = \frac{\beta}{2}\log \(  1- e^{2t/\beta} \) , 
\qquad t=Z^{-1}(x)=  \frac{\beta}{2}\log \(  1- e^{2x/\beta} \)\,,
\end{equation}
with the endpoint being located at $x_{\rm end}=t_{\rm end}= -\frac{\beta}{2} \log 2$. We obtain the velocity of the moving mirror at time $t$ as follows
\begin{equation}
Z'(t)= - \frac{e^{2t/\beta}}{1-e^{2t/\beta}}\,,
\end{equation}
which reaches the (negative) speed of light at $t=- \frac{\beta}{2} \log 2$. Using eq.~\eqref{eq:brane}, we get the corresponding EOW brane for the type D$_+$ mirror, 
\begin{equation}
v_{\rm{brane}} = -\frac{z^2}{\beta \sinh (2u/\beta)}  -\beta \log \( 2 \cosh (u/\beta) \) -2z \lambda \sqrt{ \tanh (- u/\beta)} \,,
\end{equation}
which has a positive singularity at $u=\uend=0$. 
In order to visualize the dual spacetime in AdS$_3$, we show the positions of the EOW brane at various bulk slices of constant $z$ in Fig.~\ref{fig:TypeDbrane}. The brane itself is always located on the right side of the null surface at $u=\uend$ and moves toward it with increasing $u$. 

\begin{figure}[H]
	\centering
	\includegraphics[width=3.1in]{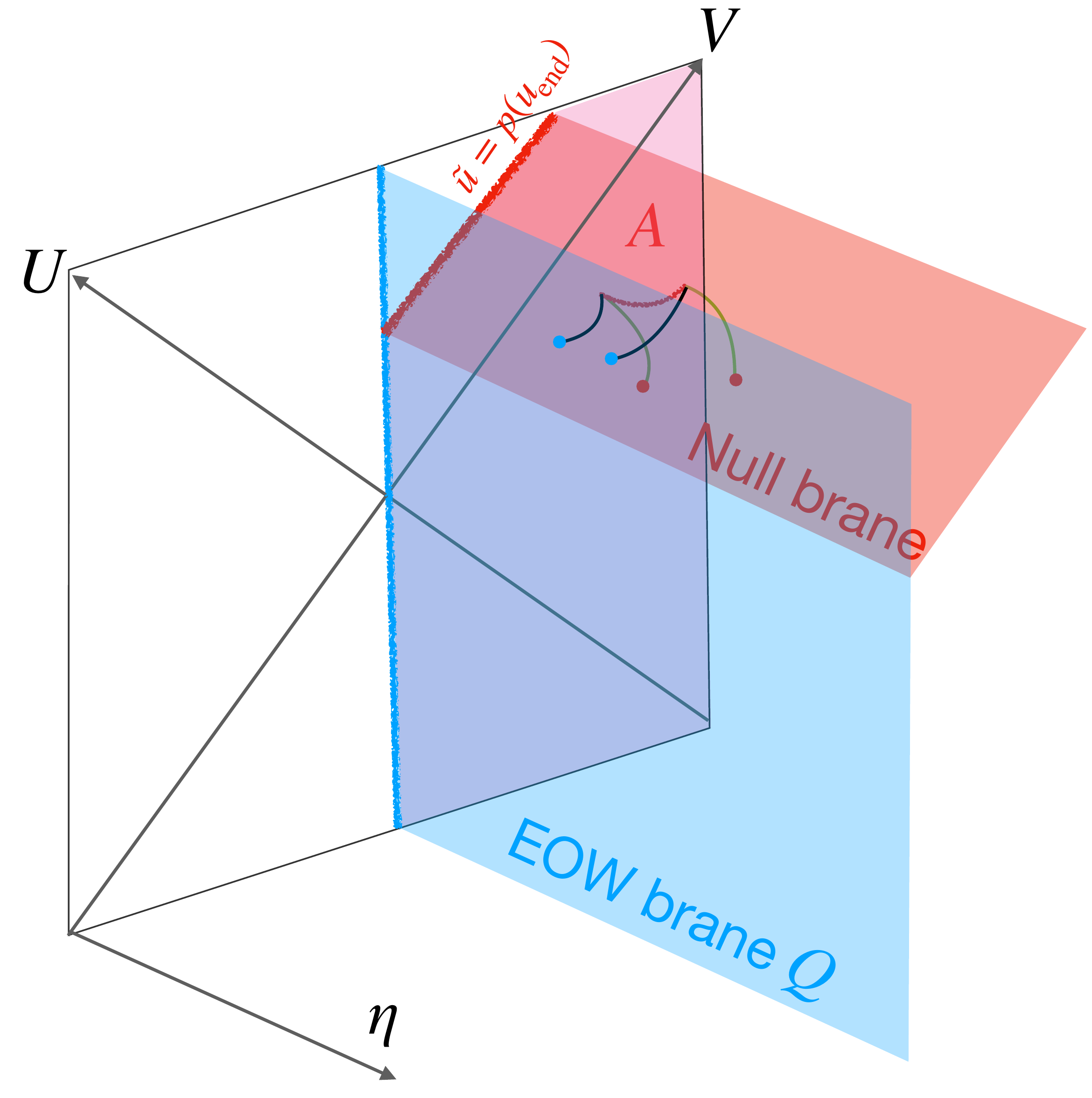}
	\caption{An extra null brane (red shaded surface) is inserted in the bulk spacetime. Different from the standard AdS/BCFT setup with one timelike EOW brane (blue surface), the appearance of a null brane at $U=p(\uend)$ would result in a different state. In this case, the disconnected extremal surface for subsystem $A$ may end on the null surface (denoted by the green curves) instead of ending on the timelike EOW brane. In this paper, we only explore the latter case with one EOW brane, where the corresponding extremal surface (black curves) anchors to the EOW brane.}
	\label{fig:NullBrane}
\end{figure}

This feature is universal for all type D$_+$ mirrors, because the position of the EOW brane in the limit $u \to \uend$ is dominated by the term $p''/p'$, 
\begin{equation}
\lim\limits_{u \to \uend} v_{\rm{brane}} \approx -\frac{p^{\prime \prime}z^{2} }{2 p^{\prime}} \sim \frac{z^2}{2} \frac{n-1}{\uend -u} \to +\infty \,. 
\end{equation}
As discussed before, this goes back to the positive infinite energy flux, $T_{uu} \to +\infty$. So the EOW brane encloses the physical spacetime and plays the role of a boundary in the asymptotically Poincar\'e AdS gravity dual.
The HEE for a semi-infinite subsystem with an endpoint close to $u=u_{end}$ can be computed via the geodesic connecting the boundary endpoint and a point on the EOW brane after mapping to the gravity dual of a static mirror, \ie half of Poincar\'{e} AdS$_3$. 
The point on the EOW brane goes beyond the null surface and takes values in $u>u_{end}$. If we transform back to  the original gravity dual in the type D$_{+}$ case, the geodesic extends from the boundary to infinity, $v\to\infty$, and this coordinate patch describes only part of the geodesic, which is complete for the static mirror coordinates.  
To understand the full gravity dual, we need to know the mirror trajectory after the null point of type D$_+$ mirrors, \ie $u>u_{end}$, though the gravity dual for $u<u_{end}$ is completely fixed by the profile of the mirrors of type D$_+$.

It might be interesting to introduce a solid null brane at $u=\uend$, as some kind of EOW brane. In the dual picture, when mapped to the right half-plane of the static mirror, this corresponds to putting a null brane at $U=p(\uend)$ in the bulk spacetime, see Fig.~\ref{fig:NullBrane}. Although the spacetime limited by $u< \uend$ looks similar, one finds that the two configurations, namely with and without a null brane, are physically different. For example, one observes that the geodesics in both cases are distinct as they can naturally end on the extra null surface, which would predict a different value for HEE.\footnote{In fact, the computation performed in such a construction with a null brane reproduces a sort of pseudo entropy but not entanglement entropy. This point will be clear after the discussion given in appendix \ref{sec:LIandDM}.} We would like to leave this possibility for future work.

\subsubsection{Type D$_-$}
\begin{figure}[h]
	\centering		
	\includegraphics[width=3in]{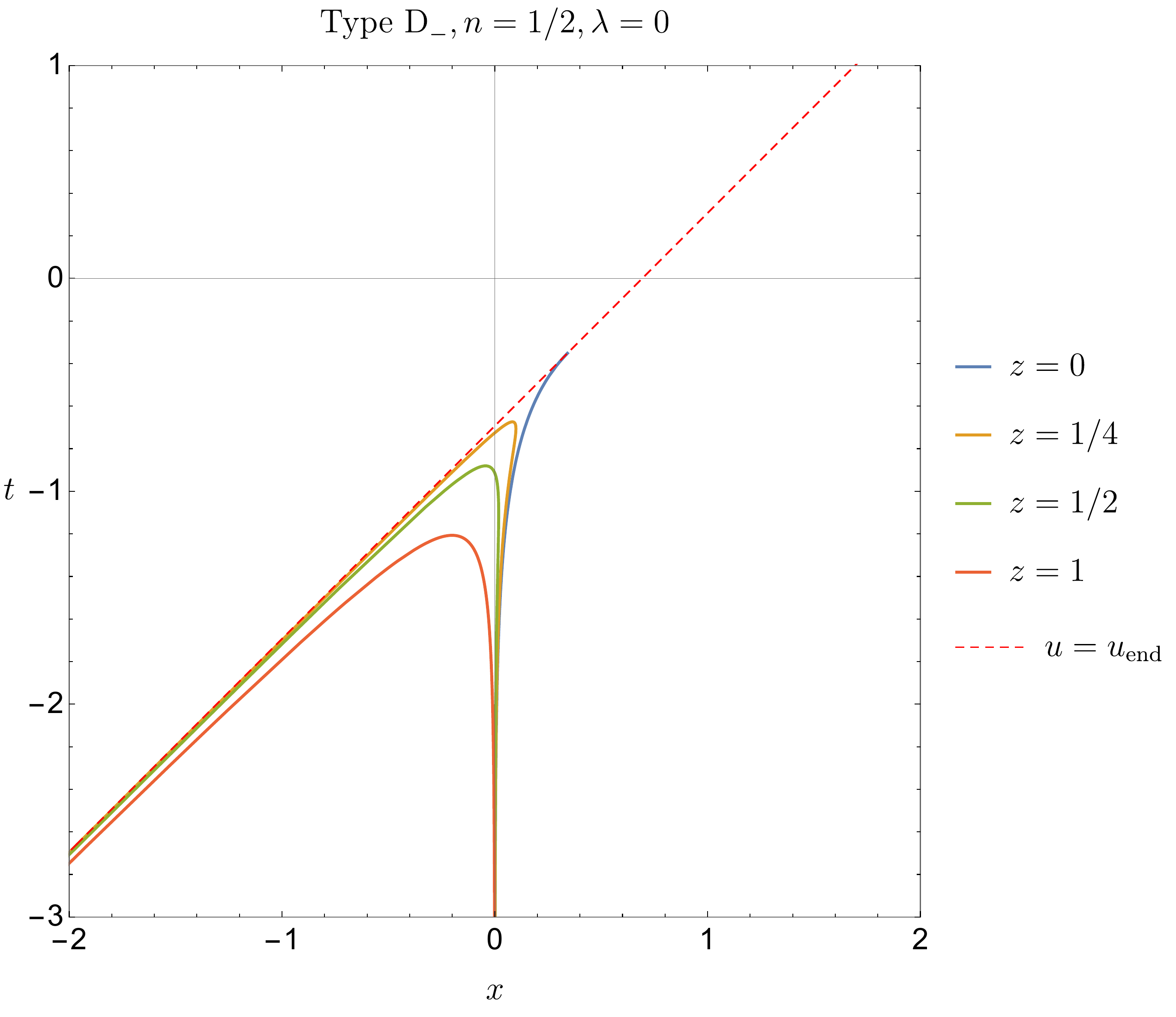}
	\includegraphics[width=3in]{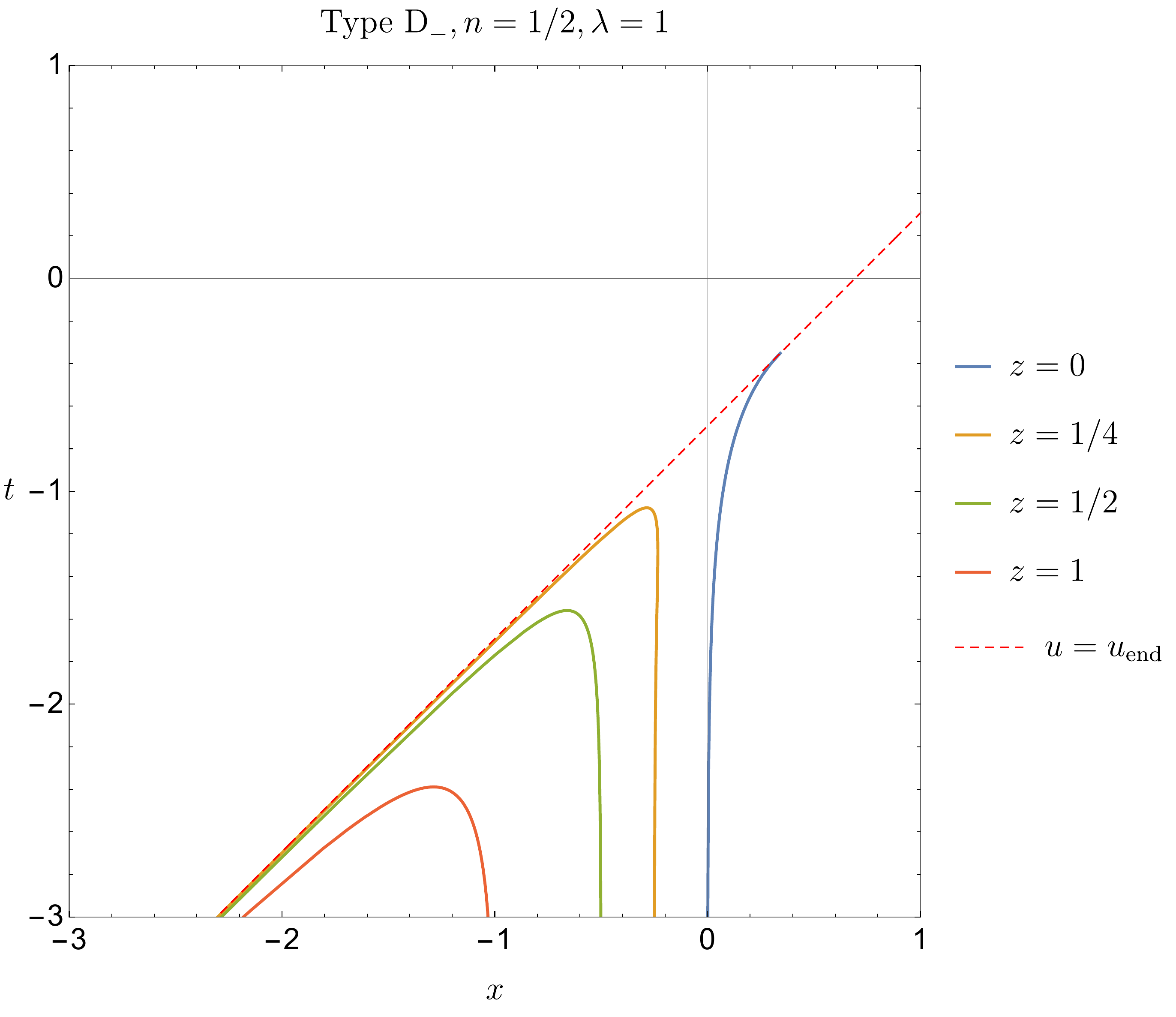}
	\caption{Profiles of the brane $Q$ associated with mirrors of type D$_-$ in coordinates $(t,x)$ for various $z$. The mapping functions are given in eq.~\eqref{eq:defineTypeDm}.}
	\label{fig:TypeDmbrane}
\end{figure}

Similar to the setup given by eq.~\eqref{eq:defineTypeDp}, we can construct a type D$_-$ mirror via the following mapping function 
\begin{equation}\label{eq:defineTypeDm}
\begin{split}
p_{\mt{D}_-}(u)&=-u+ \beta \log \( \frac{ 1-\sqrt{1- 4 e^{2u/\beta}} }{2}\)\\
&= -\beta \arcsech \(  2 e^{u/\beta} \)\,,  \qquad \text{with} \qquad n=\frac{1}{2} \,,
\end{split}
\end{equation}
where the endpoint is located at $\uend= -\beta \log 2$ and $\vend=0$. The mirror trajectory reads 
\begin{equation}\label{eq:trajectoryDm}
x=Z(t) = -\frac{\beta}{2}\log \(  1- e^{2t/\beta} \) , \qquad t=Z^{-1}(x)=  \frac{\beta}{2}\log \(  1- e^{-2x/\beta} \)  \,.
\end{equation}
Compared to the mirror trajectories in eq.~\eqref{eq:trajectoryDp} and eq.~\eqref{eq:trajectoryDm} for type D$_\pm$, one finds that mirror trajectories associated with $p_{\mt{D}_+}(u)$ and $p_{\mt{D}_-}(u)$ can be mapped to each other by replacing $x$ with $-x$. Using eq.~\eqref{eq:brane}, we find that the position of the EOW brane is related to the one for type D$_-$, as can be seen in Fig.~\ref{fig:TypeDmbrane}. The behavior shown in Fig.~\ref{fig:TypeDmbrane} is general for all mirrors of type D$_-$. Although the position of the EOW brane in the limit $u \to \uend$ is also dominated by the term $p''/p'$, similar to type D$_+$, one gets instead 
\begin{equation}
\lim\limits_{u \to \uend} v_{\rm{brane}} \approx -\frac{p^{\prime \prime}z^{2} }{2 p^{\prime}} \sim \frac{z^2}{2} \frac{n-1}{\uend -u} \to -\infty \,,
\end{equation}
as $n<1$ for type D$_-$. 

Different from the type D$_+$ case, the EOW brane for type D$_-$ mirrors does not enclose the physical spacetime. However, the physical spacetime for type D$_-$ is only specified for the region constrained by 
\begin{equation}
  u < \uend \,.
\end{equation}
As in the case of type D$_+$, the complete physical spacetime can be fixed, where the mirror of type D$_-$ may be extended to future infinity by joining it with another timelike mirror. The same is true for the corresponding gravity dual. The geodesic, which computes HEE, connects the boundary point with $u<u_{end}$ to a point on the EOW brane with $u>u_{end}$, \ie the region outside of the dual gravitational spacetime of D$_-$ (up to the null point). However, we may also consider a modified version of the original setup by introducing an extra null brane, as depicted in Fig.~\ref{fig:NullBrane}.

\section{Summary}
\label{sec:5}

In this paper, we have systematically studied moving mirror models in two-dimensional CFTs. We have focused on the case with a single mirror, though we do not expect any essential problem in extending our results to a setup consisting of multiple mirrors. The reflection of modes from moving mirrors is properly described within the framework of BCFT. We have begun our discussion by working out a classification of moving mirror models. According to the endpoints of moving mirrors, we first separate the mirror models into four types named A, B, C, and D. In addition, we further divided each type into three subgroups labeled by subscripts $0$, $+$, and $-$, by carefully examining their asymptotic behavior at late times. As a result, we end up with the following four families of mirror classes,
\begin{itemize}
	\item[] type A$_0$, A$_{+}$, A$_-$ (timelike mirrors), \qquad 
	type B$_0$, B$_{+}$, B$_-$ (escaping mirrors),
	\item[] type C$_0$, C$_{+}$, C$_-$ (chasing mirrors), \qquad 
	type D$_0$, D$_{+}$, D$_-$ (terminated mirrors),
\end{itemize}
as we have briefly summarized in table~\ref{table:AB} and table~\ref{table:CD}. 

For example, the moving mirror modeling the emission of constant Hawking radiation is denoted by type B$_-$. A particular model mimicking the formation and evaporation of a black hole,\ie the so-called kink mirror, belongs to type A$_0$. Type C mirrors, on the other hand, are examples of so-called chasing mirrors, where the mirror moves very fast in the direction of the physical system and not away from it as it is the case in the previously mentioned examples. 
Finally, type D mirrors are defined as models where the mirror terminates at a specific point in bulk spacetime. This mirror type is motivated due to the appearance of null points when, for instance, a timelike mirror trajectory changes to be spacelike. The latter segment, forming a spacelike mirror, is regarded as a model of performing a projection/preparation of a direct product state. Notably, all of these moving mirror model can be analyzed by employing conformal transformations into the setup describing a static mirror, \ie a two-dimensional BCFT defined on the upper half plane.

We also investigated the energy flux via the described conformal mapping procedure. We have found that mirrors of type A and type B lead to a finite energy flux, while mirrors of type C and type D typically encounter divergences for those quantities at the final time. The subscripts $+$, $0$ and $-$ mentioned above mean that the energy flux becomes positive, vanishing , and negative at late times, respectively. 

We have also calculated the entanglement entropy $S_A$ for a given subsystem $A$. When subsystem $A$ is taken to be a semi-infinite line, the entanglement entropy is universal in that it takes the same form for any two-dimensional CFT up to the value of the boundary entropy. 
When $A$ is a finite interval, the results generally depend on the details of the two-dimensional CFT. However, we have found an analytical formula for holographic CFTs. When the semi-infinite interval is static, we have shown that the corresponding entanglement entropy grows logarithmically, \ie $S_A\propto \log t$, for type A$_-$, B$_{\pm,0}$, and type D$_{+}$ mirrors. 
On the other hand, for other mirror types, we have ended up with a finite or vanishing entanglement entropy, where the latter scenario occurs because interval $A$ collides with the mirror at sufficiently late times. When $A$ is taken as a co-moving interval, the entanglement entropy grows logarithmically for type B$_{\pm,0}$ and type C$_{\pm,0}$ mirrors, while it approaches a constant for mirrors of type A$_{\pm,0}$. 

Finally, we have extensively studied the gravity duals for each mirror type. Based on the AdS/BCFT construction, the essential entity in the dual gravitational description of a BCFT is an EOW brane. 
For each mirror type, we have explicitly chosen an analytical profile of a conformal map and shown the shape of the corresponding EOW branes. We have found that for type A$_{\pm,0}$, B$_{\pm,0}$ and type C$_+$ mirrors, the EOW branes are given as simple bulk extensions of the mirror trajectory. 
However, for mirrors of type C$_{-,0}$ and type D$_{\pm}$, the EOW branes in the bulk AdS spacetime bend towards the direction opposite to that the original mirror trajectories heading to. 
Despite such effects, we have found that the gravity duals of type C$_{-,0}$ mirrors appear natural, when the spacetime is limited to the region $u<u_{\rm end}$. Here, $u_{\rm end}$ denotes the maximal value of $u$ for the mirror trajectory obtained after a careful consideration of the EOW brane profiles. 
On the other hand, the situation seems to be more complicated for mirrors of type D$_{\pm}$. Even though a part of the gravity dual of type D$_{\pm}$ mirrors corresponds to the asymptotic Poincar\'{e} AdS spacetime in the presence of an EOW brane, which is restricted to $u<u_{\rm end}$, we need to paste another patch for $u>u_{\rm end}$ to obtain the entire dual gravitational spacetime. However, this is reasonable since the former patch with $u<u_{\rm end}$ fully describes the gravity dual of the causal future of the mirror trajectory, whereas the other patch with $u>u_{\rm end}$ depends on how the trajectory evolves after reaching the null point at which the mirror reaches the speed of light.

\section*{Acknowledgments}
We are grateful to Yuya Kusuki, Rob Myers, and Tomonori Ugajin for useful discussions. TT and SMR are supported by the Simons Foundation through the ``It from Qubit'' collaboration. 
 This work is supported by MEXT-JSPS Grant-in-Aid for Transformative Research Areas (A) ”Extreme
Universe”, No. 21H05187. TT is also supported by Inamori Research Institute for Science and World Premier International Research Center Initiative (WPI Initiative)
from the Japan Ministry of Education, Culture, Sports, Science and Technology (MEXT) and JSPS Grant-in-Aid for Scientific Research (A) No.21H04469. ZW is supported by Grant-in-Aid for JSPS Fellows No. 20J23116.

\appendix
\section{Timelike-spacelike-timelike mirror}
\label{app:A}

In the main text, four classes of timelike moving mirrors have been explored in detail. A natural question is asking for the meaning of spacelike moving mirrors. In this appendix, we explore the effect of spacelike mirrors by considering a smooth trajectory that is timelike at the beginning, becomes spacelike, and then turns back to be timelike again. For simplicity, we call this a TST mirror. Considering the mirror parametrized by the mapping function $v=p(u)$, the spacelike regime corresponds to $p'(u)<0$. Noting the relation between the mapping function $p(u)$ and the velocity of the mirror, \ie
\begin{equation}
v_m(t) \equiv \frac{p'(u)-1}{p'(u)+1} \,, 
\end{equation}
one finds that the spacelike mirror becomes superluminal, because of $|v_m(t)| >1$ for $p'(u)<0$. 
In the following, we first construct a simple TST mirror and then study the entanglement entropy\footnote{In fact, following the discussion in appendix \ref{sec:LIandDM}, the quantity computed in this way is not guaranteed to be entanglement entropy, since $p'(u)$ has a negative region. However, we here refer to it as entanglement entropy with a slight abuse of terminology.} and also its gravity dual by using AdS/BCFT techniques.

\subsection{Construction of TST mirror}
\begin{figure}[ht!]
	\centering
	\includegraphics[width=3in]{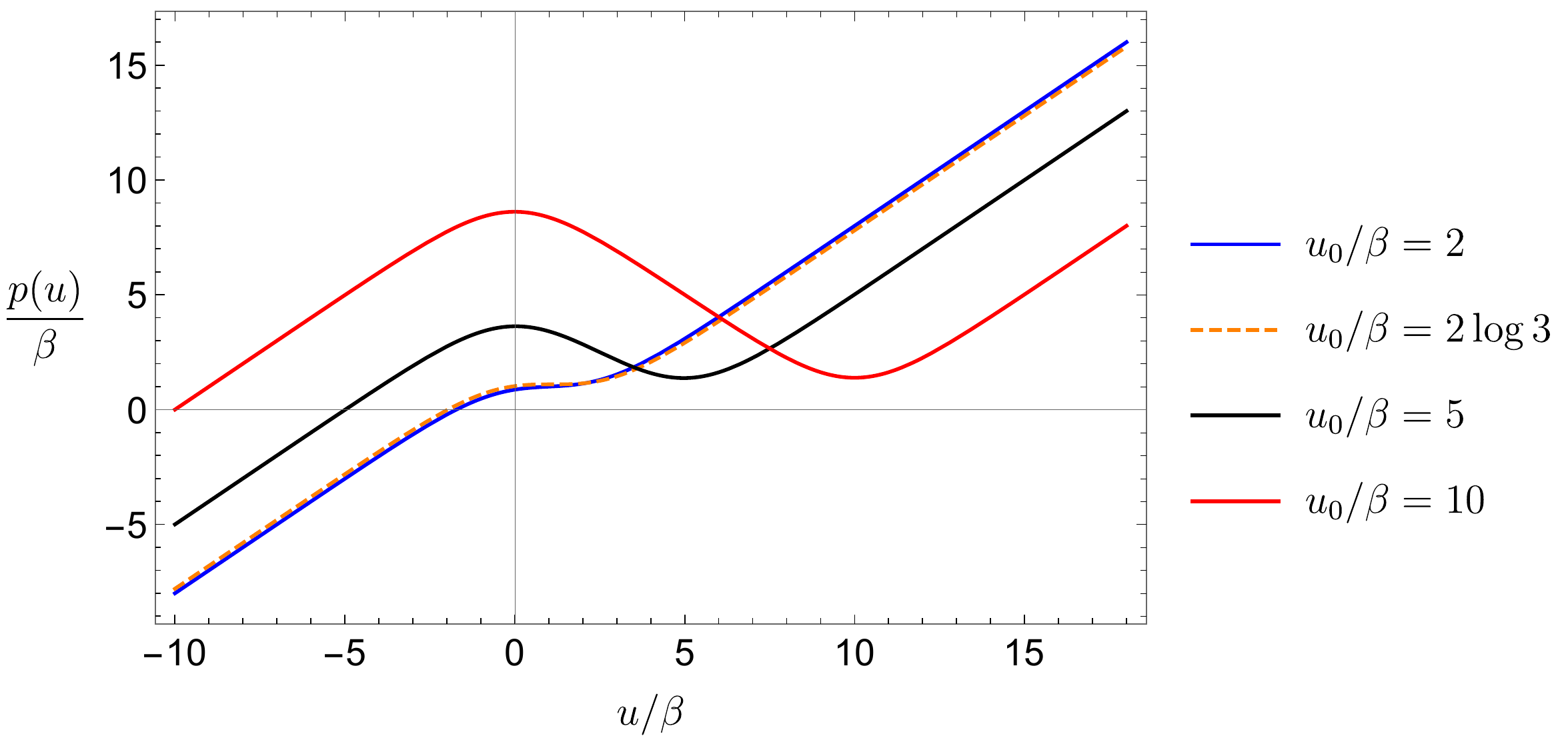}
	\includegraphics[width=3in]{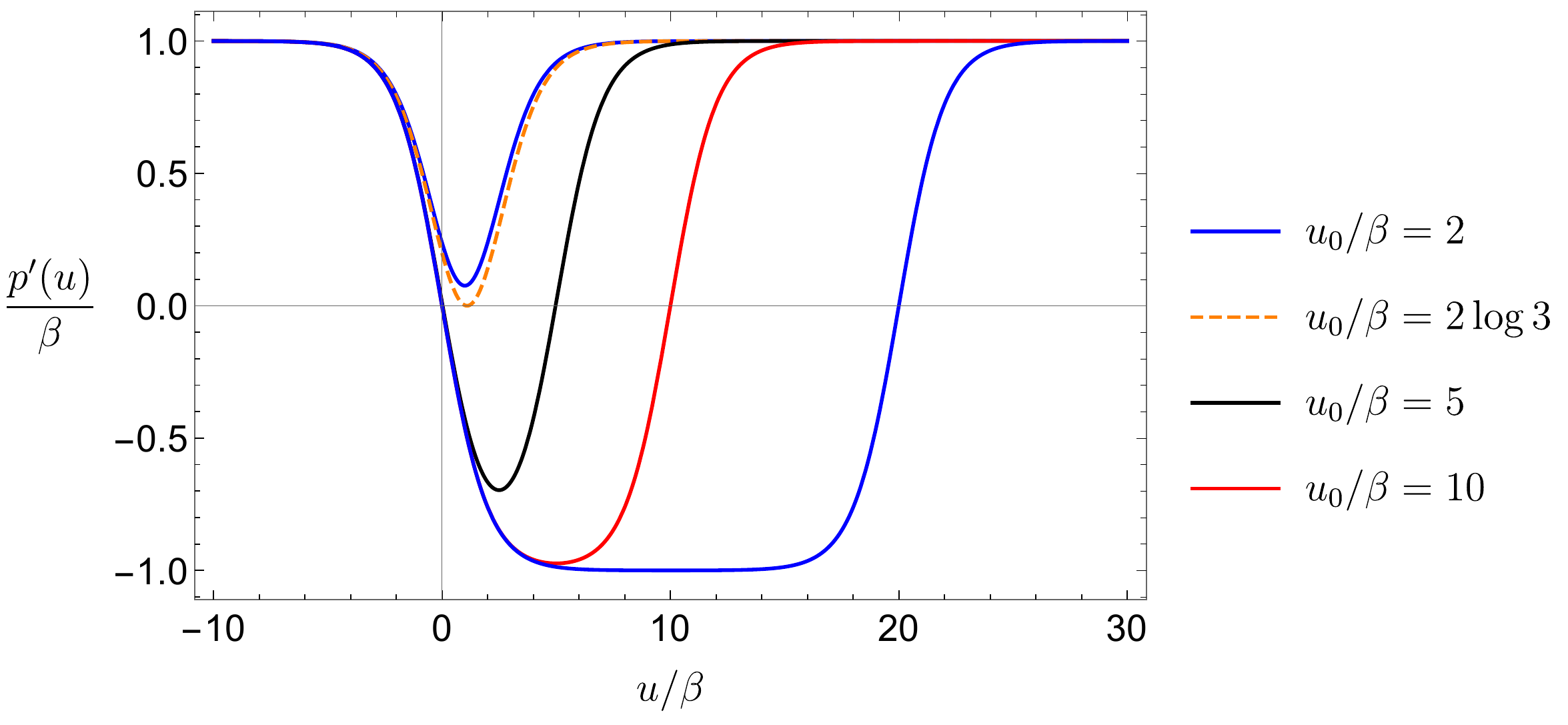}
	\caption{Left: transformation function $p(u)$ with various $u_0$. Right: first order derivative of the transformation function $p'(u)$ with various $u_0$. The TST mirror is obtained when $u_0/\beta > 2\log 3$.}
	\label{fig:TSTpdp}
\end{figure}

\begin{figure}[ht!]
	\centering
		\includegraphics[width=3.1in]{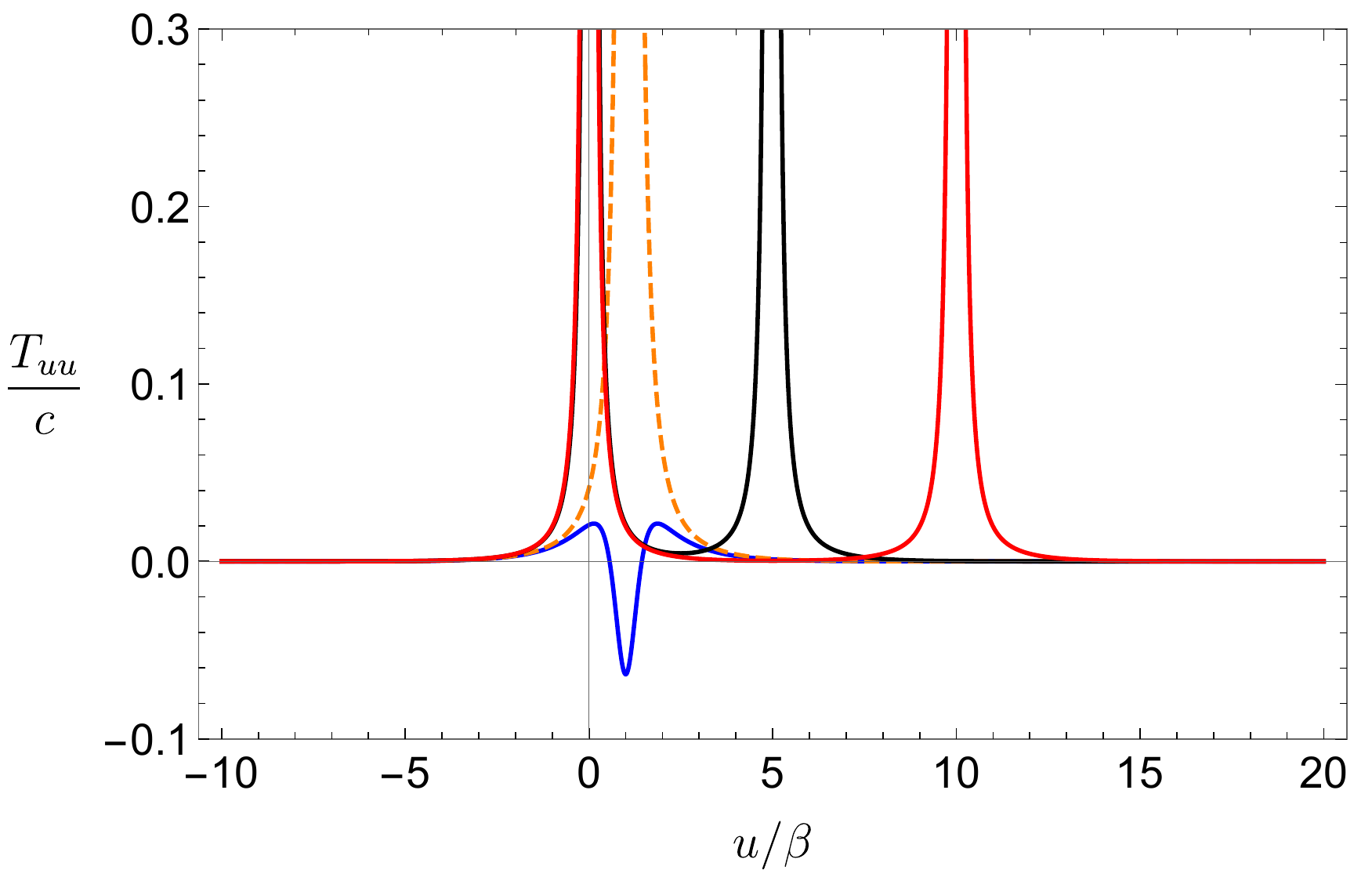}
	\includegraphics[width=2.9in]{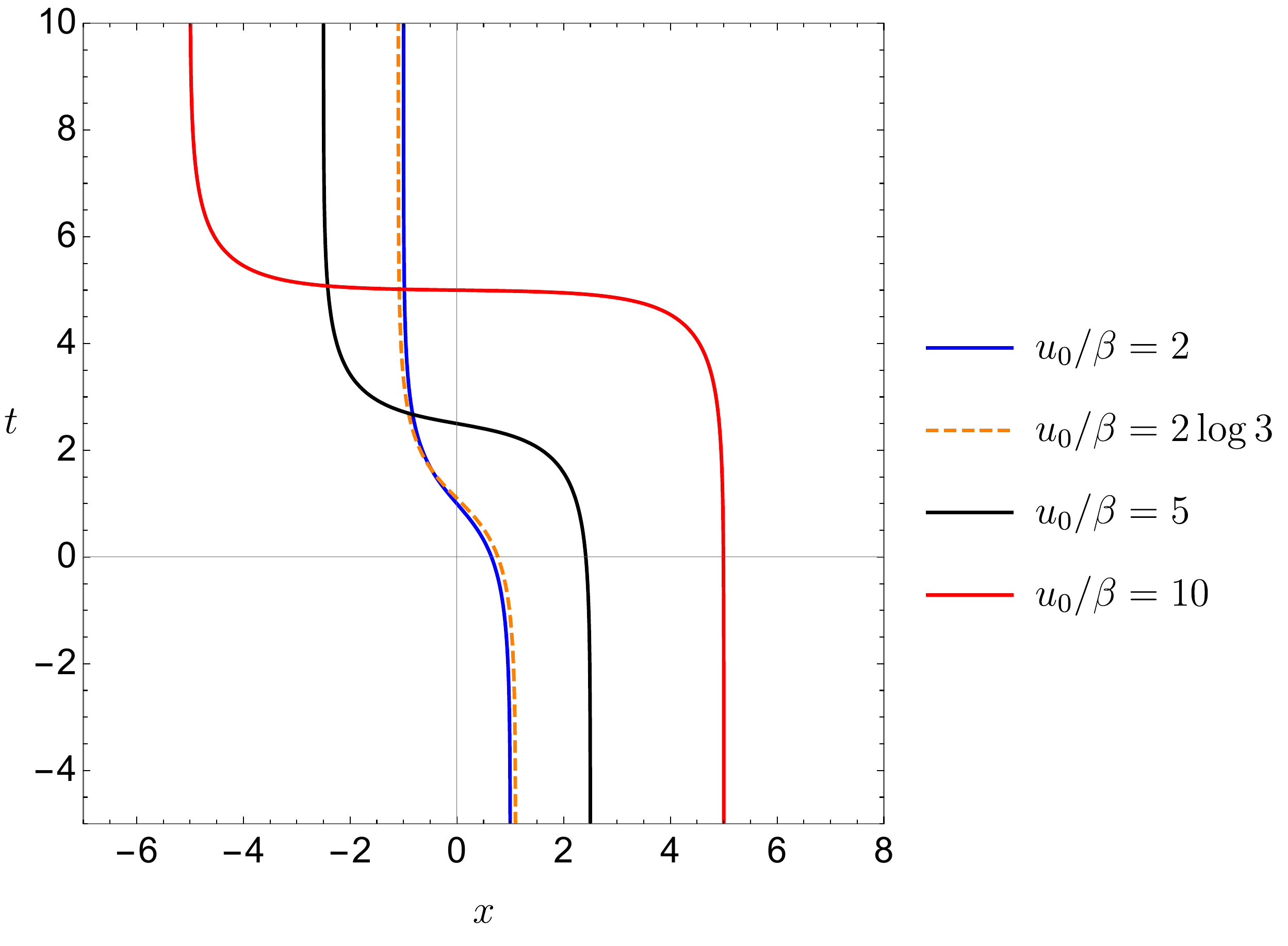}	
	\caption{Right: trajectories of various TST mirrors defined by the mapping function $p_{\mt{TST}}(u)$ in eq.~\eqref{eq:TSTp}. When the displacement parameter is larger than the critical value, \ie $u_0/\beta > 2\log 3 $, the middle part given by $u \in (u_-, u_+)\approx (0,u_0)$ of the profile is always spacelike. Left: non-zero stress tensor associated with the TST mirror. The divergences originate from the turning point $u_{\pm} $ between spacelike and timelike trajectory with $p_{\mt{TST}}'(u_{\pm})=0$.}
	\label{fig:TSTmirror}
\end{figure}

Inspired by the kink mirror studied in \cite{Akal:2020twv,Akal:2021foz}, \eg 
\begin{equation}\label{eq:kinkmirror}
\begin{split}
\text{kink mirror:} \quad p(u)&= - \beta \log \(  1+e^{-u/\beta} \) + \beta \log \( 1+e^{-(u-u_0)/\beta} \) \,,
\end{split}
\end{equation}
we can consider a new mapping function defined by 
\begin{equation}\label{eq:TSTp}
 p_{\mt{TST}}(u) = - 2 \beta \log\(   1+e^{-u/\beta} \)  +\beta \log \( 1+e^{+(u-u_0)/\beta} \)  +  \beta \log\(   1+e^{-(u-u_0)/\beta} \) \,,
\end{equation}
whose first and second order derivatives read
\begin{equation}
p_{\mt{TST}}'(u) = \frac{2}{ 1+e^{(u_0-u)/\beta}}  -\tanh \(  \frac{u}{2\beta} \) \,,\quad p_{\mt{TST}}''(u)= \frac{\text{sech}^2\left(\frac{u-u_0}{2 \beta }\right)-\text{sech}^2\left(\frac{u}{2 \beta }\right)}{2 \beta }\,,
\end{equation}
respectively.
The characteristic behaviors of $p_{\mt{TST}}(u)$ and $p_{\mt{TST}}'(u) $ are shown in Fig.~\ref{fig:TSTpdp}.
For later purpose, we note that the early-time and late-time limits of the mapping function are given by 
\begin{equation}\label{eq:linearapp}
\begin{split}
p_{\mt{TST}}(u) &\approx   u+ u_0 \,, \qquad  \( {(u-u_0)}/{\beta} \gg -1  \)\,,\\
p_{\mt{TST}}(u) &\approx   u- u_0 \,, \qquad  \( {(u-u_0)}/{\beta} \gg 1 \quad \)\,,\\
\end{split}
\end{equation}
respectively. From the second derivative $p''(u)$ one obtains the minimal value of $p'(u)$, which is located at $u=u_0/2$ with 
\begin{equation}
\min \( p_{\mt{TST}}'(u) \) =  p_{\mt{TST}}'\(\frac{u_0}{2}\)=\frac{4}{\exp\( {\frac{u_0}{2 \beta }} \)+1}-1 \,, \qquad  p_{\mt{TST}}\(\frac{u_0}{2}\) = \frac{u_0}{2} \,.
\end{equation}
As a result, we find that the entire mirror trajectory contains a spacelike region when $u_0 > 2 \beta \log 3$. In this case, the mirror profile is first timelike, becomes spacelike afterward, and then timelike again. If the displacement parameter $u_0$ is smaller than the critical value above, the mirror trajectory is always timelike and thus similar to the kink mirror defined in eq.~\eqref{eq:kinkmirror}. For the described TST mirror, we will be focusing on in the following, the transition between spacelike and timelike behavior happens at $p_{\mt{TST}}'(u_\pm )=0$, \ie 
\begin{equation}
\begin{split}
 u_\pm =2\beta \text{arcsech}\left[\frac{2 \sqrt{2}}{\sqrt{2+e^{u_0/\beta } \pm e^{-\frac{u_0}{\beta }} \left(\sqrt{\left(e^{u_0/\beta }-9\right) \left(e^{u_0/\beta }-1\right)^3}+3\right)}}\right] \,. 
\end{split}
\end{equation}
Assuming $e^{u_0/\beta} \gg 1$, the two transition points are approximated by $u_- \approx 0$ and $u_+\approx u_0$. Various trajectories of TST mirrors in the original $(t,x)$ coordinates are depicted in Fig.~\ref{fig:TSTmirror}.  

Let us remind that according our classification, truncated mirrors are of type D. One finds that the first segment of the TST mirror is nothing but a type D$_+$ mirror. We interpret the lightlike point of the type D mirror as the endpoint of its trajectory due to the insertion of an infinite energy flux. For a TST mirror, it is expected that the stress tensor $T_{uu}$ diverges when the transition between timelike and spacelike behavior happens due to $p_{\mt{TST}}'(u)=0$. 
Substituting the mapping function $p_{\mt{TST}}(u)$ for the TST mirror into eq.~\eqref{eq:stresstensor}, one can derive the corresponding energy flux. While the final expression is complicated, we show a numerical plot in the left panel of Fig.~\ref{fig:TSTmirror} for illustration. Although the TST mirror trajectory is smooth, one finds that the lightlike points play the role of a branch cut due to divergences. We also find a similar phenomenon for the entanglement entropy as well as for the holographic gravity dual. The smooth TST mirror also serves as a motivation for considering the particular type D$_\pm$ as an independent category. Besides the appearance of lightlike points in the TST mirror model, the spacelike part is rather exotic since it appears to be `unphysical'. In the following, let us first discuss how to describe the spacelike mirror by including a de Sitter brane with a large tension within the standard AdS$_3$/BCFT$_2$ framework.  

\subsection{Holography with TST mirror}
\begin{figure}[h]
	\centering		
	\includegraphics[width=3in]{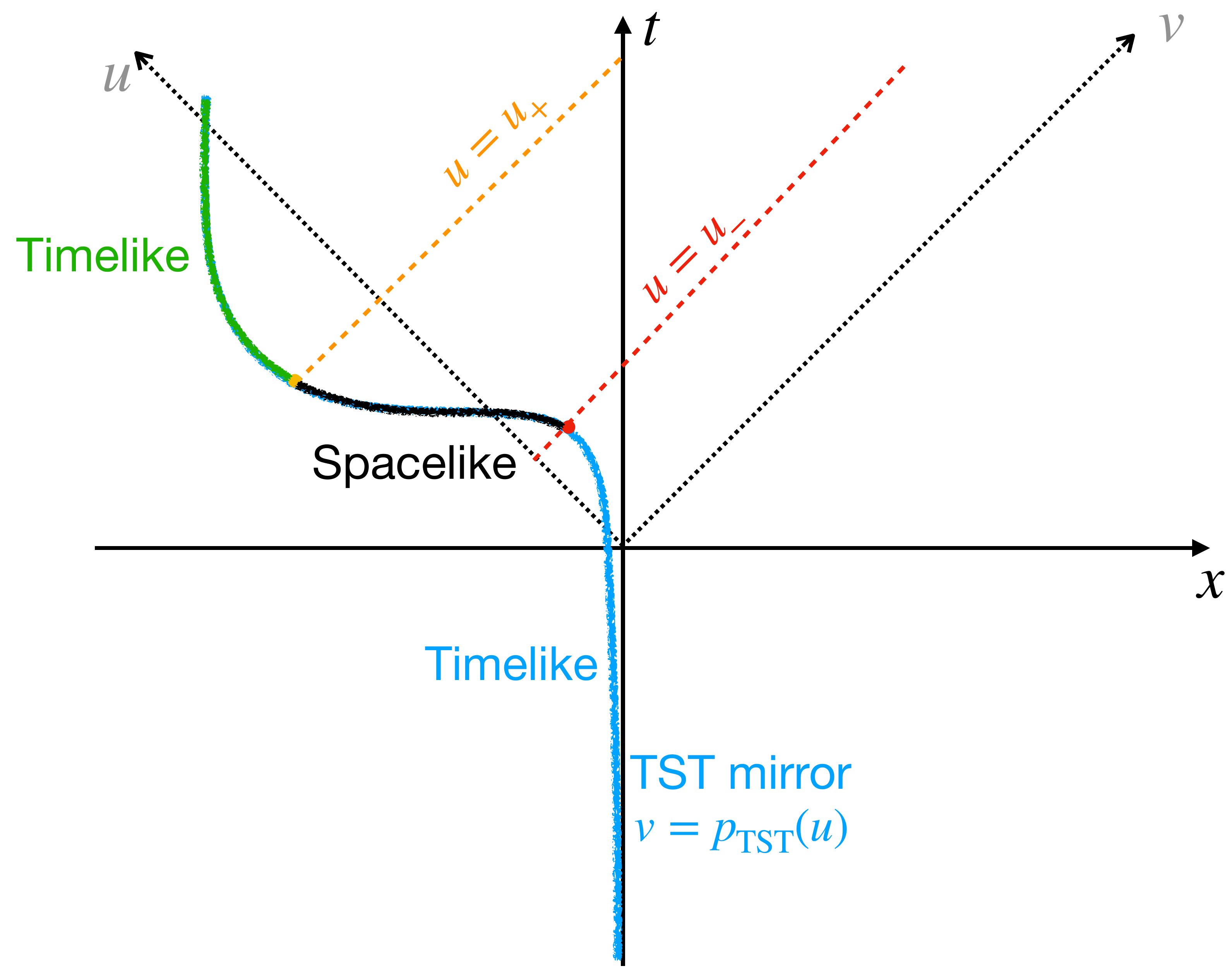}
	\includegraphics[width=3in]{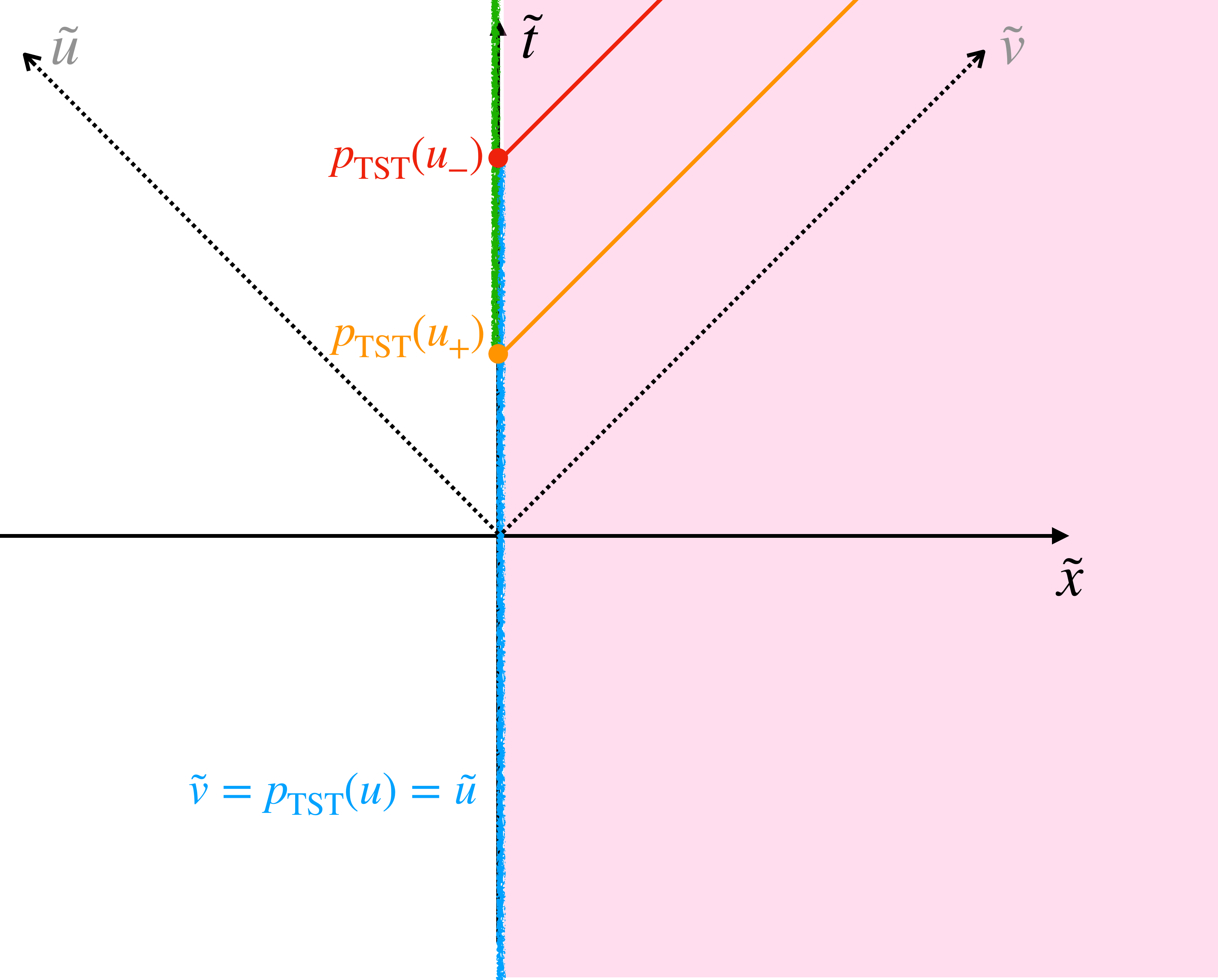}
	\caption{The map in eq.~\eqref{eq:mapTST01} defines the transformation between the TST mirror in the original coordinates, $(t,x)$, and the static mirror in $(\tilde{t}, \tilde{x})$ coordinates. The two timelike parts of the TST mirror are mapped to the static one with a doubly covered region for $\tilde{t} \in [p_{\mt{TST}}(u_+), p_{\mt{TST}}(u_-)]$.}
	\label{fig:TSTmap}
\end{figure}

By now, we have constructed a smooth mirror trajectory using the continuous function $p_{\mt{TST}}$ defined in eq.~\eqref{eq:TSTp}. In order to discuss the holographic dual of a TST mirror, a naive choice is applying the same transformations used for the other mirror types, \ie 
\begin{equation}\label{eq:mapTST01}
\tilde{u} = p(u), \qquad  \tilde{v} = v \,,
\end{equation}
which map the mirror in $(u,v)$ coordinates to the static mirror at $\tilde{x}=0$ in $(\tilde{u}, \tilde{v})$ coordinates. However, the spacelike mirror results in a problem when we use the map above. As shown in Fig.~\ref{fig:TSTmap}, the two disconnected timelike parts of the TST mirror trajectory are mapped to into two different segments of a static mirror, \ie $\tilde{u} \in (-\infty, p_{\mt{TST}}(u_-))$ and $\tilde{u} \in ( p_{\mt{TST}}(u_+), +\infty)$, respectively. Because the two lightlike points at $u_\pm$ are connected by a spacelike part, we always have $p_{\mt{TST}}(u_+) < p_{\mt{TST}}(u_-)$, which implies that the middle segment with $\tilde{u} \in (p_{\mt{TST}}(u_+), p_{\mt{TST}}(u_-))$ is doubly covered via the function $ p_{\mt{TST}}(u)$ mapping the TST mirror to the static mirror. Furthermore, one finds that the spacelike part of the TST mirror is not really necessary, since the static mirror is always timelike. The problem is, that we cannot map a (part of) spacelike mirror to the timelike boundary. In the following, we argue that we should also include a spacelike boundary rather than considering a single static mirror after the conformal map. As a consequence, the dual bulk spacetime contains an extra de Sitter brane that intersects with two EOW branes. 

\subsubsection{Gravity dual of TST moving mirror}
\begin{figure}[h!]
	\centering		
	\includegraphics[width=2.5in]{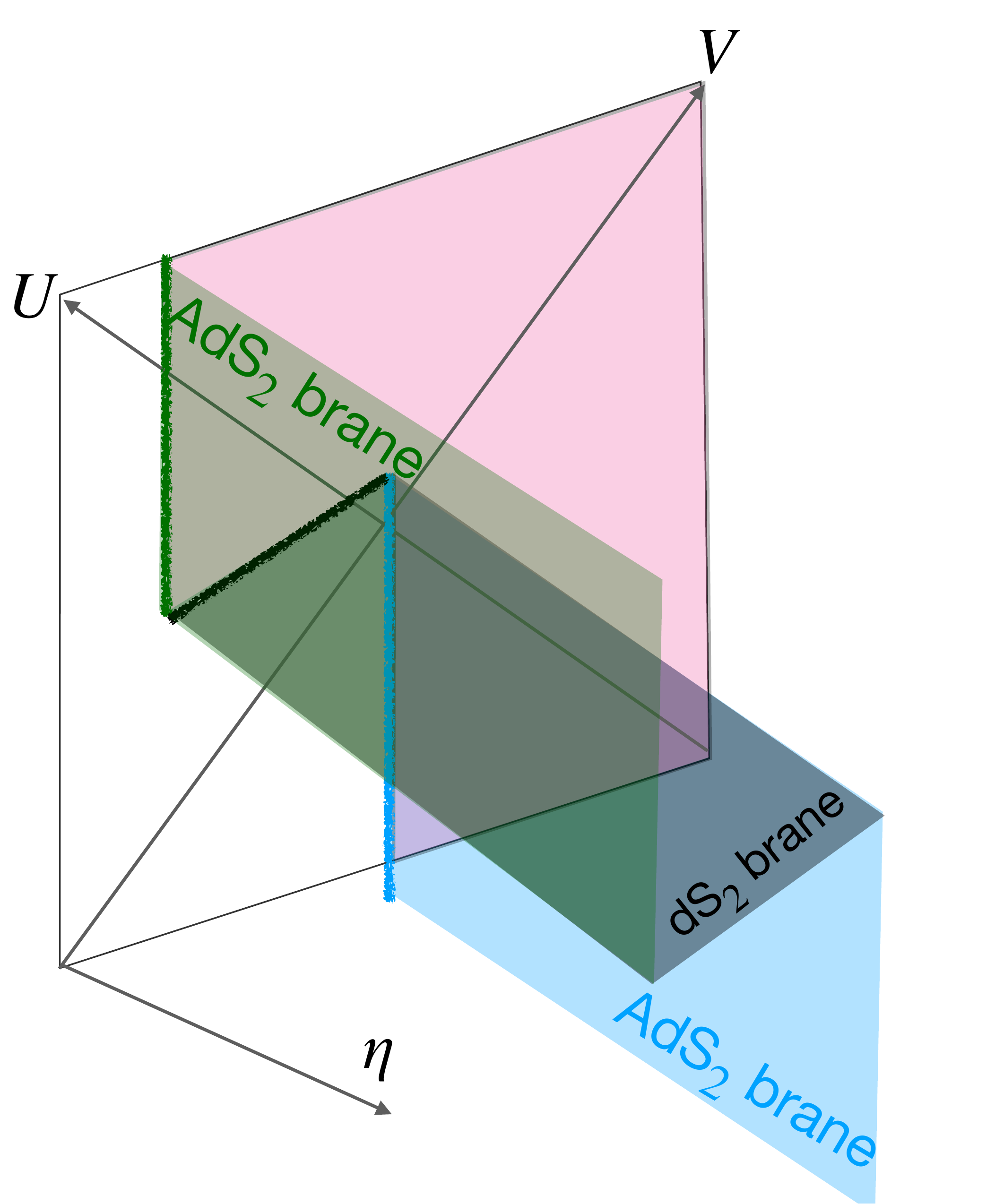}\quad
	\includegraphics[width=2.6in]{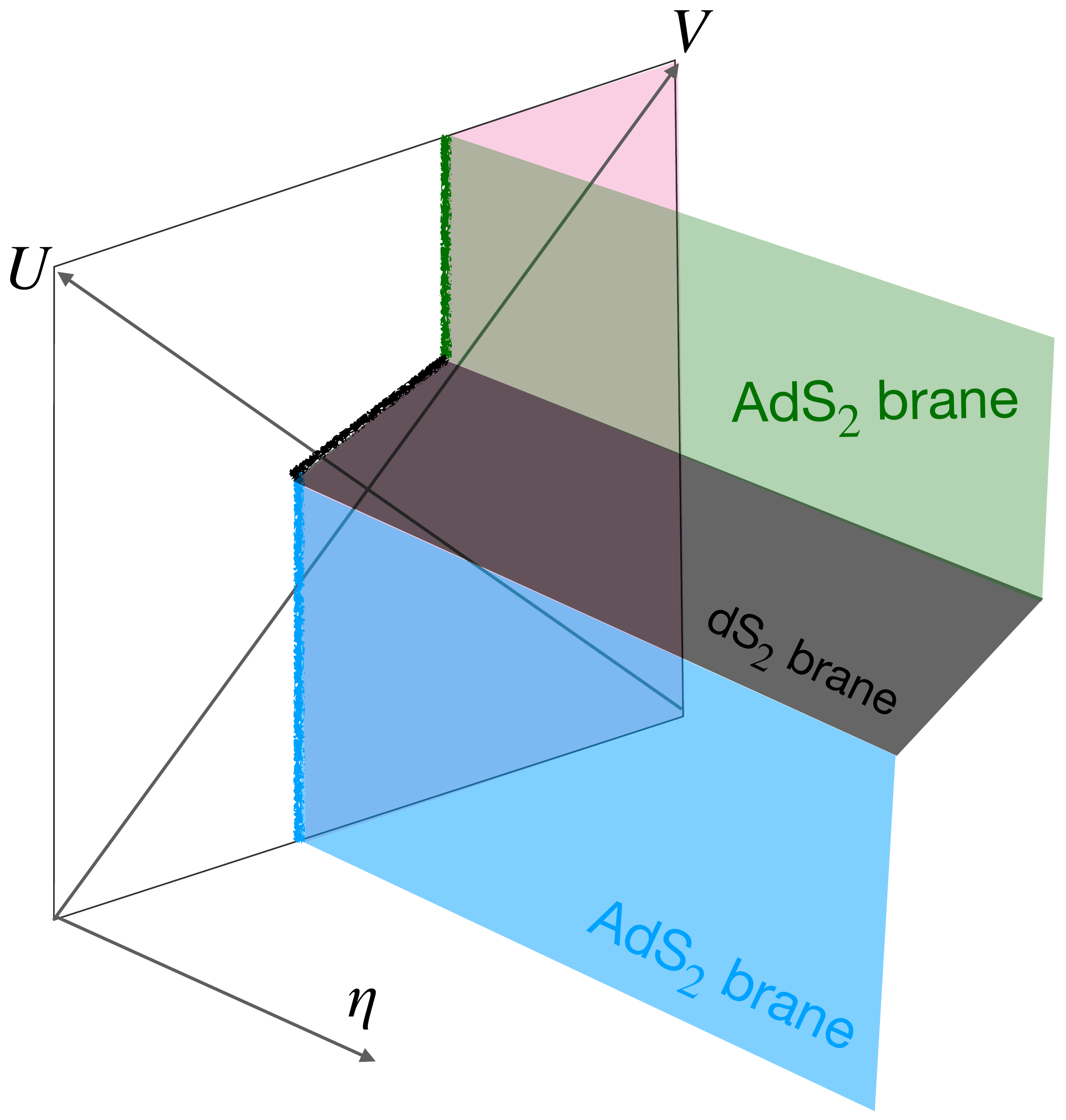}
	\caption{The spacelike boundary in BCFT corresponds to inserting a dS brane with a tension $|\mathcal{T} |>1$. Left: the expected bulk spacetime corresponding to a BCFT whose boundary follows a timelike-spacelike-timelike trajectory as in the case of the TST mirror. Right: another configuration that may mimic final state projection.}
	\label{fig:twobranes}
\end{figure}

Let us first examine a spacelike boundary in BCFT$_2$. In the standard AdS/BCFT correspondence, reviewed in section \ref{sec:AdSBCFT}, one can consider a timelike boundary, \eg $\tilde{x}=0$, in the BCFT and correspondingly obtain a timelike EOW brane, \eg $X= \- \lambda \eta$, as the boundary of the dual bulk spacetime. Particularly, the brane tension $\mathcal{T}$ satisfies $|\mathcal{T} |<1$. However, one may also find a solution by choosing the brane tension as $|\mathcal{T} |>1$.\footnote{For example, see \cite{Karch:2020iit} for various brane profiles embedded in AdS spacetime.} 

Naively, introducing the spacelike mirror can be problematic. However, we can interpret the spacelike boundary via analytical continuation (wick rotation) from Euclidean AdS/BCFT. In this case, the gravitational bulk spacetime contains a timelike dS brane whose tension satisfies $|\mathcal{T}| >1$. Introducing a real parameter $\tilde{\lambda}$ by 
\begin{equation}
 \lambda \equiv \frac{\mathcal{T}}{\sqrt{1- \mathcal{T}^2}} = (-i)  \frac{\mathcal{T}}{\sqrt{\mathcal{T}^2-1}} =i \tilde{\lambda}\,, 
\end{equation}
the EOW brane in Lorentzian Poincar\'e coordinates is given by 
\begin{equation}
T =\tilde{ \lambda} \eta \,,
\end{equation} 
with an induced metric 
\begin{equation}
d s^{2}= \frac{-\left( \tilde{\lambda}^{2} -1 \right) d \eta^{2}+d X^{2}}{\eta^{2}}\,.
\end{equation}
It is obvious that the EOW brane with $|\tilde{\lambda}|>1$, corresponding to the spacelike boundary, is nothing but a two-dimensional de Sitter spacetime. Different from the standard timelike boundary of BCFT, the boundary entropy associated with the spacelike boundary now takes complex values as follows 
\begin{equation}\label{eq:complexSbdy}
S_{\rm{bdy}} = \frac{c}{6} \log \sqrt{\frac{1+\mathcal{T}}{1-\mathcal{T}}}=  \frac{c}{6} \log \sqrt{\frac{\mathcal{T}+1}{\mathcal{T}-1}} + i \, \frac{c \pi}{12} \,. 
\end{equation}
The BCFT dual of this gravitational setup is constructed in \cite{Akal:2020wfl}.

For the sake of constructing the holographic dual spacetime for a TST mirror, a natural expectation is that the dS brane could joint the AdS branes, whose intersections with BCFT are timelike, as shown in the left panel of Fig.~\ref{fig:twobranes}. However, we need to point out several caveats in this proposal. First of all, one can find that this configuration with intersections of branes explicitly breaks the symmetry of bulk spacetime. It is also expected because the corresponding state of such a boundary theory is not described by the standard boundary state $| B \rangle $ anymore. One exciting interpretation is that the spacelike boundary in BCFT is related to a projective measurement. See \eg \cite{Akal:2021dqt} for more exploration in this direction. Furthermore, as the joint between the spacelike mirror and timelike mirror on the conformal boundary, the two types of branes also intersect in the bulk. One may wonder how to understand the meaning of the brane intersection. For example, we expect that such an intersection would cut a part of the brane, as shown in Fig.~\ref{fig:twobranes}. Finally, we expect that the dS brane inserted at distinct position would lead to different interpretations from the perspective of the boundary field theory. Taking the right panel of Fig.~\ref{fig:twobranes}, we would like to argue that it presents a similar spirit as in final state projection. It is interesting to explore this possibility as a future direction. 

\subsubsection{Entanglement entropy}
In this subsection, we study the HEE $S_{A}$ of a subsystem in two-dimensional BCFT with a TST mirror defined by the mapping function $p_{\mt{TST}}(u)$ in eq.~\eqref{eq:TSTp}.  On account of the EOW brane, the RT surface of a boundary subsystem $A$ can be defined as either the connected or the disconnected geodesic. Although we have mentioned several difficulties with the holographic dual of a TST mirror, we here leave it all aside and investigate the simplest choice by taking the results via analytical continuation. Let us again consider a semi-infinite line whose endpoint is located at $x_0$. By performing an analytical continuation from the Euclidean signature, the corresponding HEE is determined by the disconnected geodesic described by the earlier formula derived in eq.~\eqref{eq:defineSEEA}. One may worry that the entanglement entropy related to the spacelike mirror would be complex due to $p'(u) <0$. Interestingly, we can find that the imaginary part from $\log \frac{1}{\sqrt{p'(u)}}$ with $p'(u) < 0$ is exactly canceled by that in the boundary entropy $S_{\rm{bdy}}$ derived in eq.~\eqref{eq:complexSbdy}. As a result, we still arrive at a real-valued HEE, namely   
\begin{equation}\label{eq:Seemod}
\begin{split}
S_A &= \frac{c}{6} \log \frac{v_{a}-p\left(u_{a}\right)}{\epsilon \sqrt{p^{\prime}\left(u_{a}\right)}}+ S_{\mathrm{bdy}} 
=\frac{c}{6} \log \frac{v_{a}-p\left(u_{a}\right)}{\epsilon \sqrt{|p^{\prime}\left(u_{a}\right)|}}+ \Re \(  S_{\mathrm{bdy}}  \)\,. 
\end{split}
\end{equation}
In the following, we examine the entanglement entropy of the static, semi-infinite interval and the co-moving one by using the TST mirror constructed via the mapping function $p_{\mt{TST}}(u)$ in eq.~\eqref{eq:TSTp}.  

\paragraph{Static, semi-infinite interval}

\begin{figure}[ht!]
	\centering
	\includegraphics[height=1.5in]{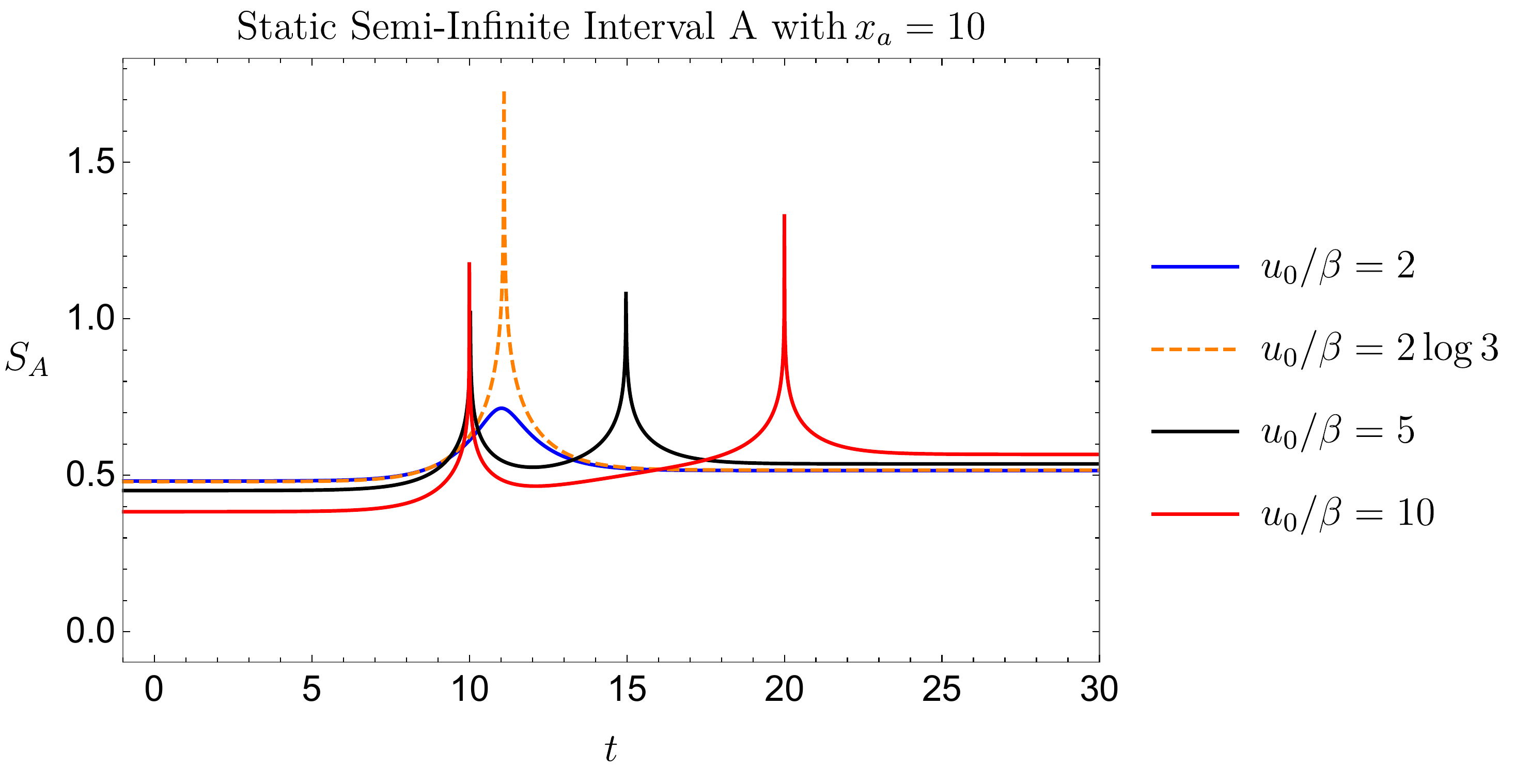}
		\includegraphics[height=1.5in]{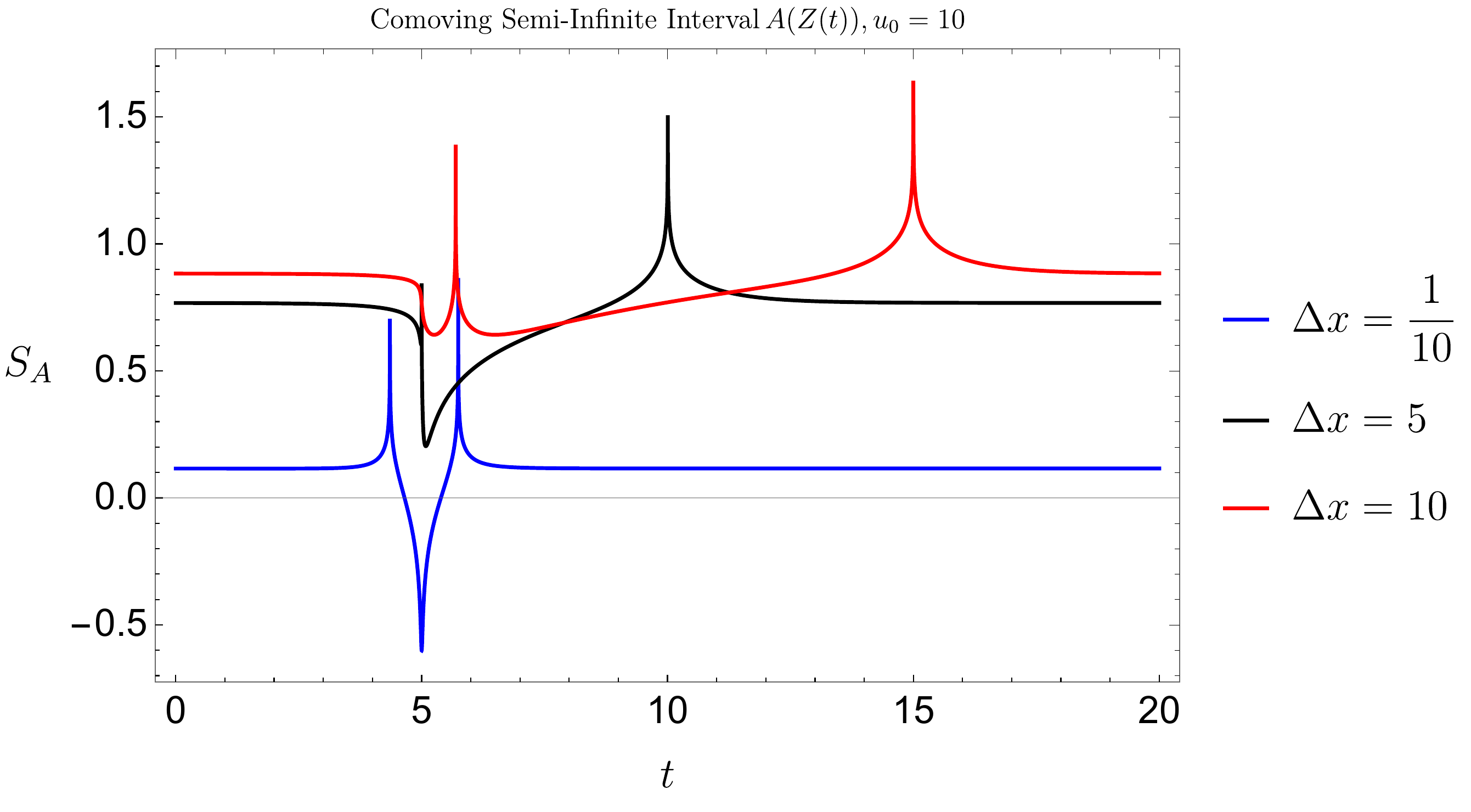}
	\caption{Time evolution of HEE $S_A$ of a subsystem $A$ in the presence of a TST mirror. Left: static, semi-infinite interval with endpoint located at $x_a=10$. Right: co-moving, semi-infinite interval at various distances to the TST mirror. Note that all peaks appearing in both plots are infinite in height.}
	\label{fig:TSTSEE}
\end{figure}
We first consider a static, semi-infinite interval $A$ with fixing its endpoint as $x_a= \text{constant}$. The time evolution of HEE for the TST mirror is plotted in Fig.~\ref{fig:TSTSEE}. By using the linear approximation of $p_{\mt{TST}}(u)$ for the TST mirror, as shown in eq.~\eqref{eq:linearapp}, we may conclude that the leading-order of the time derivative $\partial_t S_A$ vanishes, \ie the entanglement entropy is approximated by a constant at early and late times. Besides this universal property, another new feature is the appearance of divergences in the entanglement entropy. Obviously, these divergences are introduced by the two lightlike points located at $t- x_a  = u_\pm$, \ie 
\begin{equation}
t \approx  x_a \,,\qquad  \text{or} \qquad t \approx x_a + u_0 \,,
\end{equation}
where we have used the approximations $u_-\approx 0$ and $u_+\approx u_0$. 
This type of divergence is the same as the one for the type $D_+$ mirror, where the lightlike trajectory generates a positive, divergent energy flux.

\paragraph{Co-moving, semi-infinite interval}

Next, we study the time evolution of HEE for a subsystem $A$ with a co-moving endpoint $x_a(t)$. In particular, we take the distance between $A$ and the mirror trajectory to be a constant $\Delta x$, \ie $x_a(t)= Z(t)+ \Delta x $.
The time evolution of the subsystem entropy for various $\Delta x$ is shown in Fig.~\ref{fig:TSTSEE}. We observe that there is a decay around $t \approx u_0/2$. Roughly speaking, this is caused by the fast decay of the distance $V_a- U_a$, when the endpoint of the subsystem evolves in time. For example, by using the precise values $p_{\mt{TST}}\(\frac{u_0}{2}\)= \frac{u_0}{2}$ and $Z\(\frac{u_0}{2}\)=0$, and taking $\Delta x  \to 0$, we can use $v_a = u_a = \frac{u_0}{2}$, \ie $t= \frac{u_0}{2}$ and $x =Z(t)=0$, to exactly get
\begin{equation}
V_a - U_a= v_a- p_{\mt{TST}}(u_a) =t+Z(t) - p_{\mt{TST}}(t-Z(t))= 0 \,. 
\end{equation}
When we take the endpoint $x_a$ away from the trajectory, \ie $\Delta x \ne 0$, the distance $V_a-U_a$ does not exactly vanish and approaches a small value. As a result, one gets 
\begin{equation}
 \lim\limits_{t \to u_0/2} \frac{\partial S_A}{\partial t}  \sim  - \infty \,.
\end{equation}
This illustrates the decays shown in Fig.~\ref{fig:TSTSEE} for the co-moving interval $A$. 

\section{Lorentzian path integrals and density matrices}
\label{sec:LIandDM}

In this paper and previous works \cite{Akal:2020twv,Akal:2021foz}, the entanglement entropies for moving mirrors are computed by applying the RT formula \cite{Ryu:2006bv,Ryu:2006ef,Hubeny:2007xt}, or equivalently by inserting twist operators \cite{Calabrese:2004eu,Calabrese:2009qy} on Lorentzian path integrals defined on manifolds with moving mirrors. Although these computations look straightforward and their results are consistent with entanglement entropies directly computed from wave functionals of free scalars, justifications for these computations are nontrivial. 

The nontrivial point is whether our path integral gives a density matrix or not. Consider a path integral defined on a Lorentzian manifold and pick up a Cauchy slice on this path integral, then the Cauchy slice divides the whole path integral into two parts: one comes from the past and the other comes from the future. Each of them defines a quantum state, and the two quantum states are generically different. When the two quantum states are the same, the path integral defines an ordinary density matrix on this Cauchy slice. Otherwise, it will define a more general transition matrix \cite{Nakata:2021ubr}, like setups in \cite{Akal:2021dqt}. Our computation via correlation functions of twist operators turns out to lead to entanglement entropies when and only, when the path integral defines a density matrix on the Cauchy slice we are focusing on, \ie the Cauchy slice on which the twist operators are inserted. 

Based on these, the goal of this appendix is to see that the path integrals in the main text of this paper and previous works \cite{Akal:2020twv,Akal:2021foz} indeed define density matrices on the Cauchy slices we are interested in. In the following, we firstly review the case for Euclidean path integrals to help the readers grasp the idea. After that, we will present a sufficient condition for a Lorentzian path integral to define a density matrix on a given Cauchy slice. At the end, we will see that setups considered in the main text of this paper and \cite{Akal:2020twv,Akal:2021foz} indeed fall into this class. 

\subsection{Euclidean path integrals and reflection symmetries}

Given a path integral defined on a Euclidean manifold $\CM_E$, one can always introduce a codimension-1 surface $\Sigma$ which divides the original manifold into two disconnected parts.\footnote{In this appendix, we use $\Sigma$ to denote a codimension-1 surface in a path integral. This should be distinguished from $\Sigma$ in the main text, which is used to denote the manifold on which the BCFT is defined.} By regarding $\Sigma$ as a spatial slice, the path integrals on the two sides define a bra state $\bra{\alpha}$ and a ket state $\ket{\beta}$, respectively, and the partition function $Z_{\CM_E}$ computes the inner product of the two states,  
\begin{align}
     Z_{\CM_E} = \int_{\CM_E} D\phi~ e^{-S_E[\phi]} = \braket{\alpha|\beta}. 
\end{align}
Note that the states here are not necessarily normalized. If one inserts some operator $\CO$ on this codimension-1 surface $\Sigma$, the path integral computes the weak value of $\CO$ under the transition matrix $|\beta \rangle \langle \alpha|$,
\begin{align}
    \int_{\CM_E} D\phi~ e^{-S_E[\phi]}~\CO(\Sigma) 
    = \braket{\alpha| \CO |\beta} = {\rm Tr}\left(\CO |\beta \rangle \langle \alpha|\right) \,.
\end{align}
In this way, we may say that the path integral over $\CM_E$ defines the transition matrix $|\beta \rangle \langle \alpha|$ on $\Sigma$. In Euclidean formulation of QFTs, setups which are symmetric under the reflection with respect to $\Sigma$ are often considered. In these cases, $\bra{\alpha} \propto (\ket{\beta})^{\dagger}$, and therefore the path integral realizes a density matrix on $\Sigma$. This symmetry is often referred to as the time reflection symmetry, by regarding $\Sigma$ as a Euclidean time slice. 

For example, consider a Euclidean strip $[-T/2,T/2] \times \mathbb{R}$ and use $(\tau,x)$ to parameterize it. If we regard $\tau$ as the Euclidean time and impose the same boundary condition on $\tau=\pm T/2$, then the path integral on this strip is time reflection symmetric with respect to $\tau=0$. If we denote the state on $\tau=- T/2$ as $\ket{B}$ and the state on $\tau= T/2$ as $\bra{B}$, then the path integral over $\tau\in(-T/2,0)$ defines the ket state $e^{-HT/2}\ket{B}$ and that over $\tau\in(0,T/2)$ defines the bra state $\bra{B}e^{-HT/2}$. Since $\left(\bra{B}e^{-HT/2}\right)^{\dagger} = e^{-HT/2}\ket{B}$, the path integral defines the density matrix $e^{-HT/2}|{B}\rangle \langle{B}|e^{-HT/2}$ on $\tau = 0$.

Let us summarize the observations above as follows. Generally, a Euclidean path integral realizes a transition matrix on a codimension-1 surface $\Sigma$. In some special cases, the transition matrix becomes a density matrix. One sufficient condition for this to happen is the reflection symmetry with respect to $\Sigma$. 

As a result, when the Euclidean path integral is time reflection symmetric with respect to $\Sigma$, inserting twist operators on $\Sigma$ can compute entanglement entropy. Otherwise, it will be computing pseudo entropy in general. For example, pseudo entropies are computed by considering non symmetric Euclidean path integrals in \cite{Nakata:2021ubr,Mollabashi:2021xsd,Nishioka:2021cxe,Miyaji:2021lcq}.

\subsection{Lorentzian path integrals}

Let us then move on to Lorentzian cases. Similarly, given a path integral defined on a Lorentzian manifold $\CM_L$, one can always introduce a Cauchy surface $\Sigma$ which divides the original manifold into two disconnected parts. The path integrals on the two sides define a bra state $\bra{\alpha}$ and a ket state $\ket{\beta}$.

Note that a time reflection symmetric setup in Lorentzian sense does not define a density matrix in general. For example, consider a Lorentzian strip $[-T/2,T/2] \times \mathbb{R}$ and use $(t,x)$ to parameterize the time and the spatial direction, respectively. If we impose the same boundary condition on $t= \pm T/2$, then the path integral on this strip is time reflection symmetric with respect to $t=0$. Similar to but different from the Euclidean case, the path integral over $t\in(-T/2,0)$ defines the ket state $e^{-iHT/2}\ket{B}$ and that over $t \in(0,T/2)$ defines the bra state $\bra{B}e^{-iHT/2}$. Since $\left(\bra{B}e^{-iHT/2}\right)^{\dagger} = e^{iHT/2}\ket{B}$ is not proportional to $e^{-iHT/2}\ket{B}$ in general, the path integral does not define a density matrix but the transition matrix $e^{-iHT/2}|{B}\rangle \langle{B}|e^{-iHT/2}$ on $t = 0$. 

The next question we would like to ask is when does a Lorentzian path integral define a density matrix on a given Cauchy slice $\Sigma$. As explained above, by definition, 
\begin{align}
    {\rm a~given~Lorentzian~path~integral~on~a~given~Cauchy~slice~}\Sigma  \quad \Longleftrightarrow \quad (\bra\alpha)^\dagger \propto \ket{\beta}. 
\end{align}
When expressed in the language of the partition function, it means that 
\begin{align}\label{eq:DMprop}
    {\rm for~any~boundary~condition~}\phi(\Sigma) = \varphi,~\frac{\left(Z_{\CM_L^+}^{\phi(\Sigma)=\varphi}\right)^*}{Z_{\CM_L^-}^{\phi(\Sigma)=\varphi}}={\rm const}  \,,
\end{align}
where $\CM_L^+$ ($\CM_L^-$) is the upper (lower) half of $\CM_L$ separated by $\Sigma$, and 
\begin{align}
    Z_{\CM_L^\pm}^{\phi(\Sigma)=\varphi} = \int_{\CM_L^\pm}^{\phi(\Sigma)=\varphi} D\phi~ e^{iS_L[\phi]} \,,
\end{align}
is the partition function on $\CM_{L}^{\pm}$ with boundary condition $\phi(\Sigma) = \varphi$ imposed on $\Sigma$. Here, we would like to argue that the property eq.~\eqref{eq:DMprop} is preserved under the following transformations. 

First of all, we would like to argue, that for a given path integral defined on a static spacetime $\CM_{L}$, if eq.~\eqref{eq:DMprop} is satisfied on a Cauchy slice $\Sigma$ then it is also satisfied on any other Cauchy slice $\Sigma'$. Precisely, it is an assumption rather than a conclusion. Basically, what this assumption is saying is that the time translation induced by the Lorentzian path integral between $\Sigma$ and $\Sigma'$ is reversible. If we are considering a theory in which the time translations are unitary, then of course they are reversible. On the other hand, there are arguments saying that, in an expanding universe, the time translations should be regarded as an isometry but not a unitary \cite{Cotler:2022weg}. This is the reason, why we restrict this argument to a static spacetime. For example, the right half plane $\mathbb{R}^{1,1}/\mathbb{Z}_2$ is a static spacetime.

Secondly, consider a 2D CFT defined on $\CM_{L}$ which is parameterized by $(u,v) = (t-x,t+x)$ with the metric given by
\begin{align}\label{eq:CMmetric}
    ds^2 = -dudv = -dt^2 + dx^2\,.
\end{align}
Consider mapping $\CM_{L}$ to another manifold $\tilde{\CM}_{L}$ parameterized by $(\tilde{u},\tilde{v}) = (\tilde{t}-\tilde{x},\tilde{t}+\tilde{x})$ via the conformal transformation
\begin{align}
    \tilde{u} = p(u),~\tilde{v} = q(v)\,,
\end{align}
which satisfies 
\begin{align}\label{eq:positiveWeyl}
    p'(u)>0,~q'(v)>0\,,
\end{align}
with the metric given by 
\begin{align}
    ds^2 = -d\tilde{u}d\tilde{v} = -d\tilde{t}^2 + d\tilde{x}^2\,.
\end{align}
Then, for any Cauchy slice $\Sigma \subset \CM_{L}$, its image $\tilde{\Sigma} \subset \tilde{\CM}_{L}$ is also a Cauchy slice. This is because the conformal map can be regarded as a diffeomorphism on $\CM_L$ which transforms \eqref{eq:CMmetric} to 
\begin{align}
    ds^2 = -\left[\frac{1}{p'(u)q'(v)}\right]_{p(u) = \tilde{u}, q(v) = \tilde{v}} d\tilde{u}d\tilde{v}\,,
\end{align}
combined to a Weyl transformation to $\tilde{\CM}_{L}$ with the Weyl factor
\begin{align}
    e^{-2\chi(\tilde{u},\tilde{v})} = \left[p'(u)q'(v)\right]_{p(u) = \tilde{u}, q(v) = \tilde{v}}\,,
\end{align}
and both diffeomorphism and Weyl transformation (with positive Weyl factor) do not change the causal structure. We would like to argue that, in this case, if eq.~\eqref{eq:DMprop} is satisfied on $\tilde{\Sigma}\subset\tilde{\CM}_L$, it is also satisfied on $\Sigma\subset\CM_L$. To see this, note that $(\CM_L,e^{2\chi}\eta_{ab})$ can be obtained via a Weyl transformation $e^{2\chi}$ from $(\tilde{\CM}_L,\eta_{ab})$. It is known that for 2D CFTs, the measure of the path integral transforms as \cite{Ginsparg:1993is}
\begin{align}\label{eq:measure}
    [D\phi]_{e^{2\chi}\eta_{ab}} = e^{i (I[\chi,\tilde{\CM}_L]-I[0,\tilde{\CM}_L])}[D\phi]_{\eta_{ab}}
\end{align}
while the action does not change under such a transformation. Here, $I[\chi,\tilde{\CM}_L]$ is the Liouville action \cite{Polyakov:1981rd}
\begin{align}
    I[\chi,\tilde{\CM}_L] = \frac{c}{24\pi} \int_{\tilde{\CM}_L} d\tilde{u}d\tilde{v} \left[-4\partial_{\tilde{u}}\chi\partial_{\tilde{v}}\chi + \mu e^{2\chi}\right],
\end{align}
where $c$ is the central charge of the CFT we are considering.\footnote{See \cite{Caputa:2017urj,Caputa:2017yrh} for related discussions in the context of computational complexity.} 

Let us now apply the Weyl transformation to $\CM_L^+$ and $\CM_L^-$ separately. To do so, we need to understand how the Weyl factor is modified with respect to the boundary condition imposed on $\Sigma$. 

Let us explicitly write down the path integral to see it more clearly. The path integral over $\CM^{\pm}_L$ can be written as 
\begin{align}
    Z_{\CM_L^\pm}^{\phi(\Sigma)=\varphi} = \int_{\CM_L^\pm}^{\phi(\Sigma)=\varphi} D\phi~ e^{iS_L[\phi]} = \int_{\CM_L^\pm} D\phi~ e^{iS_L[\phi]} \delta[\phi(\Sigma)-\varphi] \,,
\end{align}
by using a delta wave functional $\delta[\phi(\Sigma)-\varphi]$. Under a Weyl transformation, the measure picks up an overall factor as eq.~\eqref{eq:measure}, and the $e^{iS_L[\phi]}$ part is invariant. To proceed, we need to figure out how the $\delta[\phi(\Sigma)-\varphi]$ part transforms under the Weyl transformation. This is in general a hard question, but here we would like to note that this should only depend on the quantities localized on $\Sigma$. In other words, the way of transformation for such a wave functional should be able to be specified by $\Sigma$, its image after the Weyl transformation $\Sigma_{\rm after}$, and the field configuration $\varphi$ localized on the Cauchy slice. Let us denote the prefactor associated with the Weyl transformation as $A(\Sigma,\Sigma_{\rm aft}, \varphi)$ which 
\begin{align}
    \delta[\phi_{\rm aft}(\Sigma_{\rm aft})-\varphi_{\rm aft}] = A(\Sigma,\Sigma_{\rm aft}, \varphi)~ \delta[\phi(\Sigma)-\varphi]\,,
\end{align}
without identifying its explicit form. It is easy to see that $A(\Sigma,\Sigma_{\rm aft}, \varphi) = A^*(\Sigma,\Sigma_{\rm aft}, \varphi)$ via a lattice regularization of the current path integral since the conformal weights are real. This property will turn out to be important in the following discussions. 

Based on the discussions above, if $(\tilde{\CM}_L,\eta_{ab})$ satisfies eq.~\eqref{eq:DMprop}, then for any boundary condition $\phi(\Sigma)=\varphi$, there exists a corresponding boundary condition $\phi(\tilde{\Sigma})=\tilde{\varphi}$ such that 
\begin{align}
    \frac{\left(Z_{\CM_L^+}^{\phi(\Sigma)=\varphi}\right)^*}{Z_{\CM_L^-}^{\phi(\Sigma)=\varphi}} = 
    \frac{\left(e^{i (I[\chi,\tilde{\CM}^+_L]-I[0,\tilde{\CM}^+_L])} A(\tilde{\Sigma},\Sigma, \tilde{\varphi})~ Z_{\tilde\CM_L^+}^{\phi(\tilde\Sigma)=\tilde\varphi}\right)^*}{e^{i (I[\chi,\tilde{\CM}_L^-]-I[0,\tilde{\CM}_L^-])} A(\tilde{\Sigma},\Sigma, \tilde{\varphi}) Z_{\tilde\CM_L^-}^{\phi(\tilde\Sigma)=\tilde\varphi}} 
    &= e^{-i (I[\chi,\tilde{\CM}_L]-I[0,\tilde{\CM}_L])} \times {\rm const.} \nonumber\\
    &= {\rm const.}
\end{align}
and hence $\CM_L$ satisfies eq.~\eqref{eq:DMprop} on $\Sigma$. In short, though different boundary conditions pick up different prefactors under a Weyl transformation, they are cancelled out when taking the ratio between the path integral over $\CM_L^+$ and that over $\CM_L^-$. As a result, the property \eqref{eq:DMprop} is preserved under conformal transformations in 2D CFT. 

\subsection{Path integrals on timelike moving mirror spacetime}
\begin{figure}[h]
	\centering		
	\includegraphics[width=15cm]{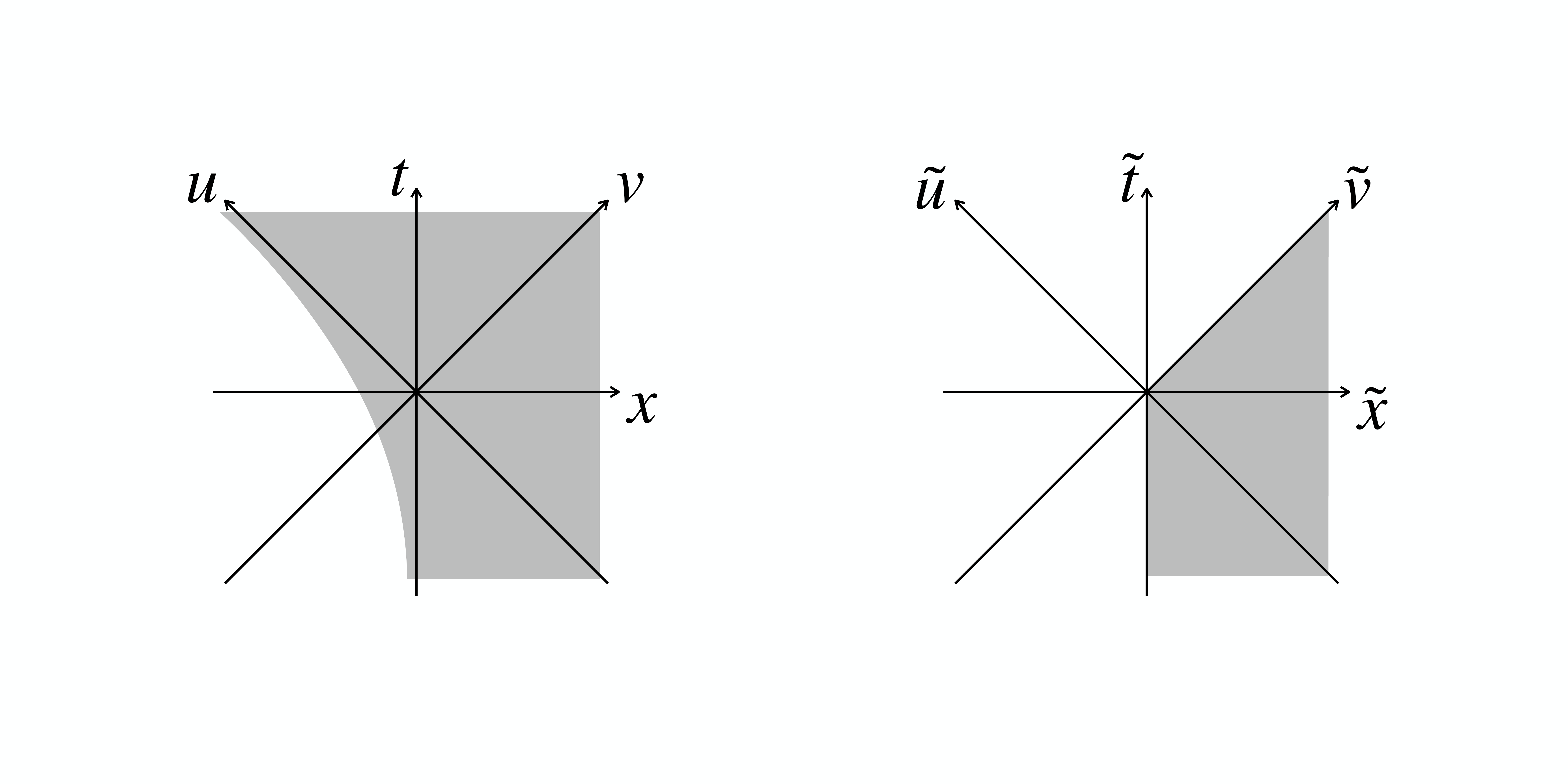}
	\caption{Sketch of a moving mirror setup (left) and its image (right). The physical region is shaded grey. Instead of the right half plane, the physical region is given by the $\tilde{u}<0$ portion of it. Therefore, one should impose the boundary condition at $\tilde{u}=0$ to perform the path integral.}
	\label{fig:EsMM}
\end{figure}

Based on the discussions above, let us then come back to moving mirror setups. For timelike moving mirrors in 2D CFTs, we analysed it in the following way. Firstly, we construct a conformal map which satisfies eq.~\eqref{eq:positiveWeyl} and use it to map the original setup to the right half plane. Then we treat the right half plane as the ground state on a half-infinite line. Therefore, according to the first argument in the previous section, the path integral of the right half plane defines a density matrix on any Cauchy slice. Moreover, according to the second argument in the previous section, the path integral on the moving mirror spacetime defines a density matrix on any Cauchy slice. This is the reason why we can compute the correlation functions and entanglement entropies successfully via the conformal map method. 

In fact, some nontrivial treatments which intrinsically help one to successfully compute the entanglement entropy are performed in the escaping mirror setup \cite{Akal:2020twv,Akal:2021foz}. Consider an escaping mirror setup with the mirror trajectory given by
\begin{align}
    x = -\beta \arcsinh \frac{e^{\frac{t}{\beta}}}{2}\,.
\end{align}
This can be mapped to a right half plane by performing the following conformal transformation 
\begin{align}
    \tilde{u} = -\beta \log(1+e^{-\frac{u}{\beta}}),~\tilde{v} = v \,.
\end{align}
More precisely, the resulting geometry is not the right half plane but the $\tilde{u}<0$ portion of it. See Fig.~\ref{fig:EsMM} for a sketch. In fact, there is a choice of the boundary condition imposed on $\tilde{u}=0$. The treatment performed in \cite{Akal:2020twv,Akal:2021foz} is to fill the $\tilde{u}>0$ portion and use the right half plane to perform computations. This treatment implicitly imposes a boundary condition on $\tilde{u}=0$, such that the path integral defines density matrices. In this sense, this treatment implicitly allowed the authors of \cite{Akal:2020twv,Akal:2021foz} to reproduce entanglement entropy in an escaping mirror. Note that the case of type D mirrors is similar to the escaping mirror explained above.

\bibliographystyle{JHEP}
\bibliography{Ref_Mirrors}

\providecommand{\href}[2]{#2}\begingroup\raggedright\begin{thebibliography}{10}

\bibitem{Davies:1976hi}
P.~Davies and S.~Fulling, \emph{{Radiation from a moving mirror in
  two-dimensional space-time conformal anomaly}}, {\emph{Proc. Roy. Soc. Lond.
  A} {\bfseries 348} (1976) 393}.

\bibitem{Birrell:1982ix}
N.~Birrell and P.~Davies, \emph{{Quantum Fields in Curved Space}}, Cambridge
  Monographs on Mathematical Physics. Cambridge Univ. Press, Cambridge, UK, 2,
  1984,
  \href{https://doi.org/10.1017/CBO9780511622632}{10.1017/CBO9780511622632}.

\bibitem{Cardy:2004hm}
J.~L. Cardy, \emph{{Boundary conformal field theory}},
  \href{https://arxiv.org/abs/hep-th/0411189}{{\ttfamily hep-th/0411189}}.

\bibitem{Hawking:1974sw}
S.~W. Hawking, \emph{{Particle Creation by Black Holes}},
  \href{https://doi.org/10.1007/BF02345020}{\emph{Commun. Math. Phys.}
  {\bfseries 43} (1975) 199}.

\bibitem{Davies:1977yv}
P.~C.~W. Davies and S.~A. Fulling, \emph{{Radiation from Moving Mirrors and
  from Black Holes}}, \href{https://doi.org/10.1098/rspa.1977.0130}{\emph{Proc.
  Roy. Soc. Lond. A} {\bfseries 356} (1977) 237}.

\bibitem{Ford:1982ct}
L.~H. Ford and A.~Vilenkin, \emph{{Quantum radiation by moving mirrors}},
  \href{https://doi.org/10.1103/PhysRevD.25.2569}{\emph{Phys. Rev. D}
  {\bfseries 25} (1982) 2569}.

\bibitem{Carlitz:1986nh}
R.~D. Carlitz and R.~S. Willey, \emph{{Reflections on moving mirrors}},
  \href{https://doi.org/10.1103/PhysRevD.36.2327}{\emph{Phys. Rev. D}
  {\bfseries 36} (1987) 2327}.

\bibitem{Wilczek:1993jn}
F.~Wilczek, \emph{{Quantum purity at a small price: Easing a black hole
  paradox}},  in \emph{{International Symposium on Black holes, Membranes,
  Wormholes and Superstrings}}, 2, 1993,
  \href{https://arxiv.org/abs/hep-th/9302096}{{\ttfamily hep-th/9302096}}.

\bibitem{Raval:1996vt}
A.~Raval, B.~L. Hu and D.~Koks, \emph{{Near thermal radiation in detectors,
  mirrors and black holes: A Stochastic approach}},
  \href{https://doi.org/10.1103/PhysRevD.55.4795}{\emph{Phys. Rev. D}
  {\bfseries 55} (1997) 4795}
  [\href{https://arxiv.org/abs/gr-qc/9606074}{{\ttfamily gr-qc/9606074}}].

\bibitem{Bianchi:2014qua}
E.~Bianchi and M.~Smerlak, \emph{{Entanglement entropy and negative energy in
  two dimensions}},
  \href{https://doi.org/10.1103/PhysRevD.90.041904}{\emph{Phys. Rev. D}
  {\bfseries 90} (2014) 041904}
  [\href{https://arxiv.org/abs/1404.0602}{{\ttfamily 1404.0602}}].

\bibitem{Hotta:2015huj}
M.~Hotta and A.~Sugita, \emph{{The Fall of Black Hole Firewall: Natural
  Nonmaximal Entanglement for Page Curve}},
  \href{https://doi.org/10.1093/ptep/ptv170}{\emph{PTEP} {\bfseries 2015}
  (2015) 123B04} [\href{https://arxiv.org/abs/1505.05870}{{\ttfamily
  1505.05870}}].

\bibitem{Hotta:2015yla}
M.~Hotta, R.~Sch\"utzhold and W.~G. Unruh, \emph{{Partner particles for moving
  mirror radiation and black hole evaporation}},
  \href{https://doi.org/10.1103/PhysRevD.91.124060}{\emph{Phys. Rev. D}
  {\bfseries 91} (2015) 124060}
  [\href{https://arxiv.org/abs/1503.06109}{{\ttfamily 1503.06109}}].

\bibitem{Chen:2017lum}
P.~Chen and D.-h. Yeom, \emph{{Entropy evolution of moving mirrors and the
  information loss problem}},
  \href{https://doi.org/10.1103/PhysRevD.96.025016}{\emph{Phys. Rev. D}
  {\bfseries 96} (2017) 025016}
  [\href{https://arxiv.org/abs/1704.08613}{{\ttfamily 1704.08613}}].

\bibitem{Good:2019tnf}
M.~R.~R. Good, E.~V. Linder and F.~Wilczek, \emph{{Moving mirror model for
  quasithermal radiation fields}},
  \href{https://doi.org/10.1103/PhysRevD.101.025012}{\emph{Phys. Rev. D}
  {\bfseries 101} (2020) 025012}
  [\href{https://arxiv.org/abs/1909.01129}{{\ttfamily 1909.01129}}].

\bibitem{Akal:2020twv}
I.~Akal, Y.~Kusuki, N.~Shiba, T.~Takayanagi and Z.~Wei, \emph{{Entanglement
  Entropy in a Holographic Moving Mirror and the Page Curve}},
  \href{https://doi.org/10.1103/PhysRevLett.126.061604}{\emph{Phys. Rev. Lett.}
  {\bfseries 126} (2021) 061604}
  [\href{https://arxiv.org/abs/2011.12005}{{\ttfamily 2011.12005}}].

\bibitem{Akal:2021foz}
I.~Akal, Y.~Kusuki, N.~Shiba, T.~Takayanagi and Z.~Wei, \emph{{Holographic
  moving mirrors}},
  \href{https://doi.org/10.1088/1361-6382/ac2c1b}{\emph{Class. Quant. Grav.}
  {\bfseries 38} (2021) 224001}
  [\href{https://arxiv.org/abs/2106.11179}{{\ttfamily 2106.11179}}].

\bibitem{Reyes:2021npy}
I.~A. Reyes, \emph{{Moving Mirrors, Page Curves, and Bulk Entropies in AdS2}},
  \href{https://doi.org/10.1103/PhysRevLett.127.051602}{\emph{Phys. Rev. Lett.}
  {\bfseries 127} (2021) 051602}
  [\href{https://arxiv.org/abs/2103.01230}{{\ttfamily 2103.01230}}].

\bibitem{Jaekel:1997hr}
M.-T. Jaekel and S.~Reynaud, \emph{{Movement and fluctuations of the vacuum}},
  \href{https://doi.org/10.1088/0034-4885/60/9/001}{\emph{Rept. Prog. Phys.}
  {\bfseries 60} (1997) 863}
  [\href{https://arxiv.org/abs/quant-ph/9706035}{{\ttfamily
  quant-ph/9706035}}].

\bibitem{Plunien:1999ba}
G.~Plunien, R.~Schutzhold and G.~Soff, \emph{{Dynamical Casimir effect at
  finite temperature}},
  \href{https://doi.org/10.1103/PhysRevLett.84.1882}{\emph{Phys. Rev. Lett.}
  {\bfseries 84} (2000) 1882}
  [\href{https://arxiv.org/abs/quant-ph/9906122}{{\ttfamily
  quant-ph/9906122}}].

\bibitem{Romualdo:2019eur}
I.~Romualdo, L.~Hackl and N.~Yokomizo, \emph{{Entanglement production in the
  dynamical Casimir effect at parametric resonance}},
  \href{https://doi.org/10.1103/PhysRevD.100.065022}{\emph{Phys. Rev. D}
  {\bfseries 100} (2019) 065022}
  [\href{https://arxiv.org/abs/1908.00835}{{\ttfamily 1908.00835}}].

\bibitem{Brevik:2000zb}
I.~H. Brevik, K.~A. Milton, S.~D. Odintsov and K.~E. Osetrin, \emph{{Dynamical
  Casimir effect and quantum cosmology}},
  \href{https://doi.org/10.1103/PhysRevD.62.064005}{\emph{Phys. Rev. D}
  {\bfseries 62} (2000) 064005}
  [\href{https://arxiv.org/abs/hep-th/0003158}{{\ttfamily hep-th/0003158}}].

\bibitem{Casadio:2002dj}
R.~Casadio and L.~Mersini-Houghton, \emph{{Short distance signatures in
  cosmology: Why not in black holes?}},
  \href{https://doi.org/10.1142/S0217751X04016453}{\emph{Int. J. Mod. Phys. A}
  {\bfseries 19} (2004) 1395}
  [\href{https://arxiv.org/abs/hep-th/0208050}{{\ttfamily hep-th/0208050}}].

\bibitem{Good:2020byh}
M.~R.~R. Good, A.~Zhakenuly and E.~V. Linder, \emph{{Mirror at the edge of the
  universe: Reflections on an accelerated boundary correspondence with de
  Sitter cosmology}},
  \href{https://doi.org/10.1103/PhysRevD.102.045020}{\emph{Phys. Rev. D}
  {\bfseries 102} (2020) 045020}
  [\href{https://arxiv.org/abs/2005.03850}{{\ttfamily 2005.03850}}].

\bibitem{Cotler:2022weg}
J.~Cotler and A.~Strominger, \emph{{The Universe as a Quantum Encoder}},
  \href{https://arxiv.org/abs/2201.11658}{{\ttfamily 2201.11658}}.

\bibitem{Cong:2018vqx}
W.~Cong, E.~Tjoa and R.~B. Mann, \emph{{Entanglement Harvesting with Moving
  Mirrors}}, \href{https://doi.org/10.1007/JHEP06(2019)021}{\emph{JHEP}
  {\bfseries 06} (2019) 021}
  [\href{https://arxiv.org/abs/1810.07359}{{\ttfamily 1810.07359}}].

\bibitem{Cong:2020nec}
W.~Cong, C.~Qian, M.~R.~R. Good and R.~B. Mann, \emph{{Effects of Horizons on
  Entanglement Harvesting}},
  \href{https://doi.org/10.1007/JHEP10(2020)067}{\emph{JHEP} {\bfseries 10}
  (2020) 067} [\href{https://arxiv.org/abs/2006.01720}{{\ttfamily
  2006.01720}}].

\bibitem{Fewster:2004nj}
C.~J. Fewster and S.~Hollands, \emph{{Quantum energy inequalities in
  two-dimensional conformal field theory}},
  \href{https://doi.org/10.1142/S0129055X05002406}{\emph{Rev. Math. Phys.}
  {\bfseries 17} (2005) 577}
  [\href{https://arxiv.org/abs/math-ph/0412028}{{\ttfamily math-ph/0412028}}].

\bibitem{Chen:2015bcg}
P.~Chen and G.~Mourou, \emph{{Accelerating Plasma Mirrors to Investigate Black
  Hole Information Loss Paradox}},
  \href{https://doi.org/10.1103/PhysRevLett.118.045001}{\emph{Phys. Rev. Lett.}
  {\bfseries 118} (2017) 045001}
  [\href{https://arxiv.org/abs/1512.04064}{{\ttfamily 1512.04064}}].

\bibitem{Hotta:2022aiv}
M.~Hotta, Y.~Nambu, Y.~Sugiyama, K.~Yamamoto and G.~Yusa, \emph{{Expanding
  Edges of Quantum Hall Systems in a Cosmology Language -- Hawking Radiation
  from de Sitter Horizon in Edge Modes}},
  \href{https://arxiv.org/abs/2202.03731}{{\ttfamily 2202.03731}}.

\bibitem{Takayanagi:2011zk}
T.~Takayanagi, \emph{{Holographic Dual of BCFT}},
  \href{https://doi.org/10.1103/PhysRevLett.107.101602}{\emph{Phys. Rev. Lett.}
  {\bfseries 107} (2011) 101602}
  [\href{https://arxiv.org/abs/1105.5165}{{\ttfamily 1105.5165}}].

\bibitem{Fujita:2011fp}
M.~Fujita, T.~Takayanagi and E.~Tonni, \emph{{Aspects of AdS/BCFT}},
  \href{https://doi.org/10.1007/JHEP11(2011)043}{\emph{JHEP} {\bfseries 11}
  (2011) 043} [\href{https://arxiv.org/abs/1108.5152}{{\ttfamily 1108.5152}}].

\bibitem{Karch:2000gx}
A.~Karch and L.~Randall, \emph{{Open and closed string interpretation of SUSY
  CFT's on branes with boundaries}},
  \href{https://doi.org/10.1088/1126-6708/2001/06/063}{\emph{JHEP} {\bfseries
  06} (2001) 063} [\href{https://arxiv.org/abs/hep-th/0105132}{{\ttfamily
  hep-th/0105132}}].

\bibitem{Ma}
J.~M. Maldacena, \emph{{The Large N limit of superconformal field theories and
  supergravity}}, \href{https://doi.org/10.1023/A:1026654312961}{\emph{Int. J.
  Theor. Phys.} {\bfseries 38} (1999) 1113}
  [\href{https://arxiv.org/abs/hep-th/9711200}{{\ttfamily hep-th/9711200}}].

\bibitem{Ryu:2006bv}
S.~Ryu and T.~Takayanagi, \emph{{Holographic derivation of entanglement entropy
  from AdS/CFT}},
  \href{https://doi.org/10.1103/PhysRevLett.96.181602}{\emph{Phys. Rev. Lett.}
  {\bfseries 96} (2006) 181602}
  [\href{https://arxiv.org/abs/hep-th/0603001}{{\ttfamily hep-th/0603001}}].

\bibitem{Ryu:2006ef}
S.~Ryu and T.~Takayanagi, \emph{{Aspects of Holographic Entanglement Entropy}},
  \href{https://doi.org/10.1088/1126-6708/2006/08/045}{\emph{JHEP} {\bfseries
  08} (2006) 045} [\href{https://arxiv.org/abs/hep-th/0605073}{{\ttfamily
  hep-th/0605073}}].

\bibitem{Hubeny:2007xt}
V.~E. Hubeny, M.~Rangamani and T.~Takayanagi, \emph{{A Covariant holographic
  entanglement entropy proposal}},
  \href{https://doi.org/10.1088/1126-6708/2007/07/062}{\emph{JHEP} {\bfseries
  07} (2007) 062} [\href{https://arxiv.org/abs/0705.0016}{{\ttfamily
  0705.0016}}].

\bibitem{Page:1993df}
D.~N. Page, \emph{{Average entropy of a subsystem}},
  \href{https://doi.org/10.1103/PhysRevLett.71.1291}{\emph{Phys. Rev. Lett.}
  {\bfseries 71} (1993) 1291}
  [\href{https://arxiv.org/abs/gr-qc/9305007}{{\ttfamily gr-qc/9305007}}].

\bibitem{Page:1993wv}
D.~N. Page, \emph{{Information in black hole radiation}},
  \href{https://doi.org/10.1103/PhysRevLett.71.3743}{\emph{Phys. Rev. Lett.}
  {\bfseries 71} (1993) 3743}
  [\href{https://arxiv.org/abs/hep-th/9306083}{{\ttfamily hep-th/9306083}}].

\bibitem{Karch:2000ct}
A.~Karch and L.~Randall, \emph{{Locally localized gravity}},
  \href{https://doi.org/10.1088/1126-6708/2001/05/008}{\emph{JHEP} {\bfseries
  05} (2001) 008} [\href{https://arxiv.org/abs/hep-th/0011156}{{\ttfamily
  hep-th/0011156}}].

\bibitem{Randall:1999ee}
L.~Randall and R.~Sundrum, \emph{{A Large mass hierarchy from a small extra
  dimension}}, \href{https://doi.org/10.1103/PhysRevLett.83.3370}{\emph{Phys.
  Rev. Lett.} {\bfseries 83} (1999) 3370}
  [\href{https://arxiv.org/abs/hep-ph/9905221}{{\ttfamily hep-ph/9905221}}].

\bibitem{Randall:1999vf}
L.~Randall and R.~Sundrum, \emph{{An Alternative to compactification}},
  \href{https://doi.org/10.1103/PhysRevLett.83.4690}{\emph{Phys. Rev. Lett.}
  {\bfseries 83} (1999) 4690}
  [\href{https://arxiv.org/abs/hep-th/9906064}{{\ttfamily hep-th/9906064}}].

\bibitem{Penington:2019npb}
G.~Penington, \emph{{Entanglement Wedge Reconstruction and the Information
  Paradox}}, \href{https://doi.org/10.1007/JHEP09(2020)002}{\emph{JHEP}
  {\bfseries 09} (2020) 002}
  [\href{https://arxiv.org/abs/1905.08255}{{\ttfamily 1905.08255}}].

\bibitem{Almheiri:2019psf}
A.~Almheiri, N.~Engelhardt, D.~Marolf and H.~Maxfield, \emph{{The entropy of
  bulk quantum fields and the entanglement wedge of an evaporating black
  hole}}, \href{https://doi.org/10.1007/JHEP12(2019)063}{\emph{JHEP} {\bfseries
  12} (2019) 063} [\href{https://arxiv.org/abs/1905.08762}{{\ttfamily
  1905.08762}}].

\bibitem{Almheiri:2019hni}
A.~Almheiri, R.~Mahajan, J.~Maldacena and Y.~Zhao, \emph{{The Page curve of
  Hawking radiation from semiclassical geometry}},
  \href{https://doi.org/10.1007/JHEP03(2020)149}{\emph{JHEP} {\bfseries 03}
  (2020) 149} [\href{https://arxiv.org/abs/1908.10996}{{\ttfamily
  1908.10996}}].

\bibitem{Sato:2021ftf}
Y.~Sato, \emph{{Complexity in a moving mirror model}},
  \href{https://doi.org/10.1103/PhysRevD.105.086016}{\emph{Phys. Rev. D}
  {\bfseries 105} (2022) 086016}
  [\href{https://arxiv.org/abs/2108.04637}{{\ttfamily 2108.04637}}].

\bibitem{Kawabata:2021hac}
K.~Kawabata, T.~Nishioka, Y.~Okuyama and K.~Watanabe, \emph{{Probing Hawking
  radiation through capacity of entanglement}},
  \href{https://doi.org/10.1007/JHEP05(2021)062}{\emph{JHEP} {\bfseries 05}
  (2021) 062} [\href{https://arxiv.org/abs/2102.02425}{{\ttfamily
  2102.02425}}].

\bibitem{BasakKumar:2022stg}
J.~Basak~Kumar, D.~Basu, V.~Malvimat, H.~Parihar and G.~Sengupta,
  \emph{{Reflected Entropy and Entanglement Negativity for Holographic Moving
  Mirrors}},  \href{https://arxiv.org/abs/2204.06015}{{\ttfamily 2204.06015}}.

\bibitem{Horowitz:2003he}
G.~T. Horowitz and J.~M. Maldacena, \emph{{The Black hole final state}},
  \href{https://doi.org/10.1088/1126-6708/2004/02/008}{\emph{JHEP} {\bfseries
  02} (2004) 008} [\href{https://arxiv.org/abs/hep-th/0310281}{{\ttfamily
  hep-th/0310281}}].

\bibitem{Akal:2021dqt}
I.~Akal, T.~Kawamoto, S.-M. Ruan, T.~Takayanagi and Z.~Wei, \emph{{On the Page
  curve under final state projection}},
  \href{https://arxiv.org/abs/2112.08433}{{\ttfamily 2112.08433}}.

\bibitem{Nakata:2021ubr}
Y.~Nakata, T.~Takayanagi, Y.~Taki, K.~Tamaoka and Z.~Wei, \emph{{New
  holographic generalization of entanglement entropy}},
  \href{https://doi.org/10.1103/PhysRevD.103.026005}{\emph{Phys. Rev. D}
  {\bfseries 103} (2021) 026005}
  [\href{https://arxiv.org/abs/2005.13801}{{\ttfamily 2005.13801}}].

\bibitem{Rajabpour:2015uqa}
M.~A. Rajabpour, \emph{{Post measurement bipartite entanglement entropy in
  conformal field theories}},
  \href{https://doi.org/10.1103/PhysRevB.92.075108}{\emph{Phys. Rev. B}
  {\bfseries 92} (2015) 075108}
  [\href{https://arxiv.org/abs/1501.07831}{{\ttfamily 1501.07831}}].

\bibitem{Rajabpour:2015xkj}
M.~A. Rajabpour, \emph{{Entanglement entropy after a partial projective
  measurement in $1+1$ dimensional conformal field theories: exact results}},
  \href{https://doi.org/10.1088/1742-5468/2016/06/063109}{\emph{J. Stat. Mech.}
  {\bfseries 1606} (2016) 063109}
  [\href{https://arxiv.org/abs/1512.03940}{{\ttfamily 1512.03940}}].

\bibitem{Numasawa:2016emc}
T.~Numasawa, N.~Shiba, T.~Takayanagi and K.~Watanabe, \emph{{EPR Pairs, Local
  Projections and Quantum Teleportation in Holography}},
  \href{https://doi.org/10.1007/JHEP08(2016)077}{\emph{JHEP} {\bfseries 08}
  (2016) 077} [\href{https://arxiv.org/abs/1604.01772}{{\ttfamily
  1604.01772}}].

\bibitem{Miyaji:2014mca}
M.~Miyaji, S.~Ryu, T.~Takayanagi and X.~Wen, \emph{{Boundary States as
  Holographic Duals of Trivial Spacetimes}},
  \href{https://doi.org/10.1007/JHEP05(2015)152}{\emph{JHEP} {\bfseries 05}
  (2015) 152} [\href{https://arxiv.org/abs/1412.6226}{{\ttfamily 1412.6226}}].

\bibitem{Affleck:1991tk}
I.~Affleck and A.~W.~W. Ludwig, \emph{{Universal noninteger 'ground state
  degeneracy' in critical quantum systems}},
  \href{https://doi.org/10.1103/PhysRevLett.67.161}{\emph{Phys. Rev. Lett.}
  {\bfseries 67} (1991) 161}.

\bibitem{Calabrese:2004eu}
P.~Calabrese and J.~L. Cardy, \emph{{Entanglement entropy and quantum field
  theory}}, \href{https://doi.org/10.1088/1742-5468/2004/06/P06002}{\emph{J.
  Stat. Mech.} {\bfseries 0406} (2004) P06002}
  [\href{https://arxiv.org/abs/hep-th/0405152}{{\ttfamily hep-th/0405152}}].

\bibitem{Calabrese:2009qy}
P.~Calabrese and J.~Cardy, \emph{{Entanglement entropy and conformal field
  theory}}, \href{https://doi.org/10.1088/1751-8113/42/50/504005}{\emph{J.
  Phys. A} {\bfseries 42} (2009) 504005}
  [\href{https://arxiv.org/abs/0905.4013}{{\ttfamily 0905.4013}}].

\bibitem{Holzhey:1994we}
C.~Holzhey, F.~Larsen and F.~Wilczek, \emph{{Geometric and renormalized entropy
  in conformal field theory}},
  \href{https://doi.org/10.1016/0550-3213(94)90402-2}{\emph{Nucl. Phys. B}
  {\bfseries 424} (1994) 443}
  [\href{https://arxiv.org/abs/hep-th/9403108}{{\ttfamily hep-th/9403108}}].

\bibitem{Karch:2020iit}
A.~Karch and L.~Randall, \emph{{Geometries with mismatched branes}},
  \href{https://doi.org/10.1007/JHEP09(2020)166}{\emph{JHEP} {\bfseries 09}
  (2020) 166} [\href{https://arxiv.org/abs/2006.10061}{{\ttfamily
  2006.10061}}].

\bibitem{Akal:2020wfl}
I.~Akal, Y.~Kusuki, T.~Takayanagi and Z.~Wei, \emph{{Codimension two holography
  for wedges}}, \href{https://doi.org/10.1103/PhysRevD.102.126007}{\emph{Phys.
  Rev. D} {\bfseries 102} (2020) 126007}
  [\href{https://arxiv.org/abs/2007.06800}{{\ttfamily 2007.06800}}].

\bibitem{Mollabashi:2021xsd}
A.~Mollabashi, N.~Shiba, T.~Takayanagi, K.~Tamaoka and Z.~Wei, \emph{{Aspects
  of pseudoentropy in field theories}},
  \href{https://doi.org/10.1103/PhysRevResearch.3.033254}{\emph{Phys. Rev.
  Res.} {\bfseries 3} (2021) 033254}
  [\href{https://arxiv.org/abs/2106.03118}{{\ttfamily 2106.03118}}].

\bibitem{Nishioka:2021cxe}
T.~Nishioka, T.~Takayanagi and Y.~Taki, \emph{{Topological pseudo entropy}},
  \href{https://doi.org/10.1007/JHEP09(2021)015}{\emph{JHEP} {\bfseries 09}
  (2021) 015} [\href{https://arxiv.org/abs/2107.01797}{{\ttfamily
  2107.01797}}].

\bibitem{Miyaji:2021lcq}
M.~Miyaji, \emph{{Island for gravitationally prepared state and pseudo
  entanglement wedge}},
  \href{https://doi.org/10.1007/JHEP12(2021)013}{\emph{JHEP} {\bfseries 12}
  (2021) 013} [\href{https://arxiv.org/abs/2109.03830}{{\ttfamily
  2109.03830}}].

\bibitem{Ginsparg:1993is}
P.~H. Ginsparg and G.~W. Moore, \emph{{Lectures on 2-D gravity and 2-D string
  theory}},  in \emph{{Theoretical Advanced Study Institute (TASI 92): From
  Black Holes and Strings to Particles}}, pp.~277--469, 10, 1993,
  \href{https://arxiv.org/abs/hep-th/9304011}{{\ttfamily hep-th/9304011}}.

\bibitem{Polyakov:1981rd}
A.~M. Polyakov, \emph{{Quantum Geometry of Bosonic Strings}},
  \href{https://doi.org/10.1016/0370-2693(81)90743-7}{\emph{Phys. Lett. B}
  {\bfseries 103} (1981) 207}.

\bibitem{Caputa:2017urj}
P.~Caputa, N.~Kundu, M.~Miyaji, T.~Takayanagi and K.~Watanabe, \emph{{Anti-de
  Sitter Space from Optimization of Path Integrals in Conformal Field
  Theories}}, \href{https://doi.org/10.1103/PhysRevLett.119.071602}{\emph{Phys.
  Rev. Lett.} {\bfseries 119} (2017) 071602}
  [\href{https://arxiv.org/abs/1703.00456}{{\ttfamily 1703.00456}}].

\bibitem{Caputa:2017yrh}
P.~Caputa, N.~Kundu, M.~Miyaji, T.~Takayanagi and K.~Watanabe, \emph{{Liouville
  Action as Path-Integral Complexity: From Continuous Tensor Networks to
  AdS/CFT}}, \href{https://doi.org/10.1007/JHEP11(2017)097}{\emph{JHEP}
  {\bfseries 11} (2017) 097}
  [\href{https://arxiv.org/abs/1706.07056}{{\ttfamily 1706.07056}}].

\end{thebibliography}\endgroup

\end{document}